# Commentari di Ipparco ai Fenomeni di Arato ed Eudosso

**Gabriele Vanin, Bruna Cusinato**

**Abstract.** This is the first translation of *Hipparchus' Commentaries on the Phenomena of Aratus and Eudoxus* into a modern language since the one, in German, published by the philologist Karl Heinrich August Manitius in 1894.
Published in 2013, it is now also available online.
This is also the first detailed astronomical commentary on the work, produced with the support of *Starry Night Pro Plus 8* software (very reliable for both precession and stellar proper motions), configured for the epochs and sites where Aratus, Eudoxus and Hipparchus worked.
The translation from Greek is by Bruna Cusinato.
The introduction and astronomical commentary are by Gabriele Vanin.

Third revised and updated edition: June 2022
(First edition: December 2013)

Composed in Times New Roman PS MT 11/12 pt

Version of 21 june 2022

We would like to thank Filippomaria Pontani, Raffaela Salvadori, Francesca Gola and Fabio Acerbi for their help with translations from Greek, Johannes Keinzel, Chiara Salata and Carlotta Poloni for help with translations from German, Emiliano Ricci for suggestions on using the Starry Night software.

for comments and communications: rheticus@tiscali.it

# Introduzione

## I contributi di Ipparco all'astronomia

Per gran parte degli storici moderni, a partire da Delambre (1817), Ipparco è stato l'autore sostanziale della maggior parte del monumentale lavoro astronomico contenuto nell'*Almagesto* di Tolomeo. Per comprendere la reale portata dei suoi contributi sembrava quindi necessario provare a scartare ciò che nell'*Almagesto* poteva essere stato aggiunto da Tolomeo. Tutto ciò ha portato a misconoscere completamente la sua opera, che tenteremo di ricostruire per sommi capi nel seguito.
Ipparco nacque a Nicea, in Bitinia, l'attuale Iznik, cittadina di 23 000 abitanti nella Turchia nordoccidentale. Non sappiamo quasi nulla della sua vita e della sua attività, se non quanto ci riferisce su di lui, in particolare sulle sue osservazioni e sulle sue opere, Tolomeo. Stando a riferimenti di autori posteriori Ipparco stesso redasse un catalogo delle sue opere, 14, ma nessuna di esse, a parte i *Commentari*, ci è pervenuta. La prima osservazione ipparchea riferita da Tolomeo è del 147 a.C., l'ultima del 127. Altre osservazioni più antiche riferite dal grande alessandrino, ma non sicuramente attribuibili ad Ipparco, risalgono al 162. Possiamo perciò ragionevolmente collocare la sua attività scientifica all'incirca fra il 165 e il 125 a.C., la sua nascita nel primo quarto del secondo secolo, la sua morte dopo il 127. Sempre secondo Tolomeo, Ipparco deve aver trascorso la prima parte della sua carriera in Bitinia, almeno fino al 141, e poi si trasferì a Rodi. Altri riferimenti a Ipparco si possono trovare ovviamente in altri autori dell'antichità ma, non trattandosi di astronomi, spesso essi hanno travisato o mal compreso i suoi apporti. Nel riferire i quali, di seguito, una certa dose di incertezza e congettura dev'essere messa in conto.
Prima di Ipparco l'astronomia greca aveva tentato di spiegare i movimenti dei corpi celesti utilizzando dei modelli qualitativi. Ipparco cercò per primo, invece, di trasformare l'astronomia in una scienza quantitativa e fu anche il primo astronomo, con tutta probabilità, a osservare il cielo in modo sistematico. Sviluppò metodi matematici in grado di usare modelli geometrici ai fini delle previsioni pratiche, e assegnò parametri numerici a questi modelli, utilizzando ampiamente a questi fini osservazioni e costanti di provenienza babilonese.
Dal punto di vista matematico Ipparco è stato il primo a costruire una tavola delle corde e quindi a provvedere una soluzione generale per i problemi di trigonometria piana. Non esistendo ancora la trigonometria sferica (usata invece da Menelao, primo sec. d.C., e da Tolomeo) riuscì a risolvere i problemi di trigonometria sulla sfera utilizzando metodi aritmetici o di geometria descrittiva: nel primo caso riuscì per esempio a stabilire i rapporti fra lunghezza massima e minima del giorno per le diverse latitudini, nel secondo quali stelle sorgono e tramontano simultaneamente per quasi tutti i luoghi del mondo. Nell'ambito delle teorie planetarie Ipparco usò sia il modello a epiciclo-deferente che quello a eccentrici e tentò di determinare parametri numerici per entrambi sulla base delle osservazioni. Avendo stabilito, per mezzo di queste, la lunghezza di ciascuna stagione, e assumendo per il Sole un modello a eccentrico con apogeo fisso, fu in grado di stabilire l'eccentricità dell'orbita solare e la posizione dell'apogeo. Per quanto riguarda la teoria lunare, Ipparco stabilì innan-zitutto le durate dei mesi siderale, sinodico, anomalistico, utilizzando sia le proprie osservazioni, sia quelle babilonesi, in particolare di eclissi, a partire già dal 747 a.C., essendo il primo astronomo greco ad accedere a questo materiale. Per rappresentare il moto lunare usò un modello a epiciclo e deferente, del quale determinò i parametri utilizzando l'osservazione dei tempi di solo tre eclissi lunari. Purtroppo, anche se elegante, il modello risultava troppo semplificato e funzionava bene solo alle sizigie (novilunio e plenilunio), mentre nelle altre posizioni lunari la discrepanza fra posizioni previste e calcolate era notevole.
Da proprie osservazioni con la diottra[1] Ipparco stabilì che la Luna, alla sua distanza media, ha lo stesso diametro apparente del Sole, che questo diametro apparente è di 0,55°, che l'ombra della Terra, alla distanza della Luna, è due volte e mezzo il diametro del satellite. Inoltre, utilizzando le osservazioni prese da due località diverse dell'eclisse di Sole del 14 marzo 190 a.C., riuscì a stabilire, con ingegnosi procedimenti geometrici e trigonometrici, le distanze della Luna e del Sole. Per la prima trovò un valore compreso fra 59 e 67⅓ raggi terrestri (il valore reale è 60,27), per il secondo 490 raggi terrestri (valore reale 23 456). Ovviamente, con questi dati in mano, l'astronomo era anche in grado di calcolare le dimensioni di Sole e Luna. Tuttavia, secondo Teone Smirneo, Ipparco pensava che la Luna fosse in volume 1/27 della Terra, valore compatibile con quello trovato per la distanza, ma che il Sole fosse in volume 1880 volte la Terra: avendo quindi un diametro 37 volte quello della Luna, doveva trovarsi ad una distanza di circa 2500 raggi terrestri.
Grazie anche a questi risultati Ipparco fu in grado di arrivare a predire le eclissi, sia di Sole che di Luna, per un determinato luogo, e probabilmente compose delle vere e proprie tavole di previsione di questi fenomeni. Plinio il Vecchio dice addirittura che aveva costruito tavole di predizione per 600 anni, ma probabilmente Ipparco aveva soltanto compilato un catalogo di tutte le eclissi registrate dai Babilonesi e altre fonti, da quella citata del 747 fino ai suoi giorni (appunto 600 anni).
Per quanto riguarda la teoria planetaria Tolomeo ci dice che Ipparco rinunciò a ogni tentativo di migliorare le ipotesi in voga ai suoi tempi, anche se faceva notare che la teoria epiciclica di Apollonio prediceva archi di retrogradazione costanti per ogni pianeta, mentre l'osservazione mostrava che essi erano variabili. Lo stesso Tolomeo ci informa però che Ipparco assemblò in una forma più consona una mole consistente di osservazioni riguardanti i pianeti, probabilmente di origine babilonese.

## Il perduto catalogo stellare di Ipparco

Nei *Commentari* troviamo alcune coordinate posizionali stellari espresse in termini di distanze polari, distanze dall'equatore celeste (declinazioni), distanze dai tropici e ascensioni rette, una serie di levate, passaggi in meridiano, tramonti delle varie parti dell'eclittica (contrassegnate da precisi valori in gradi) ad una latitudine dove il giorno più lungo dura 14,5 ore, un elenco di stelle brillanti che giacciono su o vicino a 24

1. Strumento formato genericamente da una linea di mira, costruita in metallo o legno, a cui erano applicate, ai due estremi, due alette dette *pinnule*. Le pinnule potevano essere entrambe forate, oppure la prima, quella vicina all'occhio, forata, e la seconda, quella distante, poteva essere ricavata a forma di merlo (una scanalatura a forma di "V"), oppure recare un crocicchio, ma erano possibili anche altre soluzioni. La diottra usata da Ipparco per questa osservazione in particolare doveva avere la pinnula vicina all'occhio sostituita da un piccolo cilindro, che poteva scorrere avanti e indietro, fino a coprire il diametro apparente del Sole o della Luna.

circoli orari. Le indicazioni di ascensione retta sono date in termini però inusuali; per esempio Ipparco dice che una stella “si trova sul terzo grado del Leone lungo il suo circolo parallelo”. Questo sta a significare che l’equatore, e ciascun circolo parallelo ad esso, viene diviso in 12 “segni” di 30° ciascuno: in questi termini l’ascensione retta della stella poc’anzi citata sarebbe di 123°, ovvero $8^h12^m$. In effetti anche la determinazione delle stelle orarie è un’indicazione di ascensione retta. Secondo Toomer pertanto i *Commentari* non sono un’opera giovanile di Ipparco, ma rivelano un astronomo già abbastanza esperto, e pertanto egli ritiene che siano probabilmente posteriori al famoso e celebrato catalogo stellare dell’astronomo di Nicea.
Riguardo a questo, il riferimento più esplicito è contenuto in Plinio, che riferisce come, avendo Ipparco notato una nuova stella, e avendo visto che si muoveva, cominciò a sospettare che anche le altre potessero farlo e quindi decise di enumerare e denominare le stelle per la posterità, inventando degli strumenti atti allo scopo. Ciò però potrebbe riferirsi non ad un catalogo stellare ma ad un semplice elenco di stelle e delle loro posizioni all’interno di ciascuna costellazione (senza coordinate). In effetti Tolomeo, volendo mostrare che la posizione propria delle stelle non era cambiata dai tempi di Ipparco, riportò una lista di allineamenti stellari che presumibilmente proprio Ipparco aveva determinato, per gli scopi citati da Plinio. Come già ricordato a proposito di Eratostene, nel 1892 Ernst Maass trovò due fino allora sconosciuti elenchi greci di costellazioni, uno attribuito ad Eratostene, l’altro ad Ipparco. Essi riportavano nient’altro che i nomi delle costellazioni. Ma nel 1898 Alessandro Olivieri e nel 1901 il filologo tedesco Franz Boll scoprirono degli elenchi che riportavano anche il numero delle stelle in ciascuna costellazione. In seguito ne furono scoperti altri simili, per un totale di non meno di 11 manoscritti, nove greci e due latini, alcuni anonimi, altri attribuiti ad Ipparco. Il manoscritto più completo è parte del Cod. Parisinus 2506, un manoscritto astrologico del XIV secolo. Riportiamo la lista che ne diede Boll in un suo lavoro del 1901:

| Ursa Major | 24 | Pegasus | 18 | Lupus | 13 | Virgo | 19 |
|---|---|---|---|---|---|---|---|
| Ursa Minor | 7 | Cepheus | 19 | Ara | 4 | Libra | 4 |
| Draco | 15 | Cassiopea | 14 | Corona Australis | ... | Scorpio | 15 |
| Bootes | 19 | Andromeda | 20 | Piscis Australis | 12 | Sagittarius | 16 |
| Corona Borealis | 9 | Triangulum | 3 | Cetus | 14 | Capricornus | 26 |
| Hercules | 24 | Perseus | 19 | Eridanus | ... | Aquarius | 18 |
| Ophiuchus | 17 | Auriga | 8 | Orion | 18 | Pesci | 41 |
| Lyra | 10 | Hydra | 27 | Lepus | ... | Aries | 17 |
| Cygnus | 14 | Crater | 10 | Canis Major | 21 | Taurus | 18 |
| Aquila | 4 | Corvus | 7 | Canis Minor | 3 | Gemini | 19 |
| Sagitta | 4 | Argo | 13 | Cancer | 16 | | |
| Delphinus | 9 | Centaurus | 26 | Leo | 19 | | |

In tutto le costellazioni elencate sono 46 e il numero totale delle stelle 653. Curiosamente, per 24 costellazioni il numero delle stelle è lo stesso che nei *Catasterismi* di Eratostene. Delle costellazioni tolemaiche mancano solo il Serpente e la Testa del Cavallo. Poiché sono riportati i numeri di stelle di sole 43 costellazioni se ne può arguire che, se fossero state riportate tutte, nonché anche le stelle non figurate fra una costellazione e l'altra, il totale contenuto nel catalogo originale sarebbe stato di circa 850. Boll e l'altro filologo tedesco Albert Rehm scrissero che era molto probabile la paternità ipparchea del lavoro originale da cui questa lista era tratta, e anche Otto Neugebauer sembrò, più tardi, dare credito a questa idea.
Un'altra traccia del perduto catalogo stellare di Ipparco sembra essere venuta alla luce nel 1936, quando il filologo tedesco Wilhelm Gundel trovò, in due versioni del trattato astrologico *Liber Hermetis Trismegisti*, una in francese antico, del XIV secolo, e una in latino, del 1431, una lista di 68 stelle con longitudini corrispondenti all'epoca di Ipparco. La terminologia usata, e il profilo delle costellazioni, è simile a quella dei *Commentari*, ma questi tuttavia non contengono longitudini eclittiche. Tolomeo cita anche alcune declinazioni ipparchee, anche se non è detto che queste siano state prese dal catalogo di Ipparco: per esempio potrebbero essere state desunte da uno dei suoi lavori sulla precessione. Infine, in un tardo scolio latino di Arato, pubblicato da Ernst Maass, sono stati trovati dei valori di coordinate polari, per le costellazioni circumpolari, approssimativamente corrette per l'epoca di Ipparco.
Secondo Toomer è probabile che le coordinate stellari date nei *Commentari* siano state prese dal catalogo stellare di Ipparco. Al contrario, Neugebauer ritiene che il catalogo di Ipparco sia posteriore ai *Commentari*, non solo, ma anche posteriore alla scoperta della precessione degli equinozi e che proprio per questo contenesse le coordinate stellari espresse sotto forma di longitudine e latitudine eclittiche (cosa che invece nei *Commentari* non è): infatti, poiché per la precessione le latitudini rimangono invariate e variano solo le longitudini, lo stesso catalogo può essere comodamente utilizzato anche nei secoli successivi, semplicemente aggiungendo la costante di precessione alla longitudine di ciascuna stella.

## La scoperta della precessione

Ipparco è appunto universalmente noto per la scoperta della precessione degli equinozi. A causa di questo fenomeno mutano nel tempo le longitudini eclittiche delle stelle, e l'anno tropico (ritorno del Sole sullo stesso punto equinoziale o solstiziale) risulta più corto dell'anno siderale (ritorno del Sole nella stessa posizione rispetto alle stelle). Non si sa quale dei due fatti abbia condotto per primo Ipparco a sospettare l'esistenza della precessione, ma entrambi contribuirono a confermare il sospetto. Inizialmente pensò che a muoversi fossero soltanto le stelle posizionate lungo l'eclittica, e solo in seguito ipotizzò che tale comportamento fosse comune a tutte le stelle. Ci riuscì in due modi distinti. Dapprima confrontò le posizioni di Spica prese durante misure di elongazione della stella dalla Luna in occasione di eclissi del satellite osservate da Timocari e da lui stesso. Ipparco concluse che la longitudine di Spica era aumentata di circa 2° in 160 anni (non lontano dal valore corretto di 2°13'). Trovò più o meno lo stesso valore anche dal confronto con le posizioni di altre stelle brillanti osservate ai tempi di Timocari e Aristillo.[2] Queste misure conducono a un valore precessionale di 45” per anno. Poi Ipparco cercò di misurare la lunghezza dell’anno tropico, confrontando i solstizi osservati da lui nel 135 con quelli osservati da Aristarco nel 280 e Metone nel 432. Trovò il valore di 365¼ giorni meno almeno 1/300 di giorno (valore più lungo di soli $6^m 17^s$ rispetto a quello reale); assumendo per la lunghezza dell’anno siderale il valore, mutuato dai Babilonesi, di 365¼ più 1/144 di giorno, trovò che il punto equinoziale precedeva di almeno un grado per secolo, ovvero 36” all’anno.[3] Anche se il primo valore era molto più vicino al vero (che è 50,26” all’anno), Tolomeo adottò il secondo, anche perché dalle sue proprie osservazioni trovò che la longitudine di Regolo, dai tempi di Ipparco (un intervallo di 265 anni) era aumentata di 2⅔°, ovvero proprio di 36” all’anno, valore confermato da diverse altre misure effettuate dall’astronomo alessandrino sulle altre stelle brillanti, zodiacali e non. Occorre dire, tuttavia, che già Anton Pannekoek nel 1955 mostrò che Tolomeo, delle 18 stelle le cui misure di declinazione registrate da Ipparco, Timocari e lui stesso, considerò per ricavare la costante di precessione, ne utilizzò solo sei, che consentivano di ricavare un valore di 38”. Se le avesse usate tutte, avrebbe trovato un valore di 46,2”, anche questo molto più vicino al vero.

## Altri contributi

L’unico strumento il cui uso Tolomeo assegna ad Ipparco è la diottra, con la quale l’astronomo di Nicea misurò probabilmente anche i tempi degli equinozi e dei solstizi, nonché le coordinate stellari. Forse egli inventò anche l’astrolabio planisferico, il principale strumento astronomico usato in seguito dagli astronomi arabi, con il quale risultava agevole calcolare le levate e i tramonti simultanei. Come si evince chiaramente dai *Commentari* utilizzò, e forse costruì, dei globi celesti, probabilmente in connessione col suo catalogo stellare, che gli servivano sia per fini dimostrativi, sia per il calcolo delle levate e tramonti simultanei, sia per risolvere molte altre questioni attinenti all’astronomia di posizione.

Nel campo della geografia Ipparco criticò Eratostene, sostenendo che le distanze e le relazioni fornite per i punti stabiliti sulla carta del mondo costruita da questi erano inconsistenti fra loro e con altri dati geografici. Tuttavia, la geografia astronomica di Ipparco era basata su semplici schemi aritmetici e, pur sostenendo la necessità che le longitudini fossero misurate per mezzo delle eclissi di Luna, non risulta che abbia dato applicazione a questo principio.

Secondo le fonti esistenti Ipparco era un’autorità anche in campo astrologico, soprattutto nel campo della geografia astrologica, che assegna le varie regioni del mondo all’influenza di un particolare segno zodiacale o di una sua parte. Si occupò anche di “pronostici” meteorologici collegandoli, come Arato e Gemino, al sorgere e al tramonto delle stelle fisse, di ottica, di caduta dei gravi, di aritmetica combinatoria.

Tolomeo ci ha lasciato anche dei particolari sulla personalità scientifica di Ippar-co, definendolo a più riprese “amante della verità” e sottolineando la sua apertura mentale, la sua capacità di sottoporre a verifica osservativa la sua e le altrui teorie, di esaminare criticamente tutte le ipotesi e di accettare punti di vista non ortodossi. Degno di nota fu anche il suo atteggiamento a considerare l’astronomia come una scienza bisognosa di osservazioni prolungate nel tempo per più generazioni ai fini di assicurarsi più solide basi, da cui i suoi tentativi di mettere assieme materiale osservativo di varia provenienza per uso della posterità.
Curiosamente, anche se Ipparco acquisì una grandissima reputazione nell’antichità, le sue opere non furono molto lette al di fuori della cerchia degli specialisti, essenzialmente perché erano generalmente in forma di monografie abbastanza scarne, riguardanti argomenti molto diversi fra loro e spesso anche molto tecnici, simili agli articoli referati delle moderne riviste scientifiche. Egli non elaborò mai teorie astronomiche partendo dai principi primi e non costruì sistemi astronomici, ma piuttosto li verificò ed elaborò qualche loro parte. Così, quando Tolomeo, utilizzando fra l’altro i lavori di Ipparco come fondamenta essenziali, costruì un tale sistema, inserendolo in un libro che costituiva un vero e proprio manuale di astronomia, sia pure a livello specialistico, l’interesse per i contributi originali di Ipparco declinò, e forse questo è il motivo per cui non ci sono pervenuti, a parte i *Commentari*, che devono probabilmente la loro sopravvivenza alla popolarità di Arato.

---

2. Di costoro, vissuti all’incirca fra il 320 e il 260 a.C., non sappiamo nulla se non appunto le citazioni delle misure da loro effettuate contenute nell’*Almagesto*. Operativi ad Alessandria d’Egitto furono probabilmente i primi a compiere osservazioni astronomiche posizionali abbastanza precise.

3. Infatti, essendo la differenza fra anno tropico e siderale di 1/100 di giorno, per Ipparco l’equinozio si spostava ogni anno rispetto alle stelle di sfondo di 1/36 500 del percorso del Sole (360°) sulla volta celeste, ovvero appunto circa 36”.

## I *Commentari*

L’opera che presentiamo ci è stata tramandata, parzialmente o totalmente, in 16 codici, databili fra il XIV e il XVI secolo, con l’eccezione di uno dell’XI secolo. In cinque di essi è presente un titolo, che varia anche di molto da caso a caso. Manitius, compendiandoli, ha intitolato la sua edizione critica: Ἱππάρχου τῶν Ἀράτου καὶ Εὐδόξου Φαινομένων Ἐξηγήσεως βιβλία τρία, ovvero *I tre libri del Commentario di Ipparco ai Fenomeni di Arato ed Eudosso*. Noi abbiamo scelto la dizione al plurale *Commentari* perché adottata nelle prime edizioni a stampa, dalla maggior parte degli studiosi e dallo stesso Manitius nella traduzione latina del titolo. La prima edizione a stampa di quest’opera, in greco, fu pubblicata a Firenze nel 1567 da Piero Vettori, il più importante studioso italiano di letteratura classica del suo tempo, assieme ad una edizione del catalogo stellare di Tolomeo che Vettori attribuì erroneamente a Ipparco, e al commento ai *Fenomeni* di Achille Tazio. La prima traduzione latina fu pubblicata a Parigi nel 1630 dall’insigne teologo gesuita francese Denis Pétau all’interno della sua opera *Uranologion*, assieme al commento di Tazio, all’*Introduzione ai fenomeni* di Gemino, alle *Fasi delle stelle fisse* di Tolomeo, e ad altre opere. Una seconda edizione dell’*Uranologion* fu pubblicata da Jacques Paul Migne nel 1857 a Parigi, come appendice al XIX volume della sua *Patrologia Graeca*. Inoltre fra il 1888 e il 1898 Ernst Maass pubblicò tutti i commentari ad Arato che ci sono pervenuti, quindi anche il libro primo e parte del libro secondo di Ipparco. La prima edizione critica dei *Commentari* è quella pubblicata, assieme alla prima traduzione in una lingua moderna (tedesco), dal filologo Karl Heinrich August Manitius nel 1894.
L’opera è divisa in tre libri. Il primo libro e la prima parte del secondo sono dedicati al commento e ad una critica serrata non solo dei *Fenomeni* di Arato, ma anche di quelli di Eudosso e del commento ad Arato, oggi perduto, composto dal matematico

Attalo di Rodi poco tempo prima di Ipparco. La seconda parte del secondo libro e il terzo libro sono dedicati all'esposizione propria da parte di Ipparco delle levate e dei tramonti delle principali costellazioni per una latitudine di 36° N.
Sicuramente la lettura della prima parte dell'opera lascia molto perplessi poiché, come vedremo, in molti punti è eccessivamente critica nei confronti degli autori commentati. Delambre a suo tempo affermò di non aver trovato in queste critiche niente di amaro e di ingiusto e tese a dare ragione ad Ipparco, pensando che Eudosso non avesse potuto disporre di strumenti astronomici strutturati o che si fosse limitato a parlare di cose fatte da altri, copiando. A noi sembra invece che i giudizi di Ipparco siano spesso troppo severi e i suoi rilievi inutilmente puntigliosi. Inoltre la trattazione è scarsamente lineare ed omogenea, e talvolta assistiamo a dei veri voli pindarici fra gli argomenti, che rendono difficile ritrovare il bandolo della matassa. Tanto più che lo stesso Ipparco non è immune da errori, anzi, se ne trovano così tanti che è lecito chiedersi se l'autore di questo commento e l'astronomo così celebrato da Tolomeo siano la stessa persona. È sicuramente possibile che l'eccesso di zelo abbia in qualche caso fatto passare la misura ad Ipparco, ma probabilmente molti dati sono incomprensibili se non si ammette che ci possano essere state delle interpolazioni e degli errori nella trasmissione delle copie dell'opera. È anche possibile che in diversi casi le critiche siano dovute al fatto che nelle varie epoche le costellazioni avevano forma diversa. Già nel 1936 Gundel fece notare come vi fossero diverse differenze fra le descrizioni delle costellazioni contenute nel *Liber Hermetis Trismegisti* e quelle dell'*Almagesto*. D'altra parte lo stesso Tolomeo afferma che la sua descrizione, "più naturale e proporzionata", è differente da quella dei suoi predecessori, come quella a sua volta è differente da quelle ancora più antiche. E nel commento vedremo alcune di queste differenze.
Si può senz'altro dire, invece, che alla base delle critiche di Ipparco non vi può essere la differenza nelle posizioni stellari fra l'epoca di Eudosso e quella di Ipparco dovuta alla precessione, poiché i riferimenti di Arato ed Eudosso sono piuttosto generici, non si prefiggono una grande accuratezza, la maggior parte dei fenomeni descritti non è cambiata sensibilmente fra le due epoche, e le obiezioni di Ipparco, pur se in molti casi comprendono la citazione di precise coordinate, sono su errori macroscopici, sulla cui causa, come diciamo in altra parte, ancora oggi ci interroghiamo senza trovare una spiegazione.
Il resto dell'opera di Ipparco, la seconda parte del secondo e il terzo libro, anche se non immune da errori, è certamente più godibile; in particolare, mentre gli autori precedenti fornivano soltanto elenchi di levate e tramonti di stelle e costellazioni, Ipparco fornisce le longitudini dei punti dell'eclittica che sorgono, culminano e tramontano, nonché quali stelle passano in meridiano, quando sorgono e tramontano le varie stelle poste all'inizio e alla fine di ogni costellazione. Inoltre fornisce, alla fine, un elenco di stelle brillanti che giacciono su o vicino ai 24 circoli orari ai fini di determinare più precisamente l'ora di notte.

## La datazione dei *Commentari*

Le coordinate astronomiche fornite da Ipparco sono sufficientemente precise da consentirci di datare, tenendo presente la costante precessionale, la redazione dei *Commentari*. Abbiamo utilizzato a questo scopo il software *Starry Night* che è molto affidabile sia per quanto riguarda la precessione che i moti propri stellari. Escludendo i dati relativi alle levate e ai tramonti, più soggetti per loro natura a imprecisione, e limitandoci alle indicazioni di declinazione, distanza polare e ascensione retta (96 dati), troviamo un'epoca centrata sul 137 ± 8 a.C.[4] Se però escludiamo i valori più estremi, e ci obblighiamo invece a considerare quelli compresi rispettivamente all'interno dei limiti della vita (56 dati) e dell'attività (38 dati) di Ipparco, troviamo rispettivamente 152 ± 3 e 147 ± 2 a.C. Questi ultimi dati sono molto simili a quelli trovati dall'astronomo tedesco Heinrich Vogt, che nel 1925, usando le coordinate di 77 stelle vicine all'eclittica o all'equatore celeste, trovò una datazione del 151 ± 4 a.C, ma escludendo 10 stelle con un errore di oltre 65 anni dalla media dei risultati. Gli astronomi francesi Nadal e Brunet nel 1989, utilizzando le coordinate di 78 stelle ricavate tramite un controllo incrociato fra le loro levate, culminazioni e tramonti, trovarono un valore di 141 ± 25 a.C. Pare pertanto possibile assegnare ai *Commentari* una datazione attorno al 150 a.C., confermandola quindi non un'opera giovanile, ma della maturità di Ipparco.
Se invece si prova ad utilizzare, per una datazione, le 18 declinazioni stellari di Ipparco riportate da Tolomeo nell'*Almagesto*, troviamo un valore di 122 ± 20 a.C.; escludendo le cinque più lontane dalla media si trova un valore di 132 ± 5 a.C, molto simile a quello indicato da Vogt, 131 ± 6 a.C., che utilizzò 16 declinazioni, escludendo le quattro dell'*Almagesto* con un errore di oltre 52 anni ma prendendone due dalla *Geografia* di Tolomeo e da Strabone. In tutti i casi sembra esserci un certo intervallo di tempo fra la redazione dei *Commentari* e la compilazione delle coordinate citate da Tolomeo, e questo sembra suggerire che Ipparco compilò il suo catalogo stellare successivamente, e non di poco, alla redazione dei *Commentari*. Infatti Tolomeo deve aver preso queste coordinate all'interno di un'opera scientifica di Ipparco, probabilmente proprio il suo catalogo stellare, e non certo in un testo di critica letteraria, sia pure su base scientifica.

## Osservazione astronomica e globi celesti

Riteniamo che i rilievi contenuti nell'opera sulle levate, i tramonti e i passaggi in meridiano siano stati eseguiti utilizzando un globo celeste, mentre le coordinate stellari siano frutto di osservazione diretta. Infatti, i numerosi riferimenti al grado di ciascun segno che sorge, tramonta o passa in meridiano non possono avere a che fare col cielo reale, poiché la partizione dell'eclittica in 12 parti uguali è un artificio che si può tracciare su un globo, ma che non ha controparte in cielo. Anche i riferimenti agli altri cerchi sono sicuramente meglio visualizzabili e posizionabili utilizzando un globo celeste, naturalmente costruito in modo accurato e sufficientemente grande.

## Identificazione delle stelle e precisione astronomica

Si è proceduto, quando possibile, ad identificare le stelle citate nell'opera di Ipparco, utilizzando il *software Starry Night*, ai fini di aiutare quegli studiosi che volessero

4. Qui e nei dati successivi l'incertezza è la deviazione standard della media.

controllare a loro volta i dati dell'astronomo. Le identificazioni sono riportate fra parentesi quadre con i rispettivi nomi propri, con riferimento al libro *I nomi delle stelle* di Vanin e, negli altri casi, facendo ricorso alle lettere greche di Bayer e ai numeri di Flamsteed. Johann Bayer nel 1603 nel suo atlante *Uranometria* designò le stelle contenute con lettere greche seguite dal genitivo latino della costellazione, assegnando le lettere in ordine di posizione, procedendo dall'alto in basso all'interno delle stelle della stessa magnitudine, poi facendo la stessa cosa con le stelle di una magnitudine più debole, e così via. In seguito, nell'*Historia coelestis libri duo* di John Flamsteed (fondatore e primo direttore dell'osservatorio di Greenwich), pubblicato nel 1712, le stelle furono contrassegnate da numeri seguiti dal genitivo della costellazione. Qui è usata la consuetudine delle lettere di Bayer per le più luminose, e i numeri di Flamsteed per le più deboli, seguiti dalle tre lettere che identificano facilmente, per convenzione internazionale, le costellazioni. Nei casi rimanenti è stata usata la numerazione del catalogo *Hipparchos* dell'Agenzia Spaziale Europea.
Abbiamo già detto della buona precisione delle indicazioni di coordinate e l'identificazione delle stelle in questo caso non comporta problemi. Per quelle legate ai fenomeni di levata, culminazione e tramonto, più imprecisi, le difficoltà sono state maggiori, e non sempre è stato utile il confronto con il catalogo di Tolomeo. Come accennato, infatti, la forma di diverse costellazioni, e questo si evince ripetutamente lungo il trattato, e sarà anche sottolineato in vari luoghi, dai tempi di Ipparco a quelli di Tolomeo deve essersi evoluta, e in alcuni casi non poco.
La maggior parte delle indicazioni sui gradi zodiacali in levata, al tramonto o in meridiano sono sostanzialmente corretti, con errori contenuti entro il mezzo grado o il grado. Gli errori maggiori di Ipparco, di uno o due gradi, sono soprattutto per i fenomeni all'orizzonte, più difficilmente misurabili anche su un globo celeste. Quando sono ancora maggiori di solito sono relativi ad astri di declinazione tale che le loro traiettorie appaiono più o meno parallele all'orizzonte e di cui pertanto è difficile, anche con un globo, valutare gli istanti della levata e del tramonto. Anche gli errori maggiori sui passaggi in meridiano sono collegati a fenomeni all'orizzonte di tali astri. Noi comunque abbiamo sottolineato in nota solo gli errori macroscopici, dell'ordine dei 5° e più, e questi sono probabilmente dovuti solo a sviste dei compilatori delle copie e dei codici.

## Sull'uso dei numeri cardinali e ordinali da parte di Ipparco

Già Manitius si accorse che la maggior parte delle coordinate usate nei *Commentari* presentava qualche problema di interpretazione, ma sono stati Vogt e Grasshoff a trattare esaurientemente la questione. I gradi utilizzati da Ipparco per le distanze polari, equinoziali, tropicali sono espressi come numeri cardinali. Invece i gradi utilizzati per i valori di ascensione retta dati nei termini della suddivisione dei segni di 30° dei circoli paralleli all'equatore celeste sono espressi come numeri ordinali, intesi come punti discreti, precisi, uguali in valore ai cardinali corrispondenti.
Anche i gradi utilizzati per le levate, i tramonti e i passaggi in meridiano delle varie parti dell'eclittica sono espressi come numeri ordinali, ma la situazione qui si complica ulteriormente. Infatti le espressioni che Ipparco usa per indicare i mezzi gradi, per esempio “a metà del 27° grado” sembrano sottintendere un significato dell’ordinale come insieme continuo, esteso, in questo esempio come l’insieme che va dal cardinale 26 al cardinale 27; quindi il numero indicato in questo caso corrisponde al 26° grado e mezzo, e così abbiamo tradotto tutte le espressioni di questo tipo.
Nelle espressioni usate in quest’ambito per i gradi interi è assente l’indicazione relativa al 1° grado di ogni segno: questo lascia supporre che, di nuovo, Ipparco qui intenda un uso dell’ordinale come numero discreto, preciso, ma corrispondente però al cardinale del numero inferiore. Così, per esempio, il 3° grado corrisponde al grado 2, il 2° grado al grado 1, e il 1° grado non viene indicato perché corrisponde al grado 0, indicato da Ipparco semplicemente come “inizio” del segno.
Altre conferme a questo quadro giungono da due controlli effettuati da Vogt: il primo basato sul fatto che i punti di levata-tramonto e culminazione non sono indipendenti uno dall’altro ed è possibile quindi in diversi casi effettuare un confronto incrociato, controllando e ricavando i gradi interi, quelli di attribuzione incerta, dai mezzi gradi, quelli dal significato sicuro; nei 52 casi in cui questo è stato possibile Vogt ha trovato una media di valori molto più vicini (0,09°) al cardinale diminuito di un grado che non a quello corrispondente al valore dichiarato (0,9°). Il secondo controllo effettuato da Vogt è quello basato sulla datazione e ha compreso tutte le indicazioni, sia quelle intere che i mezzi gradi, trovando che con i cardinali interi diminuiti di un grado si trova un’epoca media congruente con quanto già riferito, intorno al 150 a.C., mentre con i cardinali interi corrispondenti ai valori dichiarati si arriva ad un’epoca di 50 anni posteriore. Da parte nostra abbiamo effettuato un controllo al contrario, verificando quanti dei valori interi diminuiti di un grado, e quanti lasciati come sono, presentano i minori errori rispetto ad una datazione del 150 a.C. trovando percentuali, significative, anche se non decisive, rispettivamente del 74% e del 26%. Esse sono comunque tali, unite alle prove precedenti, da non lasciare ulteriori dubbi su questa strana scelta di Ipparco. Nella nostra traduzione, tuttavia, abbiamo lasciato inalterati i numeri ordinali interi così come nell’originale greco, con l’avvertenza che nei casi in cui il lettore volesse effettuare delle verifiche di controllo con i valori veri (e per i casi di maggiore scostamento da questi valori veri da noi indicati nelle note), essi vanno diminuiti di una unità.

## Questa traduzione

La traduzione è stata condotta di norma sul testo greco dell’edizione di Karl Manitius (riportiamo il titolo latino: *Hipparchi in Arati et Eudoxi phaenomena commentariorum libri tres*, Teubner, Lipsia, 1894). Di questa edizione sono state accettate anche la maggior parte delle integrazioni proposte, ed è stata adottata anche la numerazione dei capitoli e dei paragrafi. Da sottolineare (ma verrà ricordato anche nei singoli luoghi) che in alcuni punti, nella citazione dei versi di Arato fatta da Ipparco, alcune parole sono inserite fra parentesi graffe, poiché si tratta di aggiunte di Ipparco stesso.

# Commentari di Ipparco ai Fenomeni di Eudosso ed Arato

## Libro I

### Cap. I
### Ipparco saluta Escrione[5]

**1.** Con piacere venni a sapere dalla lettera del tuo vivo desiderio di conoscere; infatti, sia le tue personali ricerche sulla natura, sia quelle che riguardano quanto detto da Arato nelle “Levate simultanee”,[6] mi hanno rivelato un considerevole amore per le arti, tanto più apprezzabile in quanto sei stato gravato da incombenze familiari a causa della prematura fine degli ottimi fratelli. **2.** Riguardo alle altre cose ti chiarirò in seguito il mio giudizio personale; ora preferisco scriverti su quanto è stato detto da Arato nei *Fenomeni*, indicando senza riserve ciò che è stato affermato in modo corretto o scorretto nell’opera. Da questo ti sarà tutto chiaro, anche quei punti su cui ti trovi in difficoltà.
**3.** Certamente anche molti altri hanno compilato dei commentari ai *Fenomeni* di Arato; e a me sembra che Attalo,[7] l’astronomo nostro contemporaneo, abbia realizzato su di essi lo studio più accurato. **4.** Ma penso che spiegare il significato dei versi non richieda grande impegno: questo poeta infatti è semplice e sintetico, ed è inoltre facilmente comprensibile per coloro che hanno effettuato uno studio ordinario della sua opera; ma si potrebbe considerare utilissimo e scientificamente opportuno comprendere le affermazioni da lui sostenute sui corpi celesti, e quali di queste siano state scritte in accordo con i fenomeni celesti e quali no.
**5.** Osservando dunque che Arato risulta in contraddizione con i fenomeni e con la realtà su moltissimi ed essenziali punti, e che su quasi tutti questi non solo gli altri, ma anche Attalo sono d’accordo con lui, ritenni, per soddisfare il tuo amore di sapere e per la comune utilità degli altri, di esporre quello che mi sembra sbagliato. **6.** Mi proposi di far questo non per ottenere prestigio dal confutare gli altri (sarebbe davvero un misero e meschino pensiero; al contrario credo che dobbiamo essere grati a tutti quelli che si trovano ad affrontare in prima persona fatiche per l’interesse comune); ma perché né tu né altri desiderosi di sapere siate sviati nell’osservazione dei fenomeni del cosmo. **7.** Cosa che in verità è successa a molti; infatti la grazia dei poemi conferisce una certa credibilità a ciò che in essi viene affermato, e quasi tutti quelli che commentano questo poeta convengono con le sue affermazioni.
**8.** Eudosso, con maggiore competenza, ha scritto un trattato sul medesimo argomento di Arato. Giustamente quindi, anche a motivo dell’accordo di astronomi di tale levatura, la sua poesia è definita degna di fede. Perciò non è forse giusto attaccare Arato, anche se in qualche caso incorre in alcuni errori. Egli infatti ha scritto i *Fenomeni* seguendo il trattato di Eudosso, senza aver osservato personalmente il cielo o senza proporre personali considerazioni astronomiche sui fenomeni celesti, con il rischio di sbagliare.

5. Non sappiamo chi sia questo amico di Ipparco.
6. Parte conclusiva dei *Fenomeni* (propriamente detti, ovvero la parte astronomica) dal verso 559 al 732.
7. Astronomo di Rodi, di cui non si ha peraltro alcuna notizia.

**9.** Oltre all'esposizione degli errori presenti nei *Fenomeni* di Arato ed Eudosso, come pure in quei commentatori che sono d'accordo con le loro affermazioni, ho compilato per te le levate e i tramonti simultanei per tutte le costellazioni fisse, incluse le 12 zodiacali, come avvengono in realtà, affinché tu possa controllare da te stesso, in modo molto dettagliato, anche i dati di tutti gli altri commentatori. **10.** Ed espongo chiaramente non solo la levata e il tramonto simultaneo, ma inoltre quali stelle di ciascuna costellazione sorgano e tramontino per prime e per ultime, e quali delle altre stelle siano in meridiano e in quante ore equinoziali[8] sorga o tramonti ciascuna delle costellazioni fisse in simultanea con i 12 segni dello zodiaco. **11.** In più indico anche quali stelle determinino tutte le 24 ore. Ciascuno di tali [insiemi di fenomeni], come comprendi facilmente, ha delle utili applicazioni per molti programmi astronomici.

## Prima parte

### Cap. II

**1.** Che quindi Arato abbia seguito lo scritto di Eudosso sui fenomeni si può capire confrontando in più passi, per ciascuna delle affermazioni fatte, le espressioni in prosa di Eudosso con i versi del poeta. Non è inutile ricordarlo ancora, anche se brevemente, perché questo fatto è messo in dubbio da molti. **2.** Si attribuiscono ad Eudosso due libri sui fenomeni, che sono fra loro in accordo su quasi tutto, tranne pochissimi punti. Uno è intitolato *Specchio*, l'altro *Fenomeni*. Sui *Fenomeni* Arato ha composto il suo poema. **3.** In questa opera Eudosso così scrive a proposito del Drago: "Fra le Orse c'è la coda del Serpente,[9] che ha l'ultima stella sopra la testa della Grande Orsa, fa una curva vicino alla testa della Piccola Orsa e si prolunga fin sotto le sue zampe; dopo aver fatto in quel punto un'altra curva, si solleva alzando il capo di nuovo e ha la testa di fronte". **4.** Arato, come se parafrasasse questo testo, dice:

49 Ma esso ad una si estende con l'estremità della coda,
l'altra circonda intorno con una spira. L'estremità
della coda termina presso la testa dell'Orsa Elice,
Cinosura ha la testa nella spira; essa sulla stessa
testa si gira e le giunge fino alla zampa,
ed ecco che di nuovo volta indietro corre in alto.

**5.** Sul Bovaro Eudosso dice: "dietro alla Grande Orsa c'è Artofilace". Arato a sua volta:

91 Dietro Elice si muove, nell'atto di uno che guida,
Artofilace.

Ed ancora Eudosso: "sotto i piedi c'è la Vergine". E Arato:

96 Sotto entrambi i piedi del Bovaro puoi osservare
la Vergine.

8. Ore di uguale durata. Sembra oggigiorno una specificazione inutile, ma nel mondo antico venivano usati anche segnatempo ad ore di durata disuguale, più corte in inverno, più lunghe in estate.
9. La costellazione del Drago, che Ipparco chiamerà ancora erroneamente Serpente in I, 2, 6; I, 2, 11; I, 4, 4; I, 11, 1; I, 11, 3; I, 11, 10; I, 11, 16; III, 3, 7. .

**6.** Sull'Inginocchiato Eudosso dice: "presso la testa del Serpente c'è l'Inginocchiato, che ha il piede destro sopra la testa". E Arato:

69 al di sopra al centro della testa
del sinuoso Drago tiene l'estremità del piede destro.

Da tutto ciò risulta ben chiaro che entrambi sono in errore: l'Inginocchiato, in realtà, ha il piede sinistro [ι Her] sulla testa del Drago, non il destro [ν Boo].
**7.** Eudosso dice che la Corona giace sotto il dorso dell'Inginocchiato; e Arato:

74 La Corona è accanto al dorso, mentre alla sommità della testa
vicino, osserva il capo del Serpentario.

Su questi ultimi anche Eudosso afferma: "la testa del Serpentario è vicina alla testa di costui".
**8.** Riguardo alla posizione della Grande Orsa, Eudosso riporta: "sotto la testa della Grande Orsa vi sono i Gemelli, in corrispondenza del centro il Cancro, sotto le zampe posteriori il Leone. Davanti alle zampe anteriori dell'Orsa c'è una stella brillante, una più brillante sotto le ginocchia posteriori, ed un'altra sotto le zampe posteriori."
**9.** E Arato:

147 Sotto la testa i Gemelli, sotto la parte centrale il Cancro,
sotto le zampe posteriori il Leone brilla d'un bel bagliore.

E sulle stelle citate:

143 così le si muovono davanti alle zampe belle e grandi
una davanti a quelle [zampe] sotto le spalle, una davanti a quelle che scendono dai lombi,
un'altra sotto le ginocchia posteriori.

**10.** Sull'Auriga Eudosso scrive: "l'Auriga ha le spalle di fronte alla testa della Grande Orsa, essendo in obliquo sopra i piedi dei Gemelli, ha il piede destro in comune con la punta del corno sinistro del Toro." Su questo Arato scrive:

161 la sommità della testa di Elice gli
si muove di fronte.

E ancora:

174 l'estremità del corno sinistro [del Toro]
e il piede destro dell'Auriga che giace accanto
un'unica stella occupa.

**11.** Su Cefeo Eudosso così si esprime: "Cefeo ha i piedi sotto la coda della Piccola Orsa, essi formano con l'estremità della coda un triangolo equilatero; la sua parte centrale è vicina alla curva del Serpente [che si estende] fra le Orse." **12.** E Arato dice:

184 una linea di corda uguale si estende dall'estremità della coda
ad entrambi i piedi, quanta si estende da piede a piede,

ma poco sposteresti lo sguardo dalla cintura
alla prima curva del sinuoso Drago andando.

In entrambi i casi la prima affermazione è falsa; infatti la distanza fra i piedi di Cefeo [Errai e κ] è minore della distanza di ciascuno dei due dalla coda [Polare].
**13.** Su Cassiopea poi Eudosso dichiara: “Cassiopea è di fronte a Cefeo, mentre Andromeda è di fronte a lei ed ha la spalla sinistra sopra il Pesce settentrionale, la cintura sopra l’Ariete, se si prescinde dal Triangolo, che si trova fra loro; la stella sulla testa [di Andromeda] coincide con quella sul ventre del Cavallo.” **14.** E Arato:

188 davanti a lui poi si volge sventurata, non molto grande
apparendo in una notte di plenilunio, Cassiopea.

E poco sotto:

197 Là infatti si gira anche quella sventurata immagine
di Andromeda, ben distinta, sotto la madre.

E ancora:

206 comune brilla una stella
sull’ombelico di esso [cavallo], e sull’estremità della testa di lei.

E ancora sull’Ariete:

229 potresti tuttavia individuarlo dalla cintura
di Andromeda; è situato infatti poco sotto di lei.

E inoltre:

246 La spalla sinistra di Andromeda ti sia del Pesce
più settentrionale segno indicatore: infatti gli è molto vicina.

**15.** Su Perseo Eudosso dice così: “Perseo ha le spalle presso i piedi di Andromeda, il braccio destro verso Cassiopea e il ginocchio sinistro verso le Pleiadi.”
Arato su Andromeda dice:

248 entrambi i piedi indicherebbero lo sposo
Perseo, che sempre gli si muovono sulle spalle.

E ancora:

251 la sua destra è tesa verso il seggio
della suocera.

E un po’ più avanti:

254 Vicino al suo ginocchio sinistro in gruppo tutte
le Pleiadi si muovono; tutte non un grandissimo
spazio le contiene.

**16.** Sull'Uccello Eudosso afferma: "presso la mano destra di Cefeo si trova l'ala destra dell'Uccello, presso l'ala sinistra le zampe del Cavallo." Sull'Uccello Arato dice:

279 verso la destra<br>
di Cefeo l'estremità destra dell'ala mostrando,[10]<br>
mentre sull'ala sinistra è piegato il balzo del Cavallo.

E pur essendoci molti altri passi, che sembrerebbero essere stati quasi trascritti, saranno sufficienti questi, per non dilungarci troppo.

**17.** Oltre a ciò, anche la suddivisione nella descrizione delle costellazioni rende chiaro quanto affermato.[11] Ed infatti sia Eudosso che Arato prima descrivono le costellazioni a nord dello zodiaco poi, allo stesso modo, quelle a sud. Arato ha pure trattato, come Eudosso, le levate e i tramonti delle altre costellazioni che avvengono contemporaneamente con le 12 costellazioni dello zodiaco.[12]

**18.** Riguardo alle stelle che si muovono sul tropico d'estate e d'inverno e anche sull'equinozio, Eudosso sul tropico d'estate afferma: "ci sono su questo la parte centrale del Cancro e quella del corpo del Leone nel senso della lunghezza, l'area un po' al di sopra della Vergine, il collo del Serpente tenuto [in mano dal Serpentario], la mano destra dell'Inginocchiato, la testa del Serpentario, il collo e l'ala sinistra dell'Uccello, le zampe del Cavallo, poi anche il braccio destro di Andromeda, e la regione fra i piedi,[13] la spalla sinistra e la gamba sinistra di Perseo, e inoltre le ginocchia dell'Auriga e le teste dei Gemelli, poi [il tropico] si ricongiunge alle parti centrali del Cancro." **19.** Arato cominciando da questi ultimi scrive:

481 In esso si muovono entrambe le teste dei Gemelli,<br>
giacciono le ginocchia del saldo Auriga,<br>
su questo la gamba sinistra e la spalla sinistra<br>
di Perseo, e al di sopra del gomito di Andromeda, al centro<br>
il braccio destro tocca; la palma della mano gli giace sopra<br>
più vicina a borea, il gomito è inclinato su noto,<br>
gli zoccoli del Cavallo e la parte inferiore del collo dell'Uccello<br>
con la sommità del capo e le belle spalle del Serpentario.

E quello che segue.

**20.** Sul tropico d'inverno Eudosso scrive: "su questo vi sono le parti centrali del Capricorno, i piedi dell'Aquario, la coda del Mostro Marino, la curva del Fiume, la Lepre, le zampe e la coda del Cane, la poppa e l'albero di Argo, il dorso e il petto

---

10. Ipparco usa qui φαίνων ("mostrando"), ma in tutti i manoscritti aratei si trova τείνων ("tendendo").

11. Intendi: che Arato ha seguito Eudosso.

12. In greco si usava lo stesso termine, ζῴδιον, per indicare sia i segni zodiacali che le costellazioni zodiacali, e non è sempre chiaro, dai vari contesti, se si intendano gli uni o le altre. In ogni caso va subito precisato che Arato ha trattato questo argomento, come compete probabilmente ad un poeta, con riferimento alle costellazioni, non ai segni zodiacali. Questi ultimi sono una partizione dell'eclittica in 12 segni di estensione uguale, 30° ciascuno; è perciò normale che ci siano difformità, anche notevoli, riguardanti i fenomeni, ed è strano che Ipparco spesso se ne dimentichi. Per un approfondimento dell'argomento v. qui Gemino, *Introduzione ai fenomeni*, I, 3-6. Solo in un punto Arato sembra alludere ad una divisione dell'eclittica in 12 parti uguali, ai vv. 541-543 (citati anche da Ipparco in I, 9, 11), però non dice che queste parti portano il nome delle costellazioni zodiacali: pertanto quando parla dell'Ariete, del Toro, ecc., Arato senza dubbio si riferisce alle costellazioni, non ai segni.

13. In questo caso ci siamo staccati dalla lezione di Manitius, aggiungendo semplicemente una virgola dopo Andromeda, ed abbiamo riferito τὸ μεταξὺ ποδῶν ("la regione fra i piedi") a Perseo, anche se la successione delle parole è inusuale, in quanto il tropico ai tempi di Eudosso passava in mezzo ai piedi di Perseo, non di Andromeda.

del Centauro, la Fiera, l'aculeo dello Scorpione, poi esso, attraverso il Sagittario, si ricongiunge alle parti centrali del Capricorno". **21.** Così Arato:

501 Un altro [cerchio] con noto che gli va contro a metà il Capricorno
taglia e i piedi dell'Aquario e la coda del Mostro Marino;
in esso c'è la Lepre; ma del Cane non molta parte
prende, ma solo quanta occupa con le zampe, in esso Argo
e il grande dorso del Centauro, in esso l'aculeo
dello Scorpione, ed anche l'arco del debole[14] Sagittario.

In modo quasi simile a questo Arato ha riferito anche sulle costellazioni che giacciono sul circolo equatoriale.
**22.** Inoltre Arato si basa su Eudosso anche per quanto riguarda la latitudine.[15] Ed infatti Eudosso nello scritto intitolato *Specchio* dice che il tropico è tagliato in modo che le sezioni stanno fra loro nel rapporto di cinque a tre.

## Cap. III

**1.** Con quanto ho detto, credo di aver dimostrato sufficientemente che Arato ha composto i *Fenomeni* basandosi su Eudosso. Ora dimostreremo in quali errori costoro, e quanti a loro si rifanno, fra cui anche Attalo, incorrono. Esporremo senz'altro anche quali errori ciascuno di loro fa autonomamente.
**2.** Si deve premettere che Attalo sottoscrive quasi tutte le asserzioni di Arato sui fenomeni celesti, ritenendo che siano dette da lui in accordo con i fenomeni stessi, tranne una o due, che indicheremo in seguito. **3.** Egli afferma infatti nel proemio della sua opera: "perciò ti abbiamo inviato il libro di Arato, da noi emendato e il commento, avendo reso ciascuna parte in accordo con i fenomeni, e in modo conforme a quanto detto dal poeta." Ed ancora in seguito dichiara: "forse alcuni chiederanno confidando su quale argomento diciamo che la revisione del libro è stata fatta conformemente al modo di pensare del poeta; noi diamo il motivo più stringente: «l'accordo del poeta con i fenomeni»."
**4.** Dato che Attalo ha questa opinione è allora necessario che, in ogni luogo dove ravvisiamo delle contraddizioni fra le affermazioni di Arato ed Eudosso ed i fenomeni celesti, sottoponiamo anche Attalo alle nostre critiche, qualora egli concordi erroneamente sugli stessi punti.
**5.** Per prima cosa mi sembra che Arato ignori la latitudine della Grecia, pensando che essa sia tale che la durata del giorno più lungo abbia un rapporto di 5 a 3 rispetto a quella del giorno più corto.
Afferma infatti sul tropico d'estate:

497 Se questo è diviso, per quanto meglio possibile, in otto parti
cinque in cielo si volgono e sopra la Terra,
tre nell'emisfero inferiore.

14. Per la verità Arato scrive di "splendente Sagittario", come d'altra parte Ipparco riporta correttanente nel secondo luogo dove cita questi stessi versi (cfr. I, 10, 16).
15. Ipparco parla di "inclinazione del cosmo", termine con cui i Greci indicavano la latitudine.

**6.** C'è comune consenso sul fatto che dalle parti della Grecia lo gnomone ha un rapporto di 4 a 3 rispetto all'ombra dell'equinozio.[16] Là, quindi, il giorno più lungo è di $14\frac{3}{5}$ ore equinoziali circa,[17] e l'altezza del polo circa 37°.[18] **7.** Ma dove il giorno più lungo ha un rapporto di 5 a 3 col più breve,[19] lì il giorno più lungo è di 15 ore e l'altezza del polo all'incirca 41°. È chiaro dunque che quello menzionato non può essere il rapporto fra giorno più lungo e più corto valido per la Grecia, bensì per l'Ellesponto. **8.** Certo Arato non ha scritto su questi argomenti sulla base della propria opinione, ma ha seguito Eudosso anche su questo. Se anche avesse scritto basandosi sui propri dati, non avendo detto chiaramente per quali luoghi sia valida la suddetta latitudine forse, almeno su questo, non sarebbe criticabile. **9.** Perciò Attalo, che si dice d'accordo, è in errore nel dire che dalle parti della Grecia il giorno più lungo ha un rapporto di 5 a 3 col più breve. Commentando, infatti, il passo del poema sul tropico estivo, dice così: "É evidente che tutta l'opera è stata composta per la Grecia: qui, infatti, il giorno più lungo è in rapporto alla notte più breve di 5 a 3".[20]
**10.** Ancora di più ci si potrebbe stupire del fatto che non si sia accorto che nell'altro trattato Eudosso ha spiegato e scritto altrimenti. E cioè che la parte del tropico che rimane sopra la terra ha un rapporto di 12 a 7 rispetto a quella che rimane sotto.[21] È quello che scrivono anche gli astronomi della scuola di Filippo, ed anche molti altri, salvo che, quando hanno stabilito il sorgere ed il tramontare contemporaneo delle costellazioni dal punto di vista di un osservatore posto in Grecia, si sono sbagliati sulla latitudine di questa regione.
**11.** Avendo tralasciato dunque questo errore, abbiamo considerato tutta la loro opera sull'orizzonte della Grecia. Infatti non è proprio di un uomo amante della verità, ma di uno che si occupa di cose futili, attaccare questi studiosi per l'ipotesi di base errata contrastandoli su tutto anche quando vengono fatte affermazioni in accordo con i fenomeni celesti visibili in Grecia. **12.** Supponiamo quindi di prendere come orizzonte per l'osservazione quello di Atene, dove il giorno più lungo è di $14\frac{3}{5}$ ore equinoziali, e l'altezza del polo è di 37°.[22]

---

16. Ovvero, un palo alto un metro conficcato verticalmente al suolo getta un'ombra lunga 75 cm a mezzogiorno dei due giorni degli equinozi. L'inverso di questo rapporto è uguale alla tangente della latitudine: infatti, in questo caso, ¾, ovvero 0,75, è la tangente di 36,87°, circa 37°, come afferma Ipparco.

17. Attenzione a non confondere il rapporto fra lo gnomone e la sua ombra con il rapporto fra giorno più lungo e giorno più corto che, alla latitudine di 37°, vale 14/9. Passare dall'uno all'altro in modo rapido richiede l'uso della trigonometria sferica, come fa Tolomeo nell'*Almagesto* usando il teorema di Menelao. Probabilmente Ipparco ci riusciva utilizzando metodi grafici.

18. L'altezza del polo celeste è uguale alla latitudine.

19. Questo rapporto è equivalente a quello fra l'arco percorso dal Sole sopra e sotto l'orizzonte al solstizio estivo, di cui parla Arato.

20. I giorni più lunghi e più corti in un determinato luogo sono uguali alle notti più lunghe e più corte.

21. Che corrisponde ad una latitudine di 42°30'. Ovvero Ipparco afferma che Eudosso nello *Specchio* descrive i fenomeni per una latitudine di 41° e nei *Fenomeni* per una latitudine di 42,5°. Riesce difficile pensare che Eudosso non conoscesse la latitudine del luogo dal quale osservava (Cnido, a 36°41'), che poteva essere stabilita con un semplice gnomone e un po' di geometria elementare, ma non abbiamo motivo di dubitare di quanto dice Ipparco. D'altra parte, Arato potrebbe aver adottato semplicemente la latitudine della Macedonia (Pella, la capitale, si trova a 40°46' N), coerente con il valore 5/3.

22. Così abbiamo fatto anche noi, per analizzare astronomicamente i vari passi (nel senso di adottare una latitudine di 37°, non quella di Atene che è invece, contrariamente a quanto scrive Ipparco, di 38°), utilizzando l'epoca del 150 a.C., che riteniamo fra le più probabili per la compilazione dei *Commentari* (v. p. 9).

## Cap. IV

**1.** Riguardo al polo boreale Eudosso dice senza cognizione di causa: “c’è una stella ferma sempre nello stesso luogo; questa stella è il polo del cosmo”. In verità sul polo non giace nessuna stella, il luogo ne è privo; vicino ci sono tre stelle, con le quali il punto sul polo forma quasi un quadrato, come sostiene anche Pitea di Marsiglia.[23]
**2.** Di seguito, sono tutti in errore anche nella collocazione del Drago, supponendo che compia una curva intorno alla testa della Piccola Orsa. Infatti le stelle più brillanti ed occidentali[24] fra quelle che sono nel quadrilatero di questa, e delle quali la più settentrionale [Kochab] è, secondo loro, sulla testa, la più meridionale [Pherkad] sulle zampe anteriori, giacciono parallele molto vicino alla coda del Drago. Perciò non è esatto quanto segue:[25]

52 Cinosura ha la testa nella spira; essa sulla stessa
testa si gira.

Così scrive anche Eudosso. **3.** Dal canto suo Arato si sbaglia sul Drago, prima col dire “le Orse si muovono da entrambi i lati della spira”: sono infatti da entrambi i lati della coda e non della spira. Poiché essendo esse infatti rivolte in direzione opposta e stando quasi in parallelo, la coda del Drago è estesa in lunghezza fra loro; la spira circonda la Piccola Orsa, ma dista molto dalla Grande. Per lo stesso motivo anche l’espressione “tortuoso da ambo le parti” è detta in modo inesatto. Sarebbe infatti così se entrambe le Orse giacessero da entrambi i lati della spira. **4.** [Arato] è in errore sul Drago anche nei versi seguenti:

58 Obliqua è la testa, sembra proprio che si inclini
verso l’estremità della coda di Elice, sono proprio in linea retta
con la punta della coda sia la bocca che la tempia destra.

23. Questa è forse una delle poche critiche di Ipparco attribuibile alla precessione degli equinozi. Infatti intorno al 370 a.C., epoca più probabile per la compilazione dei *Fenomeni* di Eudosso, c’era una stellina di magnitudine 6,00, la HIP 63 340, molto vicina alla posizione del polo celeste, a 1°27’ di distanza. La distanza era molto minore, 32’, nel 600 a.C., e quindi Eudosso potrebbe addirittura essersi riferito non ad una sua osservazione, ma ad una tradizione consolidata da decenni. Ai tempi di Ipparco però la distanza era salita a 2°39’ e questa stella ai suoi tempi formava, con la posizione del polo e con la HIP 60 044 (m 5,46) e la HIP 59 504 (m 5,12), proprio il quadrato di cui egli parla. Questa figura geometrica, invero abbastanza regolare (lati di 2°19’, 2°29’, 2°31’ e 2°39’), doveva indubbiamente servire, agli astronomi dell’epoca di Ipparco, a trovare facilmente in cielo la posizione del polo. Strano invece che ne parli Pitea, di poco posteriore ad Eudosso, perché il quadrato non era visibile allora, a meno che questo navigatore e geografo, autore delle prime, e memorabili, esplorazioni del profondo nord europeo, fino alle regioni artiche, non intendesse un’altra figura, che però, confessiamo, non siamo in grado di reperire.
24. Per semplificare e per rendere il concetto nel modo più chiaro possibile abbiamo tradotto le espressioni εἰς τὰ ἡγούμενα (o εἰς τὰ προηγούμενα) e εἰς τὰ ἑπόμενα, che significano “verso [le parti] che precedono” e “verso [le parti] che seguono”, e i corrispondenti aggettivi (προ)ηγούμενος e ἑπόμενος, come “occidentale” e “orientale”. In queste espressioni, ovviamente, è implicato il concetto del movimento diurno della sfera celeste da est verso ovest, per cui la parte “precedente” in cielo è quella che sta davanti nel movimento, ovvero che è verso occidente: la prima a sorgere, a culminare, a tramontare; mentre la parte che sta ad oriente viene dopo, “segue”. Occorre avvertire il lettore, naturalmente, che la nostra sostituzione non funziona con le costellazioni circumpolari qualora queste siano viste in condizioni non ideali di osservazione, ovvero lontano dalla culminazione superiore. Com’è del tutto evidente, vicino alla culminazione inferiore le due direzioni vanno addirittura scambiate.
25. Non sappiamo cosa abbia scritto Eudosso su questo argomento, ma certamente Arato non ha mai detto che la stella β Umi è sulla testa dell’Orsa Minore. Per Tolomeo, β e γ Umi saranno semplicemente due stelle nel rettangolo dell’Orsa Minore, così anche per Ulugh Beg, mentre al-Sufi le collocherà nella parte posteriore, il Manoscritto di Vienna e Dürer nelle spalle, solo per citare cataloghi e iconografie antiche o da esse derivate. In ogni caso, le due stelle citate da Ipparco, qualunque parte del corpo dell’Orsa Minore esse rappresentino, sono certamente molto più vicine alla spirale che all’estremità della coda del Drago, rappresentata dalla stella λ Dra (10° contro 17°).

Non la tempia destra infatti, ma la sinistra, è in linea retta con la lingua del Serpente e con l'estremità della coda della Grande Orsa. **5.** Il dire, infatti, come fa Attalo, che [Arato] colloca la testa del Drago in senso contrario e non rivolta verso l'interno del cosmo, è del tutto inverosimile. Tutte le costellazioni, infatti, sono raffigurate dal nostro punto di vista, e come rivolte verso di noi, eccettuate quelle rappresentate di profilo. **6.** Arato stesso lo mostra con più esempi. Infatti, su tutte le costellazioni per le quali indica chiaramente il lato destro e sinistro, è in accordo con il principio formulato. D'altronde il principio si informa a criteri artistici ed armonici. Peraltro Attalo cercherà con questa interpretazione di giustificare anche l'errore sul piede sinistro dell'Inginocchiato, di cui parleremo a suo tempo.
**7.** Sul posizionamento della testa del Drago Eudosso e Arato si mostrano in accordo con i fenomeni, Attalo in disaccordo. Arato infatti, seguendo Eudosso, dice che essa si muove sul circolo sempre visibile,[26] con queste parole:

61 Quella testa qui si volge, dove appunto gli estremi
di tramonti e levate si mescolano tra loro.

Attalo dice che essa è un po' più a sud del circolo sempre visibile, sicché resta per breve tempo sotto l'orizzonte. **8.** Così si potrebbe pensare che Attalo sia in disaccordo con il fenomeno. Infatti la stella nell'estremità della bocca del Drago dista dal polo $34\tfrac{3}{5}°$,[27] il suo occhio meridionale [Rastaban] 35°, la tempia meridionale [Eltanin][28] 37°. Però il circolo sempre visibile, ad Atene e dintorni, e dove lo gnomone è in rapporto di 4 a 3 con l'ombra equinoziale, dista dal polo circa 37°. È chiaro dunque che la testa del Drago si muove nella sezione sempre visibile, poiché ha solo la tempia sinistra sullo stesso circolo e non è vero, come dice Attalo che, essendo più meridionale, tramonta per breve tempo e si leva.
**9.** Sull'Inginocchiato mi pare abbiano commesso una svista, anche se non proprio un errore, sia Eudosso che Arato, nel dire che il piede destro si trova sul centro della testa del Drago. Mi sembra che Attalo, contro l'intenzione del poeta, abbia mutato l'emistichio scrivendo così: "al di sopra del centro della testa", e immaginando che la testa del Drago sia volta all'esterno del cosmo, in modo che la parte destra della testa fosse presso il piede. Tutte le costellazioni infatti sono raffigurate da tutti e dallo stesso Arato rivolte verso la parte interna del cosmo, come dissi; e in tutti i libri[29] è scritto:

69 al di sopra al centro della testa
del sinuoso Drago tiene l'estremità del piede destro.

**10.** Inoltre, volendo Arato mostrarci chiaramente sia la figura che la collocazione dell'Inginocchiato, giustamente ci indica anche quale suo piede giace sulla testa del Drago; non per altri motivi infatti ci parla della testa del Drago, ma perchè comprendiamo la posizione dell'Inginocchiato, il che fa anche in molti altri casi. **11.** Come afferma anche sullo stesso Inginocchiato, descrivendo le altre sue parti:

---

26. Il circolo artico celeste. Le stelle che si trovano al suo interno non sorgono né tramontano, girando attorno al polo. Per una data località, il raggio del circolo artico è pari alla latitudine.
27. Stando a Tolomeo, la stella dovrebbe essere ν Dra, ma la distanza polare è sbagliata di oltre due gradi, facendo pensare ad un probabile errore dei copisti.
28. Tuttavia per Tolomeo questa stella è al di sopra della testa del Drago.
29. In tutte le copie del poema di Arato disponibili ai tempi di Ipparco.

74 la Corona è accanto al dorso, mentre alla sommità della testa
vicino, osserva il capo del Serpentario.

E ancora sulla Lira dice:

272 le è vicino[30] col ginocchio sinistro.

E pure su Perseo:

254 Vicino al suo ginocchio sinistro in gruppo tutte
le Pleiadi si muovono.

E ancora:

251 la sua destra è tesa verso il seggio
della suocera.

E sull'Uccello dice:

279 verso la destra
di Cefeo l'estremità destra dell'ala mostrando.

E in molti altri casi indica, dicendolo chiaramente, le posizioni delle costellazioni, se si tratta delle parti destre o sinistre.
**12.** Che l'affermazione riguardante il piede sia una svista di Eudosso ed Arato, e non un errore, è chiaro dalle altre informazioni date da entrambi sull'Inginocchiato. Quando infatti Arato dice sulla Lira:

270 pose {intende essa}[31] davanti ad una figura ignota
dopo averla condotta in cielo. E questa, restando accosciata,
le è vicino col ginocchio sinistro.

La gamba accanto alla Lira è posta sulla testa del Drago.
**13.** Ancora sulla levata dell'Inginocchiato dice:

612 le Chele portano alla levata solo
la gamba destra fino allo stesso ginocchio
dell'eternamente Inginocchiato, sempre inclinato sulla Lira.

Ma la gamba che sorge prima è quella posta ad ovest, e non quella sulla testa del Drago. È chiaro quindi che intende quella sinistra. **14.** In accordo con questo, anche sul sorgere del Leone scrive:

591 Tuttavia l'Inginocchiato
avendo già spinte giù le altre [parti del corpo], solo il ginocchio col piede sinistro
non volge ancora sotto l'Oceano fluttuante.

Da tutto questo è chiaro che intende che l'arto sinistro poggia sulla testa del Drago.

30. L'Inginocchiato.
31. Aggiunta di Ipparco: la Lira.

**15.** Passando ad altro, parlando del Serpentario, Arato dice autonomamente che egli, dritto, è posizionato fra gli occhi e il petto dello Scorpione. Ma in realtà sta soltanto con l'arto inferiore sinistro disteso, posto fra la fronte ed il petto dello Scorpione, mentre ha l'arto inferiore destro piegato. Neppure Eudosso afferma chiaramente che è dritto, ma scrive che il piede destro poggia sopra il corpo dello Scorpione, come è in realtà, e non sul petto.[32]
**16.** Mi sembra che Arato ignori anche la magnitudine delle stelle che sono nelle mani del Serpentario [Yed Prior e Posterior, ν e τ Oph]. Dopo aver detto infatti che quelle nelle spalle sono brillanti, sicché si potrebbero vedere anche in una notte di Luna piena, aggiunge:

79 ma le mani non son però tali,
debole splendore infatti si diffonde qua e là.

**17.** Ma le stelle nelle mani del Serpentario sono in comune anche col Serpente, come lo stesso Arato scrive:

82 Entrambe sono affaticate per [il fatto che tiene in mano] il Serpente.

Le stelle nel Serpente, che il Serpentario tiene, non sono inferiori per brillantezza a quelle delle spalle [Cebalrai, γ, ι e κ]. Invece parrebbe di capire, come anche Attalo dice, che esse non sono sufficientemente brillanti.
**18.** Successivamente sulle Chele Arato dichiara:

90 però, esse manchevoli di luci (o biasimevoli),[33] per nulla splendenti.

Attalo afferma che Arato le definisce "biasimevoli di luci" non per il fatto che le stelle nelle Chele sono deboli, ma perché in tutte sono solo quattro, e quindi non esprimono con sufficiente chiarezza la natura dell'oggetto rappresentato. Il fatto poi che dica "per nulla splendenti" è perché esse non sono per niente in grado di completare una somiglianza.[34] **19.** Tuttavia non credo che sia questo[35] il motivo per cui egli [Arato] le definisce "biasimevoli di luci" e "per nulla splendenti", ma proprio perché non sono brillanti; e infatti nelle "Levate simultanee" così dice su di loro:

607 Né le Chele sorgenti e tenuamente lucenti
potrebbero passare inosservate.

**20.** Ed aggiunge appunto subito dopo:

608 poiché la grande costellazione del Bovaro
sorge tutta intera dominata da Arturo.

Come se noi potessimo trovare le Chele, come lui suppone, di per sé non facilmente distinguibili per la loro piccolezza, partendo da Arturo; infatti è uso definire quel-

32. Nella realtà è il piede sinistro a toccare quasi la fronte e il corpo dello Scorpione, almeno secondo il catalogo dell'*Almagesto*, pertanto non è chiaro quest'ultimo rilievo di Ipparco.
33. Due varianti nel testo di Arato, già evidentemente esistenti al tempo di Ipparco.
34. Ovvero non possono somigliare, essendo così poco numerose, a qualcosa di definito.
35. Cioè per il loro essere poche di numero.

le tra le costellazioni che hanno stelle deboli per lo più come "tenui" e "per nulla splendenti".[36]

## Cap. V

**1.** Nelle affermazioni che seguono sull'Orsa mi sembra siano completamente in errore; Eudosso quando dice così: "sotto la testa della Grande Orsa giacciono i Gemelli, in corrispondenza del centro il Cancro, sotto le zampe posteriori il Leone"; e Arato:

147 Sotto la testa i Gemelli, sotto la parte centrale il Cancro,
sotto le zampe posteriori il Leone brilla d'un bel bagliore.

Si associano a loro Attalo e tutti gli altri. **2.** Ma che sbagliano è chiaro da questo:[37] infatti la testa della Grande Orsa secondo gli autori citati è la più settentrionale delle due stelle occidentali [Dubhe] che sono nel quadrilatero, mentre la più meridionale delle stesse stelle [Merak] si trova sulle zampe anteriori. **3.** Che infatti la stella in questione sia secondo loro sulla testa si desume dal fatto che dicono che la stella che si trova nell'estremità della coda del Drago [Giausar] sta di fronte alla testa dell'Orsa. **4.** Non ci sono altre stelle sotto quella nell'estremità della coda del Drago, se non la più settentrionale delle stelle occidentali del quadrilatero. Quella infatti che è nell'estremità della coda del Drago occupa il 3° grado sul circolo parallelo al Leone, mentre la citata stella nel quadrilatero si trova a poco meno del 3° grado del Leone. **5.** Che sulle zampe anteriori giaccia la più meridionale delle stelle occidentali del quadrilatero, lo chiarisce Eudosso dicendo: "davanti[38] alle zampe anteriori dell'Orsa c'è una stella brillante". E Arato:

143 così le si muovono davanti alle zampe belle e grandi
una davanti a quelle [zampe] sotto le spalle, una davanti a quelle che scendono dai lombi.

**6.** Una sola stella luminosa è ad ovest della più meridionale delle stelle occidentali del quadrilatero, quella che attualmente è segnata nelle zampe anteriori.[39] Generalmente tutti gli antichi raffigurano l'Orsa di sette stelle soltanto.[40] **7.** Diventa più chia-

---

36. Quest'ultima frase, un po' slegata dal contesto, potrebbe essere stata in origine una glossa dei versi precedenti.
37. Invece è vera l'affermazione di Eudosso ed Arato che sotto la costellazione dell'Orsa Maggiore vengono a trovarsi, nelle sere di primavera di oggi come di 2000 e anche di 2500 anni fa, le costellazioni citate, se non limitiamo l'Orsa Maggiore alle sette stelle principali, ma la estendiamo a comprendere tutte quelle elencate da Eratostene e Tolomeo. In questo modo, infatti, la costellazione si estende di più verso occidente, verso i Gemelli, e la testa non viene rappresentata dalla stella Dubhe, ma da diverse altre, l'ultima delle quali, verso occidente, è Muscida, che ne delinea la punta del muso.
38. Non è chiaro perché Ipparco travisi Eudosso affermando che egli pone sulle zampe anteriori la più meridionale delle stelle occidentali del quadrilatero, quando invece Eudosso dice "*davanti* alle zampe anteriori c'è una stella luminosa". E comunque Ipparco finge sempre di non capire che Arato ed Eudosso continuano a rifarsi ad una configurazione "estesa" dell'Orsa Maggiore, nella quale le zampe corrispondono alle stelle Talitha, κ UMa, Tania Borealis e Australis, Alula Borealis e Australis. Pertanto le stelle che stanno davanti alle zampe sono, con tutta evidenza, Dubhe, Merak e Phecda. E "davanti", fra l'altro, non significa in questo caso "ad occidente", come sembra intendere Ipparco, ma semplicemente "di fronte".
39. Non si riesce a capire di quale stella parli Ipparco, poiché la "più meridionale delle stelle occidentali del quadrilatero", Merak, secondo la configurazione "minimalista" dell'Orsa di Ipparco, è "sulle" zampe anteriori, e non c'è nessuna altra stella luminosa ad ovest di questa.
40. Tuttavia sia nei *Catasterismi*, che nel manoscritto del Cod. Paris. 2506 (v. p. 4) l'Orsa Maggiore è catalogata nel modo moderno, "esteso", con 24 stelle. Tolomeo ne elenca 27.

ro allora quanto detto da noi sulla Piccola Orsa.[41] Seguendo l'uso comune, infatti, sia la testa che le zampe sono stati da loro collocati nelle quattro stelle che comprendono il quadrilatero; infatti presso di esse[42] non vi sono altre stelle ben visibili, come appunto presso il grande quadrilatero,[43] sicché queste rappresentano la sua testa[44] e le zampe accerchiati dalla spira del Drago; così dice:

52 Cinosura ha la testa nella spira; essa sulla stessa
testa si gira e le giunge fino alla zampa.

**8.** Stando così le cose, come è possibile che i Gemelli ed il Cancro siano sotto l'Orsa, se la testa di questa e le sue zampe anteriori giacciono nel 3° grado del Leone? **9.** Bisogna certamente supporre che sotto l'Orsa ci sia solo il Leone; la stella infatti che è sulle zampe posteriori di questa, sulle quali c'è la più meridionale delle stelle orientali [Phecda] del quadrilatero[45], si trova a meno del 25° grado del Leone. **10.** Se ci si proponesse di indicare alcune figure sotto la figura completa dell'Orsa, bisognerebbe porre sotto di essa la Vergine e, per Zeus, una parte delle Chele.[46] La stella nell'estremità della coda dell'Orsa [Alkaid], che è l'ultima delle sette stelle ad oriente, si trova sul circolo parallelo all'equatore nel 4° grado delle Chele. **11.** Lo dico supponendo che i punti tropicali ed equinoziali siano all'inizio dei segni zodiacali. Se invece i detti punti giacciono al centro dei segni zodiacali, come sostiene Eudosso,[47] bisognerebbe porre sotto le zampe posteriori dell'Orsa non il Leone, ma la Vergine. Infatti la più meridionale delle stelle orientali del quadrilatero, secondo la suddetta divisione dello zodiaco, si trova nel 10° grado della Vergine, la più settentrionale [Megrez] circa nel 16° grado. **12.** Sotto la testa e le zampe anteriori bisognerebbe porre il Leone; infatti le stelle occidentali del quadrilatero si troveranno al 18° grado del Leone, secondo la suddetta divisione. **13**. È chiaro da quanto è stato detto perché, secondo Eudosso, sotto l'Orsa nella sua interezza giacciano il Leone e la Vergine e le Chele;[48] infatti la stella nell'estremità della coda dell'Orsa, secondo la medesima divisione dello zodiaco, occupa circa il 19° grado delle Chele.
**14.** Ancora Arato sull'Auriga scrive:

177 Ma sempre il Toro è più rapido dell'Auriga
a scendere dall'altra parte, pur essendo salito insieme.

Mi sembra però che sia in disaccordo con i fenomeni anche in questi versi: soltanto i piedi dell'Auriga sorgono insieme al Toro, la restante parte del suo corpo sorge con i Pesci e con l'Ariete.[49] **15.** Ed egli stesso nei versi seguenti dice:

41. Non si comprende il motivo di questa affermazione di Ipparco.
42. Le quattro della Piccola Orsa.
43. Della Grande Orsa.
44. Della Piccola Orsa.
45. Di nuovo, la stella in questione contraddistingue le zampe posteriori, secondo il modello di Ipparco.
46. Questo solo se si prendono in considerazione i segni zodiacali, non le costellazioni.
47. V. nota 137.
48. Ma non è quanto hanno scritto Eudosso ed Arato. Quello che hanno affermato sulla posizione delle costellazioni, in verità, farebbe pensare, semmai, ad uno spostamento in senso contrario rispetto a quanto ipotizzato da Ipparco.
49. Ipparco parla sempre dei segni, non delle costellazioni, zodiacali, ma spesso le differenze sono quasi inesistenti, come in questo caso. In realtà la prima parte dell'Auriga sorgeva assieme con l'ultima parte della costellazione (e del segno) dell'Ariete, l'ultima parte assieme alla prima parte della costellazione (e del segno) del Toro.

718 Ma i Capretti e la pianta del piede sinistro con la stessa Capra
si raggruppano col Toro.

Ma secondo lui tutte le parti che sorgono con il Toro sorgono con l'Ariete.[50] Riguardo all'insieme delle costellazioni, nelle "Levate simultanee" indica quali appaiono levate sull'orizzonte al sorgere dell'inizio di ciascun segno zodiacale. **16.** Molto prima del piede sinistro [dell'Auriga, ι] sorgono la spalla destra [Menkalinan] e la mano destra [θ]. È chiaro quindi che, anche secondo lo stesso Arato, solo le parti vicine ai piedi sorgono col Toro, il resto del corpo e perfino il piede sinistro sorgono con l'Ariete.[51] **17.** E neppure questo va a suo favore, perché non parla del Toro nel suo insieme, ma del fatto che il Toro tramonta prima delle parti dell'Auriga che sono sorte con lui. È falso anche questo: infatti i suoi piedi non solo non ritardano sul Toro, ma anzi tramontano prima.[52] Nello specifico la stella nel piede destro dell'Auriga [Elnath] tramonta col 27° grado del Toro. **18.** Se dunque avesse scritto in accordo con i fenomeni, come Attalo sostiene, molto meglio sarebbe stato dire, e più rimarchevole, che il piede destro dell'Auriga, sorto dopo, tramonta prima [del Toro] e non che, sorto insieme, tramonta dopo.
**19.** Inoltre, tutti sono in errore su Cefeo, dicendo che i suoi piedi[53] formano un triangolo equilatero con la stella sull'estremità della coda dell'Orsa, come anche Arato:

184 una linea di corda uguale si estende dall'estremità della coda
ad entrambi i piedi, quanta si estende da piede a piede.

Infatti quella[54] fra i piedi è inferiore alle altre due, sicché si tratta di un triangolo isoscele e non equilatero.
**20.** Nei passi successivi Arato sbaglia, dicendo di Cassiopea:

188 davanti a lui poi si volge sventurata.

Poiché Cassiopea sta ad est di Cefeo.
E sulla Lira ugualmente afferma Arato:

270 pose {questa, Ermes, dice}[55] davanti ad una figura ignota.

Invece essa giace ad oriente dell'Inginocchiato.[56] Certo non è possibile neppure dire che Arato definisca anteriori le parti che si trovano ad est; al contrario invero in molti passi chiama anteriori le parti ad ovest. Infatti, dopo aver parlato dell'Ariete, aggiunge questo sui Pesci:

50. É soltanto un'interpretazione di Ipparco, che non ci sembra suffragata dal testo arateo.
51. Anche questa è un'interpretazione soggettiva di Ipparco.
52. Tuttavia, l'affermazione di Ipparco è falsa: anche Elnath (che per gli antichi era il piede destro dell'Auriga, ma rappresentava anche la punta del corno settentrionale del Toro) tramontava sette minuti dopo l'ultima stella del Toro, ζ Tau (che rappresenta il corno meridionale). Anche considerando il segno e non la costellazione, la sua affermazione non è vera: infatti Ipparco si corregge subito dopo, quando dice che questa stella tramonta con la fine del segno del Toro, il 27° grado (in realtà 25°), non prima.
53. Ripetizione di quanto detto prima in I, 2, 12.
54. La distanza.
55. La parte fra parentesi è aggiunta da Ipparco. Intendi: Arato dice che Ermes pose la Lira…
56. Ipparco sembra interpretare questi "davanti" come se volessero significare "che sta ad ovest", ovvero nel senso di movimento della sfera celeste. Invece Arato in questi luoghi si riferisce semplicemente alla posizione reciproca delle varie figure.

239 Essi ancora più avanti, ancora sulle soglie di noto,
i Pesci.

Anche questi sono ad occidente di esso [Ariete].
**21.** In modo inesatto[57] Arato dice anche di Cassiopea:

188 non certamente grande
apparendo in una notte di plenilunio, Cassiopea;
infatti non molte stelle in successione la fanno risplendere.

In effetti la maggior parte delle sue stelle sono più brillanti di quelle nelle spalle del Serpentario, che pure, afferma, sono ben visibili anche in una notte di Luna piena, con queste parole:

77 tali le luminose spalle poste sotto la sua testa
si vedono; quelle anche in una notte di luna piena
apparirebbero visibili.

**22.** Sono quasi più brillanti anche delle stelle di Andromeda, tranne quella nella testa e la orientale di quelle sulla cintura [Mirach], sulla quale dice così:

198 non credo che tu molto
di notte debba cercare, per vederla subito bene,
così [visibile] è la testa, tali da entrambe le parti
gli omeri e le estremità dei piedi e tutta la cintura.

## Cap. VI

**1.** Inoltre si trova in Eudosso una rappresentazione erronea relativa alla testa della Grande Orsa, in verità nel solo trattato intitolato *Fenomeni*, come riportato di seguito: **2.** “Sotto Perseo e Cassiopea non molto distante c’è la testa della Grande Orsa; le stelle in mezzo sono deboli”; nello *Specchio* [dice] invece così: “dietro Perseo e non molto distante dalle anche di Cassiopea c’è la testa della Grande Orsa; le stelle in mezzo sono deboli”. **3.** Tuttavia la testa della Grande Orsa non si trova nella regione di Cassiopea e Perseo, e non ne dista poco; Cassiopea infatti giace sopra il dodicesimo[58] dei Pesci, Perseo sopra l’Ariete, la testa della Grande Orsa, secondo Eudosso, circa nel 2° grado del Leone.[59] **4.** Sarebbe stato dunque più vicino al vero dire che Perseo e Cassiopea si trovano sotto l’estremità della coda della Piccola Orsa: infatti l’ultima e più brillante stella [Polare] di questa giace all’incirca nel 18° grado dei Pesci ma, per come Eudosso divide il circolo zodiacale, nel 3° grado circa dell’Ariete.

57. Qui Ipparco al verso 188 interpreta πολλὴ, che di norma significa, “molto, numeroso, grande”, come se significasse “splendente”, mentre Arato probabilmente si riferisce alla grandezza della costellazione.
58. Nel senso di dodicesima parte del circolo zodiacale, che Ipparco utilizza come equivalente di segno zodiacale.
59. Non ha senso, per le costellazioni vicine al polo, riferirsi alle loro distanze angolari nel cielo citando solo, come fa Ipparco, le loro ascensioni rette. Infatti, la distanza tra la testa dell’Orsa Maggiore, rappresentata dalle stelle sulla sua fronte, ρ e σ UMa, e la stella di Cassiopea più vicina e sufficientemente luminosa, Navi, è di soli 34°, non di 137° come malignamente suggerisce Ipparco. Certamente Perseo, Cassiopea e l’Orsa Maggiore non sono molto vicine, ma il “non molto lontano” di Eudosso sembra abbastanza appropriato. Tra l’altro, è importante ricordare che nell’antichità non esistevano costellazioni nello spazio intermedio: Camelopardalis e Lynx furono introdotte solo nel XVII secolo.

**5.** Mi sembra che Arato sia in errore sull'Ariete, dicendo che è invisibile nelle notti di Luna piena per la piccolezza delle stelle; e che bisogna segnalare la sua posizione per mezzo delle stelle della cintura di Andromeda, e del Triangolo posto accanto ad esso, a nord. **6.** Dice infatti di questo:

225 Lì anche dell'Ariete rapidissimi sono i sentieri.

E poco dopo aggiunge:

228 Esso è invero debole e senza stelle, se con la luce della Luna
si osserva, tuttavia potresti individuarlo con la cintura
di Andromeda; è situato infatti poco sotto di lei.

E ancora:

233 Vi è poi anche un altro segno ben costruito vicino,
235 il Triangolo simile con due lati uguali
da entrambe le parti, il terzo non è altrettanto, ma è molto facile
da trovare; è rifulgente infatti per molte stelle;
un po' più meridionali di queste sono le stelle dell'Ariete.

**7.** Non è appropriato segnalare l'Ariete procedendo da queste; infatti le tre stelle poste nella sua testa [Hamal, Sheratan, Mesarthim] sono più luminose di quelle nella cintura di Andromeda [Mirach, μ e ν], simili[60] anche a quelle del Triangolo; ben visibile è anche la stella nelle sue zampe anteriori.[61]
**8.** Arato sostiene infondatamente anche quanto segue:

239 Essi ancora più avanti, ancora sulle soglie di noto,
i Pesci.

Infatti non sono entrambi più meridionali dell'Ariete, ma solo uno di essi.[62] Le [stelle] sul muso del Pesce più settentrionale [82 e σ], che sono un po' più meridionali della stella orientale [Mirach] di quelle nella cintura di Andromeda, distano dal polo settentrionale 70°; la stella occidentale di quelle nella coda [η And] dista dal polo settentrionale 78°. **9.** Delle stelle nell'Ariete la più settentrionale e posta sul muso [Hamal] dista poco meno di 78°, al pari di questa anche la più settentrionale [τ] delle stelle nella coda; le stelle poste nel corpo dell'Ariete sono tutte più meridionali di queste. È chiaro dunque che uno dei Pesci è più settentrionale dell'Ariete. **10.** Sull'Ariete anche Attalo è in errore, quando dice: "osservando che l'Ariete non è raffigurato in modo distinto né ha stelle brillanti, che possano essere viste chiaramente anche con la Luna, [Arato] tenta di indicarne la posizione attraverso le stelle che

60. Come luminosità.
61. Si tratta di η Psc, a ulteriore testimonianza che nel corso dei secoli gli astronomi hanno cambiato la disposizione delle costellazioni e delle stelle al loro interno.
62. Nel complesso la maggior parte delle stelle dei Pesci sono a sud dell'Ariete. Ma, a parte questo, la critica di Ipparco è doppiamente strana. Innanzitutto, commentando gli stessi versi poco prima (I, 5, 20), egli afferma che "avanti" in questo caso significa ad ovest ed allora Arato dice, in effetti, che i Pesci sono sia ad ovest che a sud dell'Ariete, com'è nella realtà, e non solo a sud, come fa finta di capire Ipparco qui. Inoltre, a differenza di Arato, Ipparco sullo zodiaco si esprime in termini di segni zodiacali, non di costellazioni e, da questo punto di vista, i Pesci sono una frazione di 30° di eclittica che si colloca senza alcun dubbio a sud dell'Ariete.

gli stanno vicino e attraverso quelle che si muovono sul suo stesso circolo”. **11.** Indipendentemente dal fatto che abbia stelle brillanti, quelle nella testa, come dicevamo, e quella nelle zampe anteriori, mi sembra che Attalo si sbagli pensando che il poeta dica che la posizione dell’Ariete sia ravvisabile attraverso le stelle che si muovono sul suo stesso circolo. Non per questo motivo [Arato] lo indica ma, per mostrare chiaramente ciò che accompagna l’Ariete, aggiunge:[63]

231 Percorre il grande cielo al centro, dove appunto le estremità
delle Chele e la cintura di Orione si volgono.

**12.** Nei versi seguenti è in errore Arato, nel dire su Perseo:

254 Vicino al suo ginocchio sinistro in gruppo tutte
le Pleiadi si muovono.

Infatti il ginocchio sinistro [ε] di Perseo dista molto dalle Pleiadi. Attalo dice che non bisogna intendere il termine ἄγχι come ἐγγύς, “vicino”, ma come ἐγγυτάτω, “il più vicino”; [Arato] vuole intendere, spiega: “che il ginocchio sinistro rispetto alle altre stelle è il più vicino alle Pleiadi.” **13.** Ma neanche questo è corretto: infatti, le due stelle luminose [Atik e ζ] nel piede sinistro di Perseo, e inoltre quella nella gamba sinistra [Menkib], sono molto più vicine alle Pleiadi del ginocchio sinistro.
**14.** In modo non esatto Arato dice anche che le Pleiadi comprendono soltanto sei stelle:

261 Sette {infatti dice}[64] sono quelle indicate con un nome,

258 per quanto sei sole siano visibili.

Gli sfugge però che a chi guarda intensamente in una notte serena e senza Luna appaiono sette stelle comprese nell’ammasso.[65] Perciò ci sarebbe da rimanere stupiti del fatto che Attalo, commentando questi versi, non si sia accorto dell’errore, come se Arato avesse detto bene.
**15.** Di nuovo Arato è in errore nei seguenti versi, dove afferma sull’Uccello:

276 però esso è nebuloso; alcune sue parti sono disuguali
per stelle, invero non troppo brillanti, ma neanche deboli.

Ed infatti l’Uccello ha molte e brillanti stelle, fra cui quella nella coda [Deneb] è particolarmente luminosa, quasi simile a quella brillante [Vega] nella Lira.

63. Ovvero, Arato cita l’Ariete più che altro come costellazione attraverso cui passa l’equatore celeste.
64. Fra parentesi un’aggiunta di Ipparco.
65. In realtà Arato ha detto benissimo, e questo concetto si trova espresso più o meno in questa forma in tutte le fonti dell’antichità (tanto che ha alimentato la famosa leggenda della Pleiade perduta): una persona dotata di vista normale vede sei stelle nelle Pleiadi, e sono Alcione, Atlante, Elettra, Maia, Merope, Taigete; se la stessa persona “guarda intensamente” è difficile che riesca a vedere un’altra stella. Chi è dotato di vista acuta riesce di norma a vedere non solo una settima, ma anche un’ottava e una nona stella. Prima dell’invenzione del telescopio, il primato apparteneva a Michael Maestlin, il maestro di Keplero, che nel 1579 ne vide ben 11!

## Cap. VII

**1.** Nei versi successivi Arato, dopo aver parlato del fatto che non bisogna navigare di notte, quando il Sole comincia ad attraversare il Sagittario, volendo presentare delle indicazioni di questa circostanza di tempo dice:

303 Segno di quella stagione e di quel mese ti
sarebbe lo Scorpione che sorge sul finir della notte.
Allora infatti il grande arco tende vicino al pungiglione
il Sagittario; poco più avanti di lui sta
lo Scorpione che sorge, mentre esso si leva un po' dopo.
Allora anche la testa di Cinosura al termine della notte
corre molto in alto, prima dell'aurora tramonta
tutto intero Orione, e Cefeo da una mano ad un fianco.

**2.** A questo proposito Attalo, dopo aver citato i versi, afferma: "le altre cose quindi il poeta ha detto in modo appropriato in questi versi; ma ha ignorato quello che comunemente si sa su Cefeo, avendo detto che esso tramonta in questo periodo. Infatti non tramonta allora, ma al contrario si leva. **3.** E questo risulta chiaro dagli stessi versi; dice infatti che esso inizia il tramonto nel periodo in cui le Chele, che sono in levata, stanno per essere portate su, e che tutte le sue parti che possono tramontare sono tramontate,[66] quando lo Scorpione è in levata".[67]
**4.** A mio parere certamente Attalo è in errore quando sostiene che l'affermazione di Arato riguardo alla testa della Piccola Orsa e su Orione è corretta, e che al contrario quella su Cefeo è falsa; perché, invece, proprio per quanto riguarda Cefeo, ha parlato in accordo con i fenomeni, mentre non lo ha fatto per quanto riguarda gli altri punti. **5.** Ancor prima di questo mi sembra che sbagli pensando che Arato fornisca le suddette indicazioni per la levata del Sagittario. Che infatti intenda questo, è chiaro anche dal passo prima riportato. **6.** E comunque lo dice espressamente nelle righe successive: "per consenso comune, in questo periodo, poiché tutte le sue parti,[68] eccetto quelle poste sul circolo artico, sono tramontate, è senz'altro evidente che, essendo il Sagittario in levata, non compie l'inizio del tramonto, ma al contrario della levata". **7.** Inoltre affinché sia più chiaro che al sorgere del Sagittario non corrisponde il tramonto, ma il sorgere di quello,[69] consideriamo quanto detto da lui[70] riguardo a questa circostanza di tempo; invece egli[71] interpreta tutto erroneamente. Che Arato riferisca le suddette indicazioni non alla levata del Sagittario, ma alla levata dello Scorpione, lo indica con chiarezza il poeta stesso. **8.** Poiché infatti, quando il Sole è nel Sagittario, quest'ultimo non è visibile, ci espone delle indicazioni relative a questa circostanza di tempo, che devono essere del tutto percettibili, in quanto indicazioni. Prima quindi del sorgere del Sole, quando ancora è notte e si vedono le stelle, dice:

66. Ovvero tutte le parti di Cefeo che alla latitudine della Grecia erano occidue.
67. Ai vv. 307, 310, 626, 633, 645-646, 649-652 dei *Fenomeni* Arato è molto ambiguo, perché afferma che Cefeo tramonta sia col sorgere delle Chele che dello Scorpione e, forse, del Sagittario. Il ragionamento di Attalo (desumibile tuttavia non da questo passo, come dice Ipparco, ma da quello che lo stesso Ipparco riporta in seguito) è che, avendo iniziato a tramontare con le Chele, e avendo finito il tramonto con lo Scorpione, Cefeo non può che sorgere con il Sagittario, ma purtroppo ciò non trova riscontro nel testo arateo.
68. Di Cefeo.
69. Cefeo.
70. Arato.
71. Attalo.

"lo Scorpione che sorge sia segno del finir della notte"[72]. **9.** È ragionevole pertanto che ci descriva anche le altre indicazioni della notte, se ci devono essere delle indicazioni. Questo è chiaro anche da quanto dice lo stesso Arato sull'Orsa: "allora anche la testa di Cinosura al termine della notte", come dicesse ai limiti estremi della notte; e su Orione: "tramonta prima dell'aurora", che vuol proprio dire prima dell'aurora, e non, per Zeus, quando ormai sorge il Sole nell'inizio del Sagittario.

**10.** Attalo si è ingannato, poiché ha collegato a quello che è stato detto sul Sagittario, ovvero "esso si leva un po' dopo", l'espressione "allora anche la testa di Cinosura" e il resto. Tuttavia, anche se si concordasse con lui[73] su questo, neppure così Arato può "dire correttamente", come afferma Attalo, che al sorgere di questo[74] "corre molto in alto la testa della Piccola Orsa". **11.** Quando questa infatti si muove "molto in alto" ed è in meridiano,[75] non sorge il Sagittario, ma l'Aquario. La testa della Piccola Orsa corrisponde alla parte finale dello Scorpione lungo il circolo parallelo a quello equinoziale; quando essa è in meridiano sul suo parallelo, nello zodiaco è in meridiano il 3° grado del Sagittario; quando questa è in meridiano dalle parti della Grecia e dove il giorno più lungo è di 14½[76] ore equinoziali, sorge il 17° grado dell'Aquario. **12.** Quando dunque la testa della Piccola Orsa è in meridiano o, il che è lo stesso, "corre molto in alto",[77] allora non lo Scorpione comincia a sorgere, ma è già sorta più di metà dell'Aquario, e lo Scorpione sta completamente ad ovest del meridiano. Risulta chiaro da quanto è stato detto che perciò non "correttamente" Arato si è espresso, come afferma Attalo, ma è in discordanza per più di tre segni zodiacali rispetto ai fenomeni.

**13.** Che neppure Orione tutto intero tramonti quando comincia a sorgere lo Scorpione, come dice Arato, ed Attalo è d'accordo con lui, ma piuttosto al sorgere del Sagittario, dovrebbe essere chiaro dalle seguenti affermazioni. **14.** Dalle parti della Grecia e dove il giorno più lungo è di 14½ ore, [Orione] comincia a tramontare col piede sinistro [Rigel], quando tramonta il 7° grado del Toro, e finisce di tramontare con la spalla destra [Betelgeuse], quando tramonta il 24° grado del Toro. **15.** Se però prendiamo in considerazione anche la mazza che ha nella mano destra [$\chi^1$ e $\chi^2$ Ori], tramonta insieme con l'ultimo grado del Toro. È chiaro quindi che, se tramonta insieme al Toro, tutto Orione tramonta, non quando comincia a sorgere lo Scorpione, ma il Sagittario.[78]

**16.** Che le osservazioni di Arato su Cefeo siano in accordo con i fenomeni e non in disaccordo, come suppone Attalo, dovrebbe essere chiaro da ciò che segue. **17.** Comin-

72. Veramente Arato dice che lo Scorpione che sorge alla fine della notte è segno di quella stagione.

73. Attalo.

74. Il Sagittario.

75. Tuttavia Arato non dice che è in meridiano, è una forzatura di Ipparco; il testo poetico, tutto sommato, si accorda abbastanza bene con il fenomeno.

76. Ovvero a 36° di latitudine (v. nota 220). Per gran parte dei fenomeni la differenza fra 37° e 36° di latitudine non è gran cosa, ma con levate e tramonti di astri di particolare declinazione le differenze sono sensibili.

77. V. nota 75.

78. Per la verità al verso 307 Arato non parla dell'inizio del sorgere dello Scorpione, ma usa un termine generico, per cui si deve ritenere che intenda riferirsi al sorgere della costellazione nel suo insieme. Inoltre, come al solito, Ipparco fa riferimento ai segni zodiacali, e Arato alle costellazioni. Così, quando Orione stava finendo di tramontare, all'epoca di Eudosso, lo Scorpione, che è una costellazione molto estesa in longitudine, stava ancora completando la sua levata (mancava completamente il pungiglione) e le prime stelle del Sagittario non avevano ancora fatto la loro comparsa. D'altra parte ha ragione anche Ipparco, perché la sparizione dell'ultima stella della mazza di Orione precedeva di poco, sei minuti, la levata del confine fra i segni dello Scorpione e del Sagittario (almeno nella sua epoca, mentre in quella di Eudosso mancavano ancora 21 minuti al sorgere del confine).

ciano a tramontare [di Cefeo] le parti che sono più a sud del circolo sempre visibile con il sorgere dell'8° grado delle Chele, e cessano di tramontare con il sorgere del 13° grado delle Chele. È chiaro allora che al sorgere dello Scorpione devono per forza essere tramontate le parti occidue di Cefeo. **18.** Si è sbagliato Attalo nel sostenere che Cefeo non tramonta, ma che al contrario sorge, perché ha frainteso l'idea del poeta, come ho detto, e ha supposto che Arato dicesse che al sorgere del Sagittario tramontano sia Orione che Cefeo.

**19.** Oltre a quanto è stato detto, entrambi si ingannano, Arato ed Attalo che concorda con lui, sul fatto che Cefeo tramonti con la cintola:[79]

650 sfiora la terra, mentre le parti vicino al capo tutte
immerge nell'Oceano; alle altre non è permesso,
ai piedi, alle ginocchia, al fianco: le Orse lo impediscono.

**20.** Alla latitudine della Grecia infatti Cefeo non s'immerge fino alla cintola, ma neppure fino alle spalle. Scendono sotto l'orizzonte invero solo le stelle che stanno nella testa; le spalle si muovono nell'arco sempre visibile, senza sorgere né tramontare; infatti la stella luminosa nella spalla destra [Alderamin] dista dal polo 35½°, la stella brillante nella spalla sinistra [ι][80] ne dista 34¼°. **21.** Dove il giorno più lungo dura 14,5 ore, là il cerchio sempre visibile dista dal polo 36°, ad Atene 37°. È chiaro dunque che le stelle luminose poste nelle spalle di Cefeo o, come dicono alcuni, nelle braccia di Cefeo, come se, essendo lui con le braccia tese, le sue spalle fossero senza stelle, si muovono sempre più a nord del circolo sempre visibile. **22.** L'errore sarà ancora più grande, se supporremo [maggiore] la latitudine; dove infatti il giorno più lungo è di 15 ore, là Cefeo si muove tutto intero all'interno del circolo artico.

## Cap. VIII

**1.** Arato è in errore anche sulla costellazione di Argo. Dice infatti che la parte di questa che va dalla prua fino all'albero è priva di stelle. Le sue parole:

349 In parte tutta indistinta e senza stelle fino allo stesso
albero dalla prua si volge, in parte tutta luminosa.

Infatti le stelle luminose poste nel taglio dell'imbarcazione, di cui la più settentrionale [Markeb] è nel ponte, la più meridionale [Aspidiske] nella chiglia, sono disposte a considerevole distanza verso est.[81]

---

79. La citazione di Ipparco è incompleta, e va integrata con la seconda parte del verso 649 ("… allora Cefeo con la cintola"): quindi, com'è evidente, Arato non dice che Cefeo tramonta con la cintola, ma che con la cintola sfiora la terra, due cose ben diverse, e che tramontano solo le parti vicino alla testa. Al verso 310 dice che tramonta fino al fianco e al verso 633 "di testa, di mano e di spalla". Al tempo di Eudosso la testa tramontava interamente, assieme alla mano destra e di fatto anche la spalla destra risultava invisibile a causa dell'estinzione atmosferica, raggiungendo un'altezza di solo un grado (con la rifrazione). All'epoca di Ipparco la variazione era minima e le stelle risultavano più alte solo di 40'. Osservando da Rodi (rispetto ai 37° del tempo di Eudosso) la variazione sarebbe apparsa ancora minore, impercettibile a occhio nudo, poco più di un decimo di grado. La critica di Ipparco ha parzialmente senso solo se si usa un globo celeste, ma è immotivata facendo riferimento al cielo reale.

80. Secondo Tolomeo è però nel braccio sinistro, e sulla spalla sinistra non ci sono stelle.

81. Ma Arato si rifà qui indubbiamente alla tradizione secondo la quale Argo si vede in cielo solo nella parte che va dalla poppa all'albero (v. anche Eratostene, *Catasterismi*, 35). Quindi probabilmente per Arato il taglio della nave

**2.** Nei versi successivi Arato, giunto alle restanti costellazioni a sud dello zodiaco, aggiunge così:

367 Le stelle di scarso splendore e sparse su una piccola area
fra il timone e il Mostro Marino si volgono,
poste sotto i fianchi della glauca Lepre.

In questi versi mi sembra che abbia proprio visto male (così infatti bisogna dire), se ha pensato che le stelle senza nome si trovino fra il timone e il Mostro Marino. **3.** Esse si trovano infatti fra il Fiume e il timone. A sud di Orione infatti c'è la Lepre, ancora più a sud di questa le stelle anonime, a est di queste il timone di Argo. **4.** Dal piede sinistro di Orione il Fiume si estende a ovest fino al Mostro Marino poi, voltosi a est, di nuovo si volge a sud ovest. **5.** Arato stesso dice che il Mostro Marino si trova:

358 disteso poco sopra il Fiume stellato.

È chiaro così che le stelle sotto la Lepre devono stare fra il Fiume e il timone. **6.** Anche Eudosso dice così come noi sosteniamo, in uno dei due trattati: "sotto il Mostro Marino giace il Fiume, che ha inizio dal piede sinistro di Orione; fra il Fiume e il timone di Argo, sotto la Lepre, c'è uno spazio non vasto, con deboli stelle." **7.** E nell'altro [trattato]: "fra il Fiume e il timone di Argo, sotto la Lepre, c'è uno spazio di cielo non grande, che ha deboli stelle".

**8.** Attalo non rimarca questo abbaglio, pensando che Arato abbia detto opportunamente; invece lo rimprovera riguardo alla rappresentazione inadeguata fornita nei seguenti versi:

367 Le stelle di scarso splendore e sparse su una piccola area
fra il timone e il Mostro Marino si volgono,
poste sotto i fianchi della glauca Lepre,
senza nome; infatti esse non di un'immagine ben formata
a membra somiglianti sono disposte, come molte
che, ordinate una dopo l'altra, percorrono le stesse strade
al passar degli anni. Qualcuno degli uomini che non ci son più
le osservò, né[82] pensò di chiamare tutte con un nome
dopo aver dato loro una forma precisa; non avrebbe potuto infatti di tutte
rimaste isolate da solo dire il nome, né insegnarlo;
[ce ne sono] molte infatti dappertutto, molte hanno uguali
grandezza e splendore, e tutte ruotano attorno.
E per questo anche desiderò tenere raccolte
le stelle, affinché l'una sull'altra disposte in ordine
formassero delle figure; subito nominabili divennero
le costellazioni, e non più ora stupisce una stella che sorge,
ma ben connesse in chiare immagini
appaiono; ma quelle al di sotto della Lepre che è inseguita
tutte oscure e senza nome si volgono.

---

coincideva con la posizione dell'albero. Comunque, le due stelle si potevano vedere ai tempi di Eudosso, ma Aspidiske era difficilmente visibile, solo per poche decine di minuti, presso la culminazione, alta al massimo soltanto 2°44' a Cnido, dove viveva Eudosso. Da notare, ed è curioso, che Tolomeo non la catalogherà, pur osservando da Alessandria, con la stella quattro gradi più alta.

82. Ipparco qui fa riferimento evidentemente ad un manoscritto che conteneva una negazione, non presente nella maggior parte dei codici di Arato che ci sono pervenuti, dove la frase è invece volta in senso affermativo ("e pensò di…") ed ha maggior senso logico.

**9.** Dopo aver riportato queste parole Attalo aggiunge: “in questi versi il poeta appare piuttosto confuso, spesso tornando sullo stesso pensiero, e non riuscendo a presentare un discorso ben definito. Infatti vuol dire che le stelle disposte fra il Mostro Marino e il timone, al di sotto della Lepre, non sono ordinate in una costellazione, ma sono senza nome. **10.** Poiché infatti ci sono molte stelle e talune hanno magnitudine e colore simile, colui che per primo ordinò le stelle per raggruppamenti, e diede un nome a ciascuno, non avrebbe potuto riconoscerle, se giacevano sparse qua e là, se non avesse preso fra queste stelle quelle che potevano insieme indicare qualcosa, e non avesse così dato loro un nome.”

**11.** Mi sembra invece che Attalo non abbia colto il pensiero del poeta, e non solo questo, ma anche che neppure abbia reso chiaramente il senso dei versi che si proponeva di esporre, e che lo abbia divulgato in modo incomprensibile, mentre invece Arato lo aveva reso con chiarezza. **12.** Vuole infatti dire che fra il timone e il Mostro Marino, sotto la Lepre, giacciono stelle senza nome, scarse per numero e magnitudine; in quanto tali, la loro disposizione non permette che si possano raffigurare animali, persone od oggetti, come nelle altre costellazioni, che uno degli antichi ha disegnato. **13.** E neppure [l’antico] si è impegnato a dare una forma a tutte; poiché molte stelle sono sparse e isolate, non avrebbe potuto comporle in un’unica figura “dopo aver dato loro una forma precisa”. Per questo motivo decise di dar forma e creare delle figure assemblando insieme le stelle relativamente vicine fra loro e di dar loro un nome.

**14.** In seguito Arato dice sull’Incensiere:[83]

402 Ma sotto l’aculeo ardente del grande mostro,
lo Scorpione, vicino a noto, si libra nell’aria l’Altare.
Ma questo, invero, per breve tempo in alto
osserverai; infatti si leva dall’altra parte di Arturo;
e mentre sono molto elevate nel cielo le strade di
Arturo, esso più presto s’immerge nel mare d’occidente.

In questi versi mi pare che Arato sbagli, pensando che, quanto dista Arturo dal polo sempre visibile, altrettanto disti anche l’Incensiere dal polo meridionale.[84] **15.** In modo simile sbaglia anche Attalo, dal momento che si trova d’accordo con lui. Spiegando infatti il senso dei suddetti versi, dice così: “parlando dell’Incensiere afferma che esso sta rispetto al polo invisibile così come la stella chiamata Arturo sta rispetto al polo visibile. Perciò dice che il moto dell’Incensiere sopra la terra è breve, mentre quello di Arturo è lungo”. **16.** Ma essi sono in errore pensando che Arturo abbia dal polo boreale ugual distanza di quella dell’Incensiere dal polo meridionale. Per prima cosa, infatti, le stelle nell’Incensiere non giacciono sul medesimo parallelo, ma fra loro alcune sono molto più meridionali, altre più settentrionali; se poi, a prescindere da questo, facessimo soprattutto il confronto con il centro della costellazione, nep-

83. Ipparco usa un termine diverso rispetto ad Arato ed Eratostene, θυμιατήριον (letteralmente appunto “incensiere, turibolo”), imitato in seguito da Gemino e Tolomeo. Ovviamente dove cita i versi di Arato utilizza il termine da questi adoperato, θυτήριον (“altare”).

84. Anche questa sembra una forzatura da parte di Ipparco, seguito in questo anche da Attalo. Arato non intende niente di più di quello che ha scritto, seguendo la sua ispirazione e sostanzialmente in accordo con i fenomeni: la costellazione dell’Incensiere si levava nell’epoca di Eudosso dalla parte opposta a dove tramontava la stella Arturo (la differenza di azimut è di circa 163°) e mentre quest’ultima seguiva traiettorie altissime e rimaneva visibile per un tempo lunghissimo, l’Incensiere si vedeva per poche ore, tramontando piuttosto rapidamente.

pure così sarebbe fatta salva l'affermazione; infatti Arturo dista dal polo boreale 59°, la stella luminosa al centro dell'Incensiere [α][85] dista dal polo meridionale 46°. **17.** Dunque la distanza di Arturo [dal polo] è di molto superiore a quella dell'Incensiere; e non solo dista di più rispetto all'Incensiere, ma anche allo stesso Scorpione sotto cui giace l'Incensiere; senza considerare le stelle che stanno nel suo petto [Antares, Al Niyat e τ] e nella fronte [Acrab, Dschubba e π] e il primo segmento dopo il petto [ε], il secondo segmento dopo il petto [μ] dista dal polo meridionale 59°, quanto Arturo dal polo boreale; le restanti stelle distano meno di questa e la più meridionale di loro [Girtab] dista 52⅓°.

**18.** Arato, dopo aver parlato dell'Incensiere, dicendo che quando diventa visibile da nord "soffocato da una nuvola", allora bisogna aspettarsi noto, si sofferma a parlare del Centauro in questo modo:

431 Se dal mare occidentale dista del Centauro
una spalla quanto dall'orientale, e un po' di foschia avvolge lui
stesso, mentre segni simili, dietro, mostra
la notte sull'Altare che tutto risplende, bisogna che tu non
per noto, ma per il vento euro stia in guardia.
Troverai quella costellazione sotto due altre;
infatti le sue parti simili ad uomo giacciono sotto
lo Scorpione, le parti posteriori equine le conducono sotto di sè le Chele.

**19.** In questi versi per prima cosa bisognerebbe capire di quale spalla del Centauro si tratti: infatti entrambe fanno parte della figura e non passano insieme in meridiano, sicché il discorso potrebbe riguardare sia entrambe che una delle due:[86] le spalle [Menkent e ι] del Centauro, infatti, distano abbastanza l'una dall'altra in senso est ovest. **20.** Se la spalla del Centauro avrà uguale distanza dalla levata e dal tramonto (è questo che significa l'espressione di Arato "dal [mare] occidentale"), ovvero quando essa è in meridiano, e se si verificheranno indicazioni simili a quelli sull'Altare, allora euro, non noto, bisogna aspettarsi. **21.** A prescindere da questo, è completamente in errore nel dire che il Centauro giace sotto lo Scorpione e le Chele; infatti si trova quasi tutto sotto la Vergine, tranne quanto si estende sotto le Chele: la spalla destra e il braccio destro e le zampe anteriori del cavallo. **22.** Infatti la stella orientale fra quelle nella testa [3 Cen] occupa il 29° grado della Vergine, la stella sulla spalla destra [Menkent] occupa il 4° grado delle Chele, la più meridionale delle stelle nelle zampe posteriori [Acrux] occupa circa il 13° grado della Vergine. Come quindi è possibile che le parti umane del Centauro giacciano sotto lo Scorpione, quelle equine della coda sotto le Chele? Attalo trascura anche questo, come se [Arato] avesse detto bene.

**23.** Anche nei versi immediatamente successivi ai precedenti c'è un errore:

439 D'altra parte sembra uno che tende sempre la destra
di contro all'Altare tornito.

85. Questa stella non è proprio al centro dell'Incensiere, ma sul suo margine settentrionale. D'altra parte, questa è una costellazione molto meridionale, e le due stelle più a sud nell'attuale Incensiere, la δ e la η, non furono catalogate nemmeno da Tolomeo.

86. Tuttavia è possibile che Arato intenda la spalla destra, rappresentata dalla stella più luminosa fra le due, Menkent, di seconda magnitudine.

Infatti fra la mano destra [η Cen][87] e l'Incensiere giace tutta intera la Fiera e la maggior parte del corpo dello Scorpione.[88] La mano destra è posta intorno all'8° grado delle Chele, e l'Incensiere [giace] sotto le parti estreme dello Scorpione, come lo stesso Arato dice:

402 Ma sotto l'aculeo ardente del grande mostro,
lo Scorpione, vicino a noto, si libra nell'aria l'Altare.

## Cap. IX

**1.** Nei versi successivi, parlando dei tropici e dei circoli equinoziale e zodiacale, dice così:

467 Essi [i circoli] sono privi di larghezza e ben connessi gli uni agli altri
tutti; ma per grandezza si contrappongono due a due.

E poiché è scritto in due modi,[89] in alcuni "essi privi di larghezza", in altri "essi dotati di larghezza", Attalo dice che è preferibile "essi dotati di larghezza". E infatti aggiunge: "gli astronomi suppongono che i tropici e i circoli equatoriale e zodiacale abbiano una certa estensione per il fatto che il Sole compie il solstizio non sempre sullo stesso circolo, ma talvolta più a sud, talvolta più a nord". **2.** E che sia così anche Eudosso lo sostiene. Dice infatti nello *Specchio*: "è chiaro che anche il Sole mostra qualche differenza nei luoghi delle sue conversioni, ma molto meno manifesta e in assoluto molto piccola".[90]

**3.** Mi sembra che Attalo anche in questo sia in errore, pensando che il Sole compia il solstizio ora più a sud ora più a nord, e che per questo bisogna supporre che i circoli siano dotati di larghezza. Se infatti il Sole nel suo percorso sul cerchio non si tiene strettamente al centro dei segni dello zodiaco,[91] ma si allontana verso nord e verso sud, come anche la Luna, è chiaro che anche l'ombra che proviene dalla Terra viene ugualmente deviata. **4.** Se fosse così le eclissi di Luna non si accorderebbero con quelle previste dagli astronomi, poiché essi certamente nei trattati suppongono che il centro dell'ombra si muova sul circolo al centro dei segni zodiacali; in realtà esse non si discostano rispetto alle previsioni più di due dita[92], e talvolta molto meno nella

87. Per Tolomeo questa stella è nell'avambraccio destro.
88. Ancora, Ipparco appare troppo pedante: la destra del Centauro è comunque tesa verso l'Incensiere, anche se nella destra si trova la Fiera, di cui parla Arato subito dopo, e anche se l'Incensiere è un po' staccato dal Centauro. Inoltre, lo Scorpione non è affatto interposto fra le due costellazioni, ma è del tutto a nord dell'Incensiere, come del resto riconosce subito dopo Ipparco, citando di nuovo Arato.
89. Evidentemente anche al tempo di Ipparco esistevano copie del poema di Arato non perfettamente conformi le une alle altre.
90. Probabile accenno al movimento della cosiddetta *nutazione* solare che Eudosso, nella sua teoria planetaria delle sfere omocentriche, ma anche altri prima di lui, aveva proposto per il Sole. Si trattava di un piccolo spostamento in latitudine dovuto al fatto che il Sole compiva il suo moto annuo non proprio lungo l'eclittica ma lungo un cerchio inclinato rispetto a quest'ultima di un piccolo (anche se non sappiamo quanto piccolo) angolo. Forse fu introdotto perché qualcuno aveva erroneamente osservato che il Sole, sorgendo e tramontando ai solstizi e agli equinozi, mostrava una piccola variazione di posizione. D'altra parte, va anche detto che solo per noi moderni un tale movimento è in linea teorica inammissibile, perché sappiamo che l'eclittica, il percorso del Sole in cielo, rappresenta l'orbita della terra nello spazio, e quindi un punto di riferimento assoluto che non può spostarsi, ma per gli antichi astronomi, che operavano in un sistema geocentrico, e che avevano misurato movimenti in latitudine della Luna e dei pianeti, era del tutto normale ammettere che anche il Sole ne fosse dotato. Certo, come detto, si trattava di un movimento inerente al Sole e che non aveva nulla a che fare con la larghezza dei cerchi citati.
91. Ovvero sull'eclittica.
92. Ovvero avvengono in cielo in posizioni che non si discostano più di due dita da quelle previste. Un dito è 1/24 di

trattazione svolta con la massima accuratezza. **5.** Ora, dire che anche i cerchi hanno un'ampiezza corrispondente, è uguale a non assegnare loro alcuna ampiezza;[93] del tutto a prescindere dal fatto che non sarebbe chiaro se le magnitudini delle eclissi di Luna subiscono una deviazione del suddetto ammontare in seguito al movimento del Sole o in seguito al movimento della Luna. **6.** Quindi Arato direbbe, e gli astronomi ammetterebbero, che soltanto i tropici, per questa ragione, sarebbero dotati di larghezza, il circolo equatoriale no. In generale io credo che tutti gli astronomi ammettano che i detti circoli non abbiano larghezza, i tropici, e il circolo equatoriale, e il circolo sempre visibile e quello sempre invisibile. Non è infatti possibile pensarli dotati di larghezza, perché la particolarità di ciascuno è contenere una linea immaginaria e senza ampiezza. **7.** Ed anche quando dicono che il circolo equatoriale e lo zodiaco sono fra i circoli più grandi nella sfera, e che la sfera è divisa a metà da ciascuno dei due e che lo zodiaco, il circolo equinoziale, l'orizzonte e il meridiano hanno lo stesso centro, e che lo zodiaco tocca i tropici in un solo punto e molte altre cose simili, anche attraverso questi discorsi fanno intendere che siano senza larghezza. **8.** Delle cose citate, nessuna infatti si potrebbe dire se essi avessero larghezza; e l'equinozio non verrebbe tracciato in un sol giorno, ma in più giorni; e il Sole sosterebbe più di un solo giorno nel circolo equinoziale, se esso avesse larghezza; e chi realizza gli orologi solari lo fa pensando i circoli tutti senza larghezza, come se fossero tracciati, da una parte dal centro del Sole, dall'altra dalla nostra vista.
**9.** Che anche Arato, in accordo con gli astronomi, pensi che essi siano senza larghezza, si potrebbe capire da quanto dice sul circolo equatoriale:

513 in esso i giorni sono uguali ad entrambe le notti
al cader dell'estate, e poi all'inizio della primavera.

Se esso avesse larghezza non ci sarebbe un solo giorno equinoziale nella primavera e nell'autunno, ma di più. **10.** E ancora da quanto dice sul tropico:

497 Se questo è diviso, per quanto meglio possibile, in otto parti
cinque in cielo si volgono e sopra la Terra,
tre nell'emisfero inferiore.

Se infatti esso si espandesse in larghezza, e fosse diviso in otto, sarebbe impossibile che con la posizione obliqua del cosmo vi fossero sezioni intere sopra e sezioni intere sotto. **11.** Ma certamente anche quando parlando dello zodiaco, dice così:

541 Quanto il raggio visuale dell'occhio si estende,
sei volte siffatto [raggio] percorrerebbe [il cerchio zodiacale]; ma ciascun [raggio],
ugualmente misurato, due costellazioni ricopre.

Oppure così:

553 Quanto di esso [cerchio zodiacale] sotto il concavo Oceano s'immerge
tanto sopra la terra si volge; e ogni notte
sei dodicesimi del circolo sempre tramontano,

---

cubito, antica unità di misura angolare, oltre che metrica, usata dai Babilonesi, pari a 2°. Pertanto due dita equivalgono a 10', l'angolo più piccolo che si poteva misurare a occhio nudo con gli strumenti dell'epoca.
93. Nel senso appunto che l'osservazione diretta non è in grado di percepire un angolo inferiore a 10'.

altrettanti sorgono, e tanto in lunghezza ciascuna
notte sempre si estende, quanto metà del circolo
dal cominciar della notte si eleva al di sopra della Terra.

Mi sembra evidente che egli ritenga che non abbia larghezza, ma al contrario ne sia privo. **12.** Il raggio visuale è diritto e misura sei volte[94] il circolo più grande se è privo di larghezza, ma non se è dotato di larghezza; ciascuna notte e ciascun giorno metà circolo dello zodiaco sorge e metà tramonta, muovendosi il Sole sul circolo senza larghezza e attraverso il centro dello zodiaco. Infatti non è possibile che questo accada su un circolo dotato di larghezza. **13.** Anche da numerose altre osservazioni è chiaro che Arato ipotizza che i circoli siano privi di larghezza, in accordo con gli astronomi. **14.** In seguito, dopo aver parlato del circolo latteo aggiunge che, dei quattro circoli, due sono uguali, due molto più piccoli:

477 ed invero nessun altro cerchio simile nel colore a questo
si gira, ma tanto grandi per misura su quattro ce ne sono
due, gli altri invece molto più piccoli di questi ruotano.

**15.** Non mi sembra sia detto correttamente neppure questo, che i tropici sono molto più piccoli dei cerchi equinoziale e zodiacale; infatti sono più piccoli di meno di $1/11$.[95]

## Cap. X

**1.** Riguardo alle stelle che giacciono su ciascuno dei tropici e sul circolo equinoziale, scrive Arato:

480 Uno di questi è vicino a dove scende borea;
in esso si muovono entrambe le teste dei Gemelli,
in esso le ginocchia giacciono del ben connesso Auriga.

**2.** Tuttavia le teste dei Gemelli non giacciono sul circolo tropicale estivo. Infatti il circolo tropicale è più settentrionale del circolo equinoziale di circa 24°, la testa orientale dei Gemelli [Polluce] è 30° a nord del circolo equinoziale, la testa occidentale [Castore] a 33½°; sicché l'una è più settentrionale rispetto al tropico estivo di $1/5$ di segno zodiacale, l'altra all'incirca di ⅓. **3.** E poi l'Auriga non ha neppure stelle sulle ginocchia. Se ne pone lì alcune deboli, è in errore; infatti sono i suoi piedi che giacciono nelle immediate vicinanze del tropico; infatti la stella sul piede sinistro [ι] è 27° più a nord del circolo equinoziale, la stella sul piede destro [Elnath] 23½°. **4.** E non è neanche possibile dire che egli ponga sulle ginocchia le stelle da noi collocate sui piedi; infatti quella che noi diciamo essere sul piede destro, anch'egli la colloca così nei versi:

174 la sommità del corno sinistro [del Toro]
e il piede destro dell'Auriga che giace accanto
un'unica stella occupa.

94. Tuttavia Arato dice, ai vv. 541-542, che misura un sesto.
95. Rilievo inutilmente pedante. Tutto dipende da cosa si confronta: per esempio se mangio 91 g di spaghetti invece di 100 neppure me ne accorgo; ma se due uomini sono alti 1,66 m e 1,82 m, sicuramente definiremmo basso il primo e alto il secondo.

**5.** E poi dice:

483 su questo la gamba sinistra e la spalla sinistra
di Perseo.

Anche in questi versi si allontana molto dalla verità. Infatti la stella luminosa al centro del corpo di Perseo [Mirfak] è 40° più a nord del circolo equatoriale; 16° dunque più a nord del tropico, cioè più della metà di un segno. La spalla sinistra [θ] quindi è più a nord del tropico estivo; Perseo giace infatti in una posizione tale che le sue parti vicine alla testa sono a nord, i piedi a sud, con la testa inclinata un po' verso est.[96]

**6.** Su Andromeda afferma:

484 e al di sopra del gomito di Andromeda, al centro
il braccio destro tocca.

È in errore anche su questo, e infatti le stelle nella sua spalla destra[97] sono più settentrionali del tropico. Ci sono tre stelle, la più meridionale delle quali è più a nord rispetto al circolo equinoziale di più di 30°: di queste tre stelle, poste nella mano destra, quella a sud [ι And] è 32° a nord dell'equatore.[98] Si capisce dunque che il gomito è completamente a nord del tropico, e non più a sud come dice Arato.

**7.** In seguito scrive:

487 Gli zoccoli del Cavallo e la parte inferiore del collo dell'Uccello
con la sommità del capo.

Da questo non è chiaro quali stelle ponga sugli zoccoli del Cavallo.[99] Ma da ciò che scrive sull'Uccello:

279 verso la destra
di Cefeo l'estremità destra dell'ala mostrando,
mentre sull'ala sinistra è piegato il balzo del Cavallo.

Sembrerebbe avvicinarsi alla verità. Infatti la stella nella punta dell'ala sinistra [ζ] è

96. Per la verità non si trova traccia di questa inclinazione del Perseo: quando sorge, la testa appare leggermente inclinata verso ovest, al passaggio in meridiano l'inclinazione diventa ancora più accentuata. Poi, Ipparco si dimentica della gamba sinistra, la stella Menkib, che era sicuramente più vicina al tropico della spalla sinistra, sia ai tempi di Ipparco (3° a nord) che di Eudosso (2° a nord). Menkib era sul tropico nel 755 a.C. Certo, il tropico non poteva passare vicino sia alla gamba che alla spalla sinistra, e quindi o c'è un errore dei copisti, o la costellazione di Perseo ai tempi di Eudosso era configurata in modo diverso.

97. Secondo Tolomeo sulla spalla c'è una sola stella, la π, però probabilmente Ipparco intende parlare di quelle che Tolomeo ha catalogato nel braccio destro di Andromeda, la σ, la ρ e la θ (che forse, non a caso, il Bayer pone a mezza via, una nella spalla e due nel braccio), che erano, nell'epoca di Ipparco, appena a nord del tropico. Essendo la mano di Andromeda più a nord del braccio, ovviamente le stelle sulla mano avevano declinazioni molto più settentrionali del tropico.

98. Qui ci siamo discostati leggermente dal testo, secondo il quale le prime tre stelle citate apparivano essere diverse dalle seconde; lo abbiamo fatto sulla base del riscontro astronomico che suggerisce, con tutta evidenza, trattarsi invece dei medesimi astri, ovvero la ι, la κ e la λ, posti da Tolomeo sulla mano. Tuttavia Manitius ritiene che si tratti effettivamente di stelle diverse, le prime tre essendo quelle da noi prima citate, catalogate da Tolomeo nel braccio destro di Andromeda, la σ, la ρ e la θ, e le seconde quelle della mano. Per questo Manitius nella sua traduzione sostituisce, ma senza alcun riscontro dei codici, "più di 30°" con "più di 25°". In questo caso rimarrebbe però senza spiegazione il fatto che Ipparco dica che la spalla destra si trova a nord del tropico, quando invece nella sua epoca la declinazione di π And era di 21°51'.

99. Probabilmente la π Peg, che all'epoca di Eudosso era a 50' dal tropico.

23° a nord dell'equatore. **8.** Il becco dell'Uccello [Albireo] è più settentrionale del circolo equinoziale di 25°20', la stella successiva [η], e posta quasi nella gola, è più settentrionale del circolo equinoziale di 31°. É chiaro quindi che né la testa dell'Uccello, né il collo, possono stare sul tropico d'estate.
**9.** Poi dichiara:

488 le belle spalle del Serpentario
sospinte si volgono intorno allo stesso cerchio.

In questi versi è completamente in errore; infatti delle spalle del Serpentario la destra è molto più vicina al circolo equinoziale che al tropico, la sinistra è ⅓ di segno zodiacale più a sud del tropico; infatti la spalla destra [Cebalrai] è più a nord del circolo equinoziale di circa 7°, la sinistra [κ] di circa 15°.
**10.** In seguito dice:

490 Un po' più verso noto si volge, né lo tocca,
la Vergine; sì invece il Leone e il Cancro; insieme entrambi
collocati in successione stanno, ma il circolo
l'uno sotto il petto e il ventre fin quasi alle vergogne
taglia, l'altro completamente [taglia] al di sotto del guscio
il Cancro, dove tutto spezzato in due potresti vederlo
diritto, se gli occhi andassero da una parte all'altra del cerchio.

Mi sembra che questo sia stato detto in accordo con i fenomeni: infatti la più meridionale e più brillante delle stelle [Regolo] poste nel petto del Leone, che alcuni collocano nel cuore, è poco più a sud del tropico; la stella [η] posta dopo questa, a nord di essa, è invece poco più a nord. **11.** E delle quattro stelle brillanti nelle cosce e nelle zampe del Leone, la seconda da nord [Chertan], posta nelle cosce, è più a nord del tropico, la terza [ι], posta nelle zampe, è più a sud. É chiaro pertanto che il Leone è tagliato da questo in lunghezza sotto il petto e il fianco. **12.** Delle stelle del Cancro, le quattro poste intorno alla nebulosa [il Presepe] una, la più meridionale di quelle ad ovest [θ], è più a sud del tropico di non meno di 1°, la più settentrionale di esse [η] è più a nord di non meno di 1°. Delle due stelle poste a oriente intorno alla nebulosa, la più meridionale [Asellus Australis] giace quasi sul tropico, la più settentrionale [Asellus Borealis] dista da questo circa 2½°. É chiaro dunque che anche ciò che riguarda il Cancro l'ha spiegato proprio in accordo con i fenomeni.
**13.** Però mi sembra si debba ricordare che le considerazioni sul Cancro, sul Leone, sulla Vergine e le zampe del Cavallo, ed anche quelle su Perseo e le ginocchia dell'Auriga e le teste dei Gemelli, ancor prima di Arato le ha fatte Eudosso, al quale pensiamo Arato si sia attenuto. **14.** In effetti, afferma Eudosso, la testa del Serpentario, e non le spalle, giace sul tropico, il che è pure una falsità, anche se certamente la testa è più vicina al tropico delle spalle, essendo infatti circa 7° a sud di esso. Eudosso dice che il collo e l'ala sinistra dell'Uccello sono sul tropico, così il braccio destro di Andromeda.
**15.** Perciò da quanto detto capiamo quanto in queste affermazioni l'uno si allontani dai fenomeni più dell'altro. In particolare Eudosso afferma che il collo del Serpente, che il Serpentario tiene, e la mano destra dell'Inginocchiato, stanno sul tropico, il che pure concorda col fenomeno.
**16.** Riguardo alle stelle sul tropico d'inverno Arato scrive:

501 Un altro [cerchio] che contro noto procede a metà il Capricorno
taglia e i piedi dell'Aquario e la coda del Mostro Marino;
in esso c'è la Lepre; ma del Cane non molta parte
prende, ma solo quanta occupa con le zampe, in esso Argo
e il grande dorso del Centauro, in esso l'aculeo
dello Scorpione, ed anche l'arco dello splendente Sagittario.

Tutto ciò è in accordo con i fenomeni, con l'eccezione delle stelle nel pungiglione dello Scorpione,[100] che sono più a sud del circolo tropicale invernale di più di 8°. Sono piuttosto le parti centrali dello Scorpione che giacciono sul tropico d'inverno.
**17.** Eudosso poi fa anche altre affermazioni allo stesso modo di questi.[101] Dice che giacciono sul tropico la curva del Fiume, e le zampe e la coda del Cane. Giace su di esso anche la Fiera che il Centauro tiene. Ma su questo sbaglia: la Fiera infatti è molto più a sud del tropico d'inverno.[102]
**18.** Riguardo al circolo equinoziale Arato così si esprime:

515 Come sua indicazione l'Ariete e le ginocchia del Toro stanno,
l'Ariete esteso in lunghezza attraverso il cerchio,
delle gambe del Toro, quanto appare a ginocchia piegate.

In questi versi sono errate le osservazioni sull'Ariete; esso è infatti tutto più a nord del circolo equinoziale: solo la stella posta nelle sue zampe posteriori [μ Cet] si muove su di esso.
**19.** E poi dice:

518 Su di esso la cintura dello splendente Orione
la curva dell'infiammata Idra, su di esso anche la debole
Coppa, in esso il Corvo.

Dunque, la cintura di Orione giace sul circolo equinoziale,[103] ma la spira del Drago[104] e la Coppa e il Corvo[105] sono molto più a sud di esso, con l'eccezione della zona intorno alla coda del Corvo, che vi si avvicina.

520 in esso, {dice},[106] le non molte stelle
**20.** delle Chele, in esso stanno le ginocchia del Serpentario.

---

100. Qui ci sono molti errori macrosopici di Arato, che inspiegabilmente Ipparco non segnala. Il tropico attraversava le parti centrali del Capricorno intorno al 980 d.C., i piedi dell'Aquario intorno al 1130 d.C., la coda del Mostro Marino intorno all'80 d.C., la Lepre intorno al 900 a.C., le zampe del Cane in nessuna epoca, Argo verso il 460 d.C., il dorso del Centauro verso il 2640 a.C., l'arco del Sagittario verso il 1100 a.C. Forse questa parte è stata manipolata.
101. Di Arato.
102. In realtà sbaglia anche sul Fiume (il tropico vi passava nell'821 d.C.) e sul Cane (v. sopra).
103. La cintura di Orione non si è mai trovata sull'equatore celeste durante l'attuale ciclo precessionale. La sua stella più a nord, Mintaka, raggiungerà una minima distanza verso il 2600 di 7' a sud. I cicli successivi, a causa dell'effetto della variazione dell'inclinazione dell'asse terrestre, che si somma alla precessione, porteranno la cintura un po' più a nord, ma ugualmente mai centrata sull'equatore (con Mintaka a nord dell'equatore di poco, ma le altre due stelle decisamente a sud), verso il 28 140, 53 900, 79 700, e così via. Ai tempi di Ipparco la cintura si trovava fra i 4°30' e i 5°40' a sud dell'equatore (ai tempi di Eudosso ancora 40' più a sud), pertanto non si capisce perché Ipparco non rilevi l'errore.
104. L'Idra.
105. La Coppa e il Corvo sì, ma la curva dell'Idra, almeno quella che Tolomeo contraddistingue con le stelle ι, $\tau^1$ e $\tau^2$, era a nord dell'equatore ai tempi di Ipparco.
106. Aggiunta di Ipparco.

Tuttavia delle Chele solo la stella brillante nella Chela settentrionale è vicina al circolo equinoziale, le altre sono molto più a sud di questo.[107] E le ginocchia del Serpentario sono più a sud del circolo equinoziale, il sinistro [ζ] di 3½°, il destro [Sabik] più di 10°.
**21.** In seguito scrive:

522 Non è privo neppure dell'Aquila, ma molto vicino
gli vola il grande messaggero di Zeus; presso questo
la testa del Cavallo e la parte inferiore del collo si volgono.

Anche questo corrisponde quasi alla realtà.
**22.** Eudosso ha detto le altre cose in modo simile, e afferma però che le parti centrali delle Chele giacciono sul circolo equinoziale,[108] e l'ala sinistra dell'Aquila e anche le reni del Cavallo e inoltre il più settentrionale degli stessi Pesci. **23.** Ma l'Aquila non tocca il circolo equinoziale, e la stella sulle reni del Cavallo [Algenib] è più di 3½° a nord del circolo. Ed invero il più settentrionale dei Pesci è più a nord dell'equatore all'incirca di 10°.
**24.** Attalo tralascia di trattare in modo specifico le stelle che si trovano nei cerchi di cui si è parlato e scrive così: "egli in questi versi, nei quali espone con precisione attraverso quali stelle ognuno dei tre circoli paralleli passa, fa un'affermazione del tutto scorretta, perché i circoli non possono assolutamente passare attraverso le stelle che egli indica, ma ho omesso di dirlo qui perché te ne sei convinto attraverso osservazioni con la diottra". **25.** Tuttavia, innanzitutto sarebbe stato necessario, se aveva capito [la natura dell'errore di Arato], indicare anche per gli altri studiosi quali di queste [stelle] si trovavano più a sud e più a nord del tropico estivo[109] e di quanto.
**26.** Inoltre, per comprendere il semplice fatto che i cerchi non passano attraverso le stelle citate, non c'è bisogno di diottra: alcune sono infatti molto più a sud, altre molto più a nord, e basta guardare in alto per accorgersi immediatamente che non è possibile che attraverso le dette stelle siano tracciati o il circolo equinoziale o un circolo parallelo a questo.

## Cap. XI

**1.** Oltre [a quelle sui] tre circoli citati Eudosso elenca chiaramente anche le stelle poste sul circolo artico. E dice che su di esso giacciono la spalla sinistra di Artofilace, l'area al di sopra della Corona, e la testa del Serpente fra le due Orse, l'area al di sopra della Lira e dell'ala destra dell'Uccello, il petto di Cefeo, la parte superiore di Cassiopea; e poi questo prosegue sotto le zampe anteriori della Grande Orsa, fra l'Orsa e il Leone, fino alla spalla di Artofilace.
**2.** Comunque la spalla sinistra di Artofilace [Seginus] è più a sud del circolo sempre visibile in Grecia di oltre 4°. Dista infatti dal polo settentrionale 41¼°, mentre il circolo ne dista 37°. "L'area al di sopra della Corona e della Lira" è una locuzione

107. Tuttavia ai tempi di Eudosso l'affermazione ci poteva stare: anche se l'equatore non passava in mezzo alle Chele, tuttavia ne attraversava una buona porzione della parte settentrionale.
108. Questo era vero due secoli e mezzo prima di Eudosso, nel 625 a. C.
109. Così nel testo; in realtà avrebbe dovuto dire: "dei tropici e del circolo equinoziale".

molto approssimativa; infatti queste[110] non sono vicine al circolo sempre visibile, ma l'una e l'altra sono molto più meridionali: la più settentrionale delle stelle della Lira dista dal polo boreale 49°.[111] **3.** La testa del Serpente che è fra le Orse è invece più settentrionale, muovendosi vicino al circolo sempre visibile; infatti la stella più meridionale e posta sulla tempia sinistra dista dal polo 37°, proprio gli stessi del circolo. **4.** Il petto di Cefeo [Kurhah] è più a nord del circolo sempre visibile, come sopra abbiamo spiegato.[112] L'area al di sopra dell'ala destra dell'Uccello e il piede di Cassiopea si muovono sul circolo sempre visibile: infatti la più settentrionale fra le stelle [κ] nell'estremità dell'ala destra dista dal polo boreale 40°; la più settentrionale di quelle in Cassiopea [ε], che è nei piedi,[113] dista dal polo 38°. **5.** Anche le zampe della Grande Orsa si muovono più a nord del circolo sempre visibile. Quella occidentale infatti [Merak] dista dal polo settentrionale circa 24°, l'orientale [Phecda] circa 25°.[114] A ragione dunque dice che il circolo sempre visibile "passa fra l'Orsa e il Leone".[115]

**6.** Riguardo al circolo sempre invisibile,[116] le costellazioni che vi si muovono non si possono vedere; "molto vicino a questo" dice[117] "vi sono le parti estreme dell'acqua del Fiume, il fondo e il timone di Argo; e poi la Fiera e l'Incensiere, inoltre le zampe posteriori del Sagittario. Vi si trova la stella che si vede dall'Egitto."[118]

**7.** Riguardo alle altre, poiché si è parlato di vicinanza, non sarebbe da mettere in dubbio,[119] ma non è detto correttamente che la stella chiamata Canopo si muove nel cerchio invisibile: questa infatti è la stella luminosa più meridionale di quelle nel timone [di Argo]. Essa dista dal polo meridionale 38½° circa. **8.** Il cerchio sempre invisibile ad Atene dista dal polo circa 37°, a Rodi circa 36°. É chiaro quindi che questa stella è più a nord del cerchio invisibile in Grecia e può essere vista muoversi sopra la Terra. E certamente si vede anche nei luoghi intorno a Rodi.[120]

**9.** Inoltre, Eudosso tratta chiaramente anche le stelle che stanno sui circoli chiamati coluri e afferma che su uno di essi giacciono la parte centrale della Grande Orsa, e la parte centrale del Cancro e il collo dell'Idra, e la parte di Argo fra la poppa e l'albero. **10.** Poi, dopo il polo invisibile, la coda del Pesce Australe, la parte centrale del Capricorno, la parte centrale della Freccia; e che esso è tracciato attraverso il collo dell'Uccello, la sua ala destra, la mano sinistra di Cefeo, la curva del Serpente e presso la coda della Piccola Orsa.

---

110. Probabilmente queste costellazioni.

111. La più settentrionale delle stelle catalogate da Tolomeo nella Lira è la ε, ma essa era distante ai tempi di Ipparco quasi 51° dal polo; forse Ipparco intende la HIP 92 831, di m 5,43, che era distante 49½°.

112. Ai tempi di Eudosso comunque soltanto di 1°, mentre erano 2° ai tempi di Ipparco.

113. Secondo Tolomeo è nella gamba. L'astronomo di Alessandria elenca per la verità un'altra stella più a nord della ε, la ι, anche se l'identificazione è dubbia.

114. Come già detto, Ipparco fa riferimento ad una configurazione "minimale" dell'Orsa Maggiore, limitata a sette stelle, che non trova riferimenti nella letteratura antica. Nella tradizione consolidata dal catalogo di Tolomeo "le zampe anteriori della Grande Orsa" sono Talitha e la κ, che distavano dal Polo, ai tempi di Eudosso, rispettivamente 37° e 37½°.

115. Ma questo avrebbe senso con un'Orsa di sette stelle, non ne ha alcuno, invece, con un'Orsa al completo.

116. Il circolo antartico celeste, che comprende le costellazioni che non sorgono mai sopra l'orizzonte di un luogo. Anche il suo raggio è pari alla latitudine della località da cui si osserva.

117. Sempre Eudosso.

118. Canopo.

119. Qui Ipparco vuol dire che Eudosso afferma correttamente che le costellazioni citate nel periodo precedente sono vicine al cerchio invisibile, mentre sbaglia dicendo che Canopo è proprio sul cerchio.

120. La distanza della stella dal polo ai tempi di Eudosso era di 37°10' e a quelli di Ipparco 37°18', troppo poco, anche tenendo conto della rifrazione, per essere vista culminare da Atene, che si trova, come detto, a 38° di latitudine. L'osservazione era sicuramente impossibile anche da 37° di latitudine. Va ricordato, infatti, che l'estinzione atmosferica indebolisce moltissimo la luce delle stelle basse sull'orizzonte. Anche contando la rifrazione, l'altezza di Canopo alla culminazione risultava inferiore al grado, e ad 1° di altezza l'indebolimento medio è di ben otto magnitudini. Pertanto la luminosità della stella, che è di -0,6, diventava di 7,4 a 1° sull'orizzonte, cioè diventava invisibile a occhio nudo (a occhio nudo si possono vedere stelle fino alla magnitudine 6). Perfino a Rodi (lat. media 36°10') Canopo poteva esser vista solo nelle notti molto limpide e/o da luoghi elevati, come del resto ricorda Gemino ancora un secolo dopo (*Introduzione ai fenomeni*, III, 15).

**11.** Tuttavia riguardo all'Orsa è completamente in errore: le sue stelle occidentali infatti, quella nella testa e quella nelle zampe anteriori, giacciono entrambe nel dodicesimo del Leone. Come dunque è possibile che il centro dell'Orsa giaccia nell'inizio del Cancro?[121] **12.** Le stelle occidentali della testa dell'Idra si trovano oltre il 10° grado del Cancro. É chiaro pertanto che il collo dell'Idra rimane ancora molto più ad est del detto circolo. Riguardo ad Argo poi non è stato detto correttamente. **13.** La più occidentale delle stelle nella coda del Pesce Australe [γ Gru] si trova oltre il 23° grado del Capricorno. Tanto dunque si trova ad est del suddetto circolo e non si muove su di esso. **14.** La Freccia nella sua interezza rimane ad est del detto circolo, quindi non è tagliata a metà da questo: infatti la sua stella occidentale [α] è situata nel 1° grado del Capricorno; non è possibile, così, che sia tagliata in due dal circolo.[122]

**15.** Ugualmente poi anche l'Uccello tutto intero si trova ad est del circolo: infatti in esso la stella occidentale e sulla punta del becco [Albireo] occupa il 1° grado e mezzo del Capricorno; delle stelle nella punta dell'ala destra quella posta più ad ovest [κ] si trova oltre il 6° grado e mezzo del Capricorno; non è dunque possibile che giacciano sul detto cerchio né il collo dell'Uccello, né la sua ala destra. **16.** Certamente la mano sinistra di Cefeo oltrepassa notevolmente il cerchio verso est: infatti le stelle nella sua testa, che sono più occidentali, occupano oltre il 10° grado dell'Aquario, la stella brillante nel braccio sinistro [ι], che alcuni pongono sulla spalla, occupa il 25° grado dell'Aquario, quasi pari alla differenza [in longitudine] fra due segni.[123] Ma invero, riguardo alla curva del Serpente e alla coda della Piccola Orsa, ha mostrato bene.

**17.** Nell'altro coluro dice che giacciono prima la mano sinistra di Artofilace e le parti centrali, intese nel senso della lunghezza, poi le parti centrali delle Chele, intese nel senso della larghezza, il braccio destro del Centauro e le sue ginocchia anteriori; e, dopo il polo invisibile, l'ansa del Fiume, la testa del Mostro Marino, il dorso dell'Ariete inteso nel senso della larghezza, la testa [τ Per] e il braccio destro [η Per] di Perseo.

**18.** Tuttavia la mano sinistra di Artofilace si trova ad est del suddetto circolo di circa mezzo segno zodiacale e non si muove su di esso, come Eudosso afferma; la sua stella più occidentale [κ] si trova oltre il 13° grado delle Chele. **19.** È in errore anche dicendo che il suo corpo è tagliato a metà in lunghezza dal circolo; infatti la stella sulla testa [Nekkar] si trova circa sul 16° grado e mezzo delle Chele,[124] la stella nel piede destro [ζ] giace all'incirca sul 24° grado e tre quarti delle Chele,[125] la stella brillante nella cintura [Izar] è situata circa sul 14° grado e un terzo delle Chele.[126]

121. È possibile assumendo una molto maggiore estensione dell'Orsa verso occidente, come già rilevato.

122. Rilievo corretto all'epoca di Ipparco. Ma al tempo di Eudosso, a causa della precessione degli equinozi, la Freccia risultava spostata di 2½° verso ovest, tanto da giustificare l'affermazione dell'astronomo di Cnido (per la precisione, la costellazione risultava bisecata esattamente qualche decennio prima, verso il 458 a.C.).

123. Ovvero la distanza della stella dall'inizio del Capricorno, dove passa il coluro, è di 55°, quasi pari all'estensione di due segni (60°).

124. Per la verità sul 25° grado delle Chele.

125. Sul 14° grado e due terzi.

126. Sul 17° grado e mezzo. Sembra possibile che questi ultimi tre valori possano essere stati scambiati fra loro da qualche amanuense distratto.

**20.** La stella nella mano destra del Centauro [η] è ad est del circolo all'incirca di un quarto di segno zodiacale: occupa infatti l'8° grado delle Chele. La testa del Mostro Marino è ad est del circolo non di molto;[127] il Nodo dei Pesci [Alrescha], infatti, che giace nella zona della testa del Mostro Marino, sulla sua cresta, occupa il 3° grado e un quarto dell'Ariete. **21.** Neppure il dorso dell'Ariete giace sul circolo in questione, ma è più di un terzo di segno zodiacale ad est; infatti la stella al centro del suo dorso [ν] giace all'11° grado e mezzo dell'Ariete. Similmente, sia il braccio destro [η] di Perseo, sia la testa [τ], sono ad est del circolo di circa un terzo di segno zodiacale.

127. Per la verità, ai tempi di Ipparco di quasi 14° (il coluro bisecava la testa nel 1250 a.C.).

# Libro II

## Cap. I

**1.** Alle osservazioni sopra esposte, o Escrione, su quanto scritto da Arato ed Eudosso nei *Fenomeni*, aggiungiamo ora il discorso sulle levate e i tramonti simultanei delle costellazioni, mostrando quanto è stato detto correttamente da loro, e in quali passi invece vi sia disaccordo con i fenomeni.
**2.** Per prima cosa dunque Arato, volendo insegnare come, attraverso la levata e il tramonto delle costellazioni zodiacali, si possa conoscere l'ora della notte,[128] dice così:

559 Sarebbe proficuo per chi attende il giorno
osservare le parti, quando ciascuna sorge;
sempre infatti con una sola di esse si leva insieme lo stesso
Sole. Potresti certamente considerarle,
guardandole direttamente; ma se oscure per nubi
fossero o coperte da un monte sorgessero,
cerca indizi ben connessi con quelle che sorgono.
Certamente da entrambi i lati te ne darebbe lo stesso
Oceano, tutti quelli di cui si fa corona,
quando dal fondo ciascuno di quelli conduce.

Afferma quindi che in particolare noi potremo conoscere l'ora se osserveremo il sorgere di una delle 12 costellazioni dello zodiaco. **3.** E che colui che sa in quale costellazione zodiacale o in quale parte di essa sia il Sole, e poiché in ogni notte sorgono sei costellazioni, facilmente dedurrà l'ora della notte dalla costellazione zodiacale che sta sorgendo. Se poi, a causa delle montagne o delle nubi, non fosse visibile la costellazione che sta sorgendo, dalle altre stelle esterne al circolo zodiacale, poste vicino all'orizzonte, noi potremo riconoscere la costellazione che sorge, se faremo attenzione a quali costellazioni sorgono insieme o tramontano in opposizione[129] con ciascuna costellazione zodiacale.
**4.** Innanzitutto, [Arato] sbaglia supponendo che sia sufficiente, per conoscere l'ora della notte, il considerare quante costellazioni ancora manchino al sorgere del Sole: questo sarebbe possibile infatti se ciascuna costellazione sorgesse in un tempo uguale. Poiché invece si dà il caso che ci sia una grande differenza nelle levate delle 12 costellazioni, non saprà l'ora della notte chi si serve di tale ragionamento.
**5.** Ci si può certamente meravigliare come mai anche Attalo concordi con lui; dice infatti così: "nei [versi] successivi tenta di mostrare come si potrebbe conoscere l'ora della notte attraverso le costellazioni. Poiché infatti l'inizio della notte coincide col tramonto del Sole, e il Sole è sempre in una delle 12 costellazioni zodiacali, è chiaro che per chi sa in quale costellazione è il Sole e in quale grado del segno, è facile capire anche quale costellazione dello zodiaco sorga all'inizio della notte e con quale grado. **6.** Infatti a sorgere sarà il grado che si trova diametralmente opposto a quello occupato dal Sole all'inizio della notte. Se questo è stato stabilito in

128. Tuttavia è evidente, dal v. 559, che ad Arato non interessa osservare le costellazioni zodiacali per conoscere l'ora di notte, ma solo per sapere quando sorgerà il Sole. Gran parte delle critiche successive di Ipparco appaiono pertanto fuori luogo.
129. Cioè tramontano in corrispondenza al sorgere di una determinata costellazione.

precedenza, e inoltre acquisito dall'esperienza, poiché ogni notte sei costellazioni si levano, si saprà anche quanta parte della notte è trascorsa e quanta ancora rimane fino al sorgere del Sole."

**7.** Oltre a ciò, è sfuggito a tutti e due che, neppure vedendo sorta per intero la costellazione zodiacale, sarebbe possibile ricavare esattamente l'ora della notte, seguendo il modo sopra esposto. Se anche infatti ciascuna delle costellazioni zodiacali visibili riempisse un dodicesimo[130] esatto del circolo zodiacale, sbaglieremmo a causa della disuguaglianza dei tempi nelle levate. **8.** Poiché né le costellazioni zodiacali sono uguali ai dodicesimi, né tutte giacciono all'interno del proprio dodicesimo, ma alcune di esse sono più piccole di questo, altre molto più grandi, per esempio il Cancro non occupa neppure un terzo del proprio dodicesimo,[131] mentre la Vergine sconfina in una parte del Leone e in una delle Chele, e il più meridionale dei Pesci è quasi tutto nel dodicesimo dell'Aquario,[132] come sarebbe possibile da un tale sorgere delle 12 costellazioni zodiacali capire l'ora della notte? **9.** Poiché alcune di esse non si trovano neppure ad essere interamente nel circolo zodiacale, ma sono molto più a nord, come il Leone e il più settentrionale dei Pesci, è chiaro che, prendendo come riferimento il sorgere delle figure zodiacali visibili, l'errore che si commette nel calcolare l'ora della notte sarà notevole. **10.** Dunque, sorta la costellazione del Cancro, poiché la testa del Leone è portata su nell'orizzonte sottostante, e sorge insieme a questa l'8° grado del Cancro, Arato, vedendo sorgere la testa del Leone, e credendo che si levi il dodicesimo del Leone, sbaglierà di un'intera ora e mezza.[133] **11.** Un altro esempio: lo Scorpione comincia a tramontare al sorgere del 12° grado e mezzo dell'Ariete, ed è tutto sotto l'orizzonte al sorgere del 6° grado del Toro.[134] Abbiamo dimostrato tutte queste cose nell'opera sulle levate simultanee.[135] **12.** É dunque chiaro che, indipendentemente dal fatto che i 12 segni dello zodiaco si levano in tempi disuguali, non coglierà l'ora esatta chi fa il calcolo sull'apparente disposizione delle 12 costellazioni, proprio perché esse sono di disuguale grandezza e disposte in modo irregolare.

**13.** É evidente dalle cose dette che è veramente impossibile conoscere con esattezza o con buonissima approssimazione l'ora della notte guardando alla costellazione in levata. **14.** E di seguito poi, avendo supposto che le levate simultanee e i tramonti in opposizione delle costellazioni siano calibrate sui segni del circolo zodiacale e non sulle costellazioni,[136] consideriamo anche ciò che in esse è stato detto in modo conforme o difforme rispetto ai fenomeni celesti.

**15.** Si deve considerare, però, innanzitutto, che Arato ha fatto la divisione del circolo zodiacale cominciando dai punti tropicali ed equinoziali, sicché questi punti sono l'inizio dei segni zodiacali, Eudosso invece ha fatto la divisione in modo che i detti

---

130. Fosse cioè estesa esattamente per 30° di longitudine eclittica, come ciascun dodicesimo.
131. In effetti, anche se per Tolomeo il Cancro va da μ ad Acubens, per quasi 14°, per Ipparco va solo da β ad Acubens, poco più di 9°.
132. In verità solo parte del Pesce più meridionale, esclusa la coda ed il cordone, si trovava all'epoca di Ipparco nel dodicesimo dell'Aquario, per poco più di 11° di longitudine.
133. Ipparco esagera un po': l'ultima stella del Cancro, Acubens, sorgeva con il 17° grado del Cancro e assieme a questa faceva la sua comparsa la testa del Leone, con le stelle κ, Alterf, Rasalas ed ε. Il dodicesimo del Leone iniziava a sorgere un'ora più tardi.
134. Ipparco intende dire che, se considerassimo i segni, lo Scorpione comincerebbe a tramontare quando inizia a sorgere la parte dell'eclittica posta in opposizione, a 180° di longitudine, ovvero l'inizio del Toro, e terminerebbe quando finisce di sorgere il 30° grado del Toro.
135. Forse si tratta dell'opera di Ipparco citata da Pappo, matematico greco del quarto secolo: Ἐν τῷ περὶ τῆς τῶν ιβ' ζῳδίων ἀναφορᾶς ("Sulla levata dei dodici segni dello zodiaco").
136. Come detto ripetutamente, Arato si riferisce invece alle costellazioni, e non ai segni.

punti sono al centro, gli uni del Cancro e del Capricorno, gli altri dell'Ariete e delle Chele.[137]
**16.** É ben evidente quindi, da quanto detto, che Arato pone i detti punti all'inizio dei dodicesimi, in accordo con queste [considerazioni]. Parlando infatti dei quattro circoli, i due tropici, l'equatore e lo zodiaco, dice che tre sorgono e tramontano in parallelo fra loro, e ciascuno di essi secondo uno e il medesimo punto sorge e tramonta, mentre lo zodiaco compie le levate e i tramonti nella zona dell'orizzonte compresa fra il sorgere del Capricorno e il sorgere del Cancro. **17.** Scrive:

534 E alcuni sorgono e di nuovo nel profondo tramontano
tutti parallelamente, e uno solo è di ciascuno di essi
ordinatamente da entrambe le parti il tramontare e il sorgere.
Ma questo[138] si muove lungo tanta acqua dell'Oceano,
quanta dal Capricorno che prorompente sorge
al Cancro che si leva si volge.

Da questo è dunque chiaro che i punti tropicali sono supposti all'inizio, l'uno all'inizio del Cancro, l'altro del Capricorno;[139] il circolo zodiacale, infatti, sorge nell'arco di orizzonte compreso fra le levate di questi segni. **18.** Potrebbe sembrare che dichiari questo anche dalle affermazioni sul Leone, quando dice:

149 Dove sono caldissimi percorsi del Sole,
e vuote ormai di spighe appaiono le terre coltivate
al primo sorgere del Sole con il Leone.

Infatti intorno alla levata[140] del Cane ci sono più spesso anche le vampe di calore: essa avviene circa 30 giorni dopo il solstizio d'estate. In questo momento, dopo circa altrettanti giorni, secondo lui, il Sole è all'inizio del Leone. Nel solstizio allora occupa l'inizio del Cancro. **19.** E da tutti gli antichi astronomi, o dalla maggior parte, il circolo dello zodiaco è stato diviso in questo modo.
**20.** Che Eudosso invece abbia posto i punti tropicali al centro dei segni zodiacali,[141] è chiaro dalle affermazioni seguenti: "c'è un secondo circolo nel quale ha luogo il solstizio estivo; in esso si trovano le parti centrali del Cancro". E ancora dice: "poi c'è un terzo cerchio in cui ha luogo l'equinozio; vi si trovano sia le parti centrali dell'Ariete che delle Chele; e c'è un quarto [cerchio] in cui avviene il solstizio d'inverno, sono in questo le parti centrali del Capricorno." **21.** Nel seguito inoltre lo spiega in modo ancora più chiaro; infatti riguardo ai circoli chiamati coluri, quelli che sono descritti attraverso il polo e i punti tropicali ed equinoziali, dice così: "ci sono altri due circoli, che si bisecano ad angolo retto, attraverso i poli del cosmo. In questi si

137. Nonostante quanto dice, e nonostante i tentativi che farà per motivare queste affermazioni nel seguito, Ipparco non riesce ad essere convincente su questo punto: v. appena sotto. Incidentalmente, poi, occorre notare che, almeno dalle citazioni fornite da Ipparco, non si riesce assolutamente a desumere che Eudosso abbia mai suddiviso lo zodiaco in dodici parti uguali.
138. Il circolo zodiacale, l'eclittica.
139. Certamente qui e anche ai versi successivi Ipparco attribuisce a queste affermazioni di Arato un significato che esse non hanno), tanto più che il poeta si riferisce, come sempre, alle costellazioni, e non ai segni zodiacali.
140. Si parla qui di levata eliaca (v. Gemino, *Introduzione ai fenomeni*, XIII).
141. Al centro delle costellazioni zodiacali, non dei segni. Fra l'altro i termini τὰ μέσα o τὸ μέσον "centro, punto mediano" o "centri, punti mediani", usati da Eudosso, probabilmente hanno qui un significato generico, non astronomicamente preciso, e noi li abbiamo infatti tradotti con "parte centrale", "parti centrali". Che Eudosso si riferisca sempre alle costellazioni è evidente anche da altri luoghi commentati da Ipparco (I, 2, 18-20).

trovano le seguenti costellazioni: per primo il polo sempre visibile del cosmo, poi la parte centrale dell'Orsa nel senso della larghezza e la parte centrale del Cancro." E poco dopo aggiunge: "e sia la coda del Pesce Australe che la parte centrale del Capricorno". **22.** Nelle righe successive dice che nell'altro dei due cerchi che passano attraverso i poli si trovano, con altre [costellazioni] che egli enumera, sia le parti centrali delle Chele nel senso della larghezza che il dorso dell'Ariete nel medesimo senso.
**23.** Occorre preliminarmente capire anche che Arato, facendo coincidere la levata di ciascuna costellazione zodiacale con l'inizio del segno, indica chiaramente anche quante delle costellazioni esterne allo zodiaco siano sorte insieme alla detta costellazione zodiacale o siano tramontate in opposizione; il che ha per scopo soprattutto di ricavare dalle costellazioni esterne il dodicesimo dello zodiaco che sta per sorgere.
**24.** Lo riafferma anche sulla levata dei Pesci: dopo aver parlato infatti dell'Idra, [dicendo] che le parti che vanno dalla testa fino alla prima spira non sono più sopra l'orizzonte al levarsi dell'Aquario, mentre il resto del suo corpo lo spingono giù i Pesci, aggiunge:

704 Così anche le braccia sofferenti, e le ginocchia e le spalle
di Andromeda tutte [divise] in due parti, alcune in avanti, altre dietro
si stendono, quando da poco dall'Oceano compaiono
entrambi i Pesci. La parte destra
trascinano questi, quella sinistra dal basso tira
l'Ariete che sorge.

Infatti l'espressione "quando da poco compaiono i Pesci" non significa altro che "quando cominciano a sorgere i Pesci". **25.** Inoltre lo riafferma anche su Argo. Dice infatti che, al sorgere della Vergine, sorge anche la poppa; e che le parti restanti di essa, fino all'albero, sorgono quando sorge tutta intera la Vergine; dicendo infatti:

596 Non certamente poche [stelle] sotto le profondità della terra spinge
la Vergine sorgendo.

E poco dopo dice:

603 Il Cane, sorgendo prima, alza su le altre zampe
trascinandosi dietro la poppa di Argo dalle molte stelle.
Ed essa corre attraverso la terra tagliata in due lungo lo stesso albero,
appena la Vergine tutta intera è sull'orizzonte.

**26.** Anche Eudosso, che Arato ha seguito, fa coincidere allo stesso modo nelle "Levate simultanee"[142] la levata di ciascuna costellazione zodiacale con l'inizio del segno. Sicché anche da questo l'argomento è chiaro.

## Cap. II

**1.** Ciò detto, consideriamo ora le levate simultanee e i tramonti in opposizione da loro esposti.
**2.** Arato, supponendo che l'inizio del Cancro sia in levata, dice che metà della Coro-

142. Probabilmente parte dell'opera di Eudosso *Fenomeni*.

na è tramontata, del Pesce Australe la parte fino alla spina dorsale, dell'Inginocchiato la parte sopra il ventre, il Serpentario fino alle spalle, il Serpente, che il Serpentario tiene, fino al collo, di Artofilace più della metà. **3.** Dice che sorgono la cintura di Orione e le spalle. Le dette costellazioni, quindi, alcune sorgono, secondo lui, con i Gemelli,[143] altre tramontano in opposizione.

**4.** Eudosso, invece, elencando quante [costellazioni] sono sopra la terra quando comincia a sorgere il Cancro, dice che è tutto visibile l'Inginocchiato, metà della Corona, e la testa di Artofilace, e la testa del Serpentario e la coda del Serpente che tiene in mano, delle costellazioni a sud tutto Orione e del Pesce Australe la parte vicino alla testa.[144]

**5.** Per quanto riguarda la Corona l'indicazione di Arato è quasi corretta; infatti dalle parti della Grecia essa comincia a tramontare quando sorge il 23° grado dei Gemelli, e tramonta tutta al sorgere del 3° grado e mezzo del Cancro. Certamente è logico che sia in disaccordo con Eudosso, poiché l'inizio del Cancro per lui è a metà del dodicesimo dei Gemelli. **6.** Per non doverlo indicare in continuazione è da sottolineare una volta per tutte che, dato che essi non fanno le loro ipotesi alle stesse condizioni,[145] non è possibile che il fenomeno si accordi con entrambi. Comunque le loro descrizioni delle levate sono fondamentalmente concordi, tranne pochi casi; Arato infatti ha seguito quanto detto da Eudosso; ma certamente le levate simultanee da loro esposte si accordano più con la divisione dello zodiaco di Arato che con quella di Eudosso. Questo sarà evidente dalle affermazioni che esporremo punto per punto su questo argomento.

**7.** La maggior parte del Pesce Australe deve trovarsi sopra la Terra, e non solo la zona della testa, come dicono: comincia infatti a tramontare quando sorge il 24° grado dei Gemelli, e tramonta completamente al sorgere del 17° grado e mezzo del Cancro. **8.** Dell'Inginocchiato sono tramontate solo la testa [Rasalgethi] e la spalla destra [Kornephoros] con il braccio [γ]; invece la spalla sinistra [δ] e il resto del corpo sono ancora sopra la Terra; mentre infatti tramonta la spalla destra sorge il 27° grado dei Gemelli, e mentre tramonta la spalla sinistra sorge l'8° grado del Cancro. Inoltre dopo la spalla sinistra tramontano le stelle nei fianchi [ζ e ε]. Non è tramontata dunque la parte che va dal suo ventre fino alla testa, come afferma Arato.[146] **9.** Invece sia Arato che Eudosso riguardo al Serpentario danno informazioni in accordo con ciò che appare; infatti le sue spalle tramontano in opposizione con i Gemelli, mentre la stella nella sua testa [Rasalhague] tramonta al sorgere del 2° grado del Cancro. **10.** Del Serpente, che il Serpentario tiene, la coda soltanto è sopra la Terra, come afferma Eudosso, e non è tramontata soltanto la testa, come [dice] Arato. Infatti la sua testa tramonta proprio al sorgere del centro dei Gemelli, la stella sull'estremità della coda [Alya] tramonta al sorgere del 9° grado del Cancro.

**11.** Su Artofilace mi sembra sbaglino completamente; dicono infatti che questo tramonta in opposizione con quattro segni zodiacali, Ariete, Toro, Gemelli, Cancro. **12.**

143. Poiché col sorgere dell'inizio del Cancro già alcune parti di queste costellazioni sono sorte e tramontate, Ipparco intende dire che esse hanno cominciato a sorgere e tramontare con la levata dei Gemelli.

144. Nel complesso, su questo punto Eudosso è un po' più in accordo con i fenomeni rispetto ad Arato.

145. Si riferisce al fatto che, secondo lui, Eudosso ha posto i punti tropicali al centro dei segni, non all'inizio, come Arato.

146. In verità Arato (v. versi 575-576 e qui II, 2, 16) dice che l'Inginocchiato è tramontato solo con le parti superiori, non fino al ventre. E comunque, a differenza di quanto dice Ipparco, quando sorgeva la prima stella del Cancro, la $\sigma^3$, era tramontata solo γ, non Rasalgethi e Kornephoros.

Arato infatti scrive che quando il Toro è in levata:

721 Tramonta Artofilace già con la prima parte
delle quattro che lo trascinano giù, esclusa la mano.

Sicché esso comincia a tramontare in opposizione con l'Ariete che sorge.[147] In linea con questo presupposto, anche quando il Cancro comincia a sorgere, dice così:

579 Neppure Artofilace sarebbe grande da entrambi i lati,
più piccolo nel tempo diurno, più grande già in quello notturno;
infatti il Bovaro che scende con quattro parti insieme
Oceano accoglie.

**13.** Eudosso scrive: "Quando sorge l'Ariete, delle [costellazioni] a nord tramontano i piedi di Artofilace, di quelle a sud la parte restante della Fiera." Soffermandosi, all'inizio della trattazione sulle levate, su quante [costellazioni] sono in tramonto e in levata quando comincia a sorgere il Cancro, dice poi: "Quando sorge il Cancro, non sorge nessuna delle costellazioni a nord, di quelle a sud la Lepre, le zampe anteriori del Cane, e Procione, e la testa dell'Idra; delle [costellazioni] a nord tramonta la testa di Artofilace." É chiaro dunque che anche[148] secondo Eudosso le prime parti di Artofilace tramontano in opposizione all'Ariete, le ultime al Cancro.
**14.** Ma Artofilace non tramonta in opposizione con quattro segni, come costoro dicono, ma con solo due e mezzo scarsi;[149] né comincia a tramontare in opposizione all'Ariete, ma inizia il tramonto[150] quando sorge il 6° grado del Toro, e tramonta tutto intero,[151] tranne la mano sinistra [κ, ι e θ] e il gomito [λ], al sorgere del 18° grado e mezzo del Cancro. **15.** Si deve poi affermare che su Artofilace Arato parla in accordo con i fenomeni, anche se è completamente sbagliato che tramonta in opposizione a quattro segni zodiacali; sorto infatti l'Ariete, dice che esso tramonta in opposizione al Toro e ai Gemelli, ma quando comincia a sorgere il Cancro, che è l'ultimo dei quattro segni zodiacali con i quali anche secondo lui Artofilace tramonta in opposizione, correttamente egli stesso dice che non più molta parte di esso, e sopra la Terra e sotto, è intercettata dall'orizzonte, come anche era al sorgere dei Gemelli, ma che la maggior parte è sotto la Terra, la minore apparendo sopra; l'aggettivo *diurno* infatti indica la parte di cielo sopra la Terra, l'aggettivo *notturno* la parte di cielo sotto

147. Qui Ipparco sembra contraddire se stesso: ha appena detto che Arato fa tramontare la prima parte del Bovaro con il Toro sull'orizzonte orientale, e qui afferma invece che ciò accade con la comparsa dell'Ariete. Per inciso, la cosa avveniva col Toro, con il piede sinistro [Muphrid, τ e υ] che tramontava con le Pleiadi ben alte e Arturo che tramontava con le Iadi già sorte. Ma naturalmente occorre dire che qui e in altri luoghi si gioca anche sull'ambiguità dei vari fenomeni descritti da Arato, nel senso che quando scrive che una costellazione sorge (e lo stesso vale per i tramonti) non è del tutto chiaro se intenda che sta per sorgere (e non se ne vede perciò ancora alcuna stella), che inizia a sorgere (e se ne vedono perciò le prime stelle), se è sorta a metà, per tre quarti o tutta intera. E Ipparco, invece di riconoscere lealmente la giusta approssimazione cogente ad ogni scrittura poetica, qui e in molte altre parti assegna arbitrariamente dei punti fermi dove fermi non sono per nulla. Non essendoci inoltre pervenuta l'opera di Eudosso, non siamo in grado di giudicare in che termini l'astronomo di Cnido parlava delle levate e dei tramonti, ma anche qui è forte l'impressione che Ipparco vada a interpretare questi passi in modo fortemente arbitrario, e non mancheremo di farlo notare nel seguito.
148. Per la verità avrebbe dovuto dire: "solo secondo Eudosso".
149. Ricordiamo sempre che Eudosso e Arato si riferiscono alle costellazioni, e non ai segni. Sotto questo profilo, possiamo dire che per Eudosso effettivamente il Bovaro tramonta con quattro costellazioni che sorgono, mentre per Arato soltanto con tre, Toro, Gemelli e Cancro.
150. Col piede sinistro.
151. Cioè con la testa, Nekkar.

la Terra. **16.** Egli chiarisce questo[152] in diversi casi, come per esempio sull'Inginocchiato, scrivendo:

575 Ma egli volto all'indietro, benché non ancora
con la parte inferiore del ventre, ma nelle parti superiori, è portato via dalla notte.

**17.** I versi seguenti non fanno riferimento al suo stesso tramonto,[153] ma contengono una sua caratteristica:

582 e quando si sia saziato di luce,
impiega più di metà della notte a staccare i buoi dall'aratro,
quando tramonta col calar del Sole.
Quelle notti prendono il nome dal suo tardo tramonto.

Quando, infatti, comincia a tramontare con le Chele quando il Sole è in esse, nel tempo nel quale ha luogo il suo tramonto le notti prendono il nome da Arturo. **18.** Esso tutto il giorno si muove sopra la Terra, per il fatto che sorge con la Vergine e impiega più di metà della notte a tramontare, poiché tramonta insieme con quattro segni. É evidente che secondo lui sorge con la Vergine, nei seguenti versi:[154]

607 Né le Chele sorgenti e tenuamente lucenti
potrebbero passare inosservate, poiché la grande costellazione del Bovaro
sorge tutta intera dominata da Arturo.

**19.** Non solo Arato ed Eudosso, ma anche quasi tutti gli altri astronomi che abbiano parlato del Bovaro, alla stregua di costoro si sono ingannati, pensando che esso tramonti con quattro segni. **20.** Attalo, anche sul fatto che inizia a tramontare non trascinato giù con lo Scorpione, come noi abbiamo dimostrato,[155] ma con le Chele, conferma questa impressione, dicendo in questo modo: "taluni infatti, né prestando attenzione ai fenomeni, né essendo in grado di capire le parole dette dal poeta, hanno fatto affermazioni discordi su di esso. **21.** E alcuni affermano che prima comincia a tramontare con la Vergine, poi con i successivi tre segni, altri che comincia [a tramontare] con lo Scorpione, e termina con l'Aquario. Anche tu ti sei imbattuto in alcuni di costoro confutati da noi e che hanno mutato parere. **22.** In generale, poiché il fenomeno prova, e il poeta dice chiaramente, che [Artofilace] è completamente tramontato quando il Leone è in levata, e comincia a tramontare quando il Toro sta per sorgere; e che il segno zodiacale posto in diagonale[156] con esso, lo Scorpione, è sul tramonto, e sono tramontate poco prima le Chele, con le quali Artofilace comincia a declinare, penso sia chiaro che coloro che sostengono una cosa diversa da noi su questo sono del tutto inesperti dei fenomeni."

**23.** Ma sarebbe stato meglio se non solo avesse detto, con un'affermazione non provata, che quelli che si opponevano, confutati da lui, avevano cambiato parere, ma anche se avesse fornito la prova per iscritto, affinché fossero persuasi anche gli op-

152. Probabilmente intende dire che "notturno", "portati via dalla notte" sono sinonimi di "tramonto".
153. Sempre del Bovaro.
154. Confessiamo che ci sfugge il ragionamento di Ipparco.
155. Ipparco lo ha dimostrato indirettamente, sottintendendo che se il tramonto del Bovaro avviene con la levata del Toro, avverrà anche col tramonto del segno in opposizione, posto a 180° di longitudine, lo Scorpione.
156. Posto in una configurazione tale che, mentre lo Scorpione tramonta, Artofilace sta per cominciare il suo tramonto lungo l'orizzonte nordoccidentale.

positori restanti, che [Artofilace] comincia a tramontare insieme con le Chele, e non con lo Scorpione. **24.** Peraltro, poiché così non ha fatto, non sarà inutile ricordare brevemente attraverso quali argomenti ci siamo convinti che esso inizia a tramontare con lo Scorpione; ne abbiamo trattato più ampiamente nell'opera sulle levate simultanee.[157]
**25.** Si consideri quindi, sull'orizzonte occidentale, la stella più meridionale di quelle nel piede sinistro di Artofilace [υ]. Questa stella si trova 27⅓° a nord dell'equatore, su un punto posto sul cerchio di 360° che passa attraverso i poli. **26.** Per questo motivo il circolo parallelo all'equatore, passante attraverso tale stella, sottende sopra la Terra un arco di 15 parti meno circa 1/20,[158] mentre tutto il circolo è di 24 parti.[159] La metà di detto arco, compresa fra il meridiano e l'orizzonte ovest, è di quasi sette parti e mezzo. **27.** Ma la stella di cui si parla giace sul circolo parallelo a circa 1° delle Chele. Quando essa tramonta deve essere in meridiano, sul circolo parallelo,[160] il 23° grado e mezzo del Capricorno; quando questo [grado] è in meridiano sul parallelo, nello zodiaco dev'essere in meridiano il 22° grado del Capricorno. Ma quando nello zodiaco è in meridiano il 22° del Capricorno, nell'orizzonte sottostante deve sorgere certamente il 6° grado del Toro. **28.** Tutto quanto abbiamo detto è in effetti dimostrato da noi, con l'ausilio di rappresentazioni grafiche, nelle opere composte su tali argomenti.[161] Dunque il piede sinistro del Bovaro tramonta in opposizione al 6° grado del Toro. **29.** Che questo piede tramonti anche per primo, oltre che essere dimostrato da quanto affermato, sarebbe chiaro anche dalla stessa posizione del Bovaro: pur essendo infatti eretto, il Bovaro è in posizione inclinata, cosicché il piede sinistro, che è molto più a sud della testa, è oltre ¾ di segno zodiacale più ad ovest del circolo tracciato tra i poli e la sua testa.
**30.** Su Orione Eudosso è in accordo con i fenomeni, affermando che sorge tutto intero con i Gemelli.
**31.** Arato sostiene che quando comincia a sorgere il Leone, le altre [costellazioni, che calavano] con il Cancro [che sorgeva], e l'Aquila, sono tramontate, sono visibili il piede sinistro e il ginocchio dell'Inginocchiato, mentre le altre parti[162] sono tramontate. E sorgono la testa dell'Idra, e la Lepre e Procione, e le zampe anteriori del Cane.
**32.** Lo stesso concetto di Arato afferma anche Eudosso. Dunque è chiaro che, secondo loro, delle dette costellazioni alcune sorgono insieme al Cancro, altre tramontano in opposizione. Lo stesso si consideri riguardo alle altre costellazioni zodiacali.
**33.** In effetti nelle altre affermazioni sono in accordo con i fenomeni, ma dell'Inginocchiato non solo il ginocchio sinistro [θ] e il piede [ι] sono sopra la Terra, ma anche il ginocchio destro [τ]; questo infatti tramonta al sorgere del 16° grado del Leone. **34.** La Lepre non sorge solo insieme al Cancro, come pensa Arato, ma anche con i Gemelli: comincia infatti a sorgere quando si leva il 27° grado dei Gemelli, cessa di sorgere quando

157. Vedi nota 135.
158. Ovvero $14^h\ 57^m$.
159. Se la stella fosse sull'equatore il circolo sottenderebbe 12 parti sopra la terra e 12 sotto, ovvero la stella passerebbe 12 ore sopra l'orizzonte e 12 sotto. Essendo a nord dell'equatore, la stella rimane molto più tempo sopra che sotto. Il calcolo, compiuto da Ipparco probabilmente con metodi grafici, è davvero straordinariamente preciso, anche se per la latitudine di 36°, non per quella della Grecia: la trigonometria ci dice oggi che una stella con le coordinate assegnate a υ Boo (anche quelle comunque eccezionalmente precise, essendo le coordinate reali 27°18'40") rimaneva a quel tempo sopra la Terra $14^h\ 56^m\ 27^s$.
160. Oggi diremmo sullo stesso parallelo di declinazione.
161. Riferimento ad opere di cui purtroppo non sappiamo nulla.
162. Dell'Inginocchiato.

si leva assieme ad essa l'11° grado e mezzo del Cancro. **35.** Il Cane intero sorge insieme al Cancro, tranne la stella nella coda [Aludra], e non, come dice Arato, soltanto le sue zampe anteriori [Mirzam].[163]
**36.** Quando comincia a sorgere la Vergine

597 Allora {dice}[164] la Cillenia Lira
e il Delfino tramontano e la Freccia ben costruita;
e con essi l'estremità delle ali dell'Uccello fino alla stessa
coda e si coprono d'ombra le rive del Fiume;
tramonta la testa del Cavallo, tramonta anche il collo.
Sorge l'Idra quasi tutta fino alla stessa
Coppa. Il Cane, sorgendo prima, alza su le altre zampe
trascinandosi dietro la poppa di Argo dalle molte stelle.
Ed essa corre attraverso la terra tagliata in due lungo lo stesso albero,
appena la Vergine tutta intera è dalla parte opposta.

Le stesse cose di Arato dice anche Eudosso. E da questo è evidente che Arato riprende quanto è stato detto da Eudosso.
**37.** Per la maggior parte delle affermazioni tuttavia sono in accordo con i fenomeni. E però la Freccia non tramonta solo in opposizione al Leone,[165] ma anche al Cancro; comincia infatti a tramontare al sorgere del 26° grado e mezzo del Cancro, cessa di tramontare al sorgere del 1° grado e mezzo del Leone. **38.** Il Fiume che ha inizio da Orione non comincia a tramontare in opposizione al Leone, come dice Eudosso, ma alla Vergine;[166] infatti la sua stella occidentale e più meridionale [Acamar], che è anche la più brillante, tramonta al sorgere del 7° grado della Vergine.[167]
**39.** Afferma[168] che solo la poppa di Argo sorge insieme al Leone,[169] e che essa si porta in su, tagliata in due lungo l'albero, finché sorge tutta intera la Vergine. **40.** Certamente Attalo, non afferrando il pensiero del poeta, ma supponendo egli dica che, quando sta per sorgere la Vergine, la poppa si è levata fino al centro dell'albero, sostiene che l'aggettivo "tutta" è aggiunto inutilmente all'ultimo verso. **41.** Spiegando infatti gli ultimi dei suddetti versi, aggiunge così: "in questi versi l'aggettivo «tutta» ha funzione pleonastica; infatti sia che si riferisca ad Argo, facendo un simile discorso: «tutta corre attraverso la terra tagliata in due lungo lo stesso albero, mentre la Vergine sull'orizzonte in quell'istante si leva», e quindi secondo il consenso generale questo «tutta» avrebbe funzione riempitiva; sia alla Vergine, e sarà la stessa cosa,[170] a meno che non si accetti che dica che tutta la Vergine è sorta; **42.** Ma non vuole certo affermare questo; se infatti la Vergine intera è sorta, è chiaro che le Chele sono sul punto di levarsi; in

163. Mirzam faceva capolino quando tutte le stelle del Cancro erano già abbondantemente sorte (il 15° grado del Cancro come segno). Col Cane tutto sorto mezza costellazione del Leone era sopra l'orizzonte (5° grado del Leone come segno).
164. Aggiunta di Ipparco.
165. Però Arato (e secondo Ipparco anche Eudosso) parla della Vergine, al verso 596 e prima parte del 597, da Ipparco non riportati. E quindi Ipparco non s'accorge che l'errore dei due è ben maggiore, perché effettivamente la Freccia tramontava in opposizione alla costellazione del Leone.
166. Come dice del resto Arato (con riferimento ai versi prima ricordati).
167. 27° grado del Leone. D'altra parte Acamar era una stella che era alta al massimo 4° perfino a Rodi, e quindi le sue misure di posizione erano pesantemente affette dall'estinzione atmosferica, che la rendeva appena visibile a occhio nudo.
168. Arato.
169. Arato non cita minimamente il Leone a proposito di Argo.
170. Cioè avrà ugualmente funzione riempitiva.

accordo con questa circostanza sostiene che tutta Argo è sorta, scrivendo così:

607 Né le Chele sorgenti e tenuamente lucenti
potrebbero passare inosservate, poiché la grande costellazione del Bovaro
sorge tutta intera dominata da Arturo.
Argo poi ben tutta alta nel cielo ormai sta.”[171]

**43.** Mi pare che Attalo si sia perso, non comprendendo che l’espressione “tagliata in due lungo lo stesso albero” è usata dal poeta in modo esornativo e non vuol dire che “sorge fino al centro dell’albero”. **44.** Il seguito del discorso è tale: quando comincia a sorgere la Vergine il Cane e la Poppa di Argo sono già sorti;[172] questa invero, essendo mezza fino all’albero, si porta su, fin tanto che sorge tutta intera la Vergine. Arato si serve di questa forma costruttiva anche sull’Inginocchiato, dicendo così:

618 Di lui dunque soltanto
una gamba appare insieme con entrambe le Chele,
egli girato fino alla testa dall’altra parte
attende lo Scorpione che sorge e il Saettatore con l’Arco;
essi infatti lo portano su.

**45.** Afferma che quando cominciano a sorgere le Chele, il Bovaro è tutto sorto e pure Argo, dell’Idra solo le parti vicine alla coda sono sotto la Terra, dell’Inginocchiato è sorta la gamba destra solo fino al ginocchio, il resto del suo corpo, tranne il braccio sinistro e la testa, è sopra la Terra quando lo Scorpione è in levata, il braccio sinistro e la testa appaiono in alto quando sorge il Sagittario. **46.** Dice che insieme con le Chele sorgono anche metà della Corona e la coda del Centauro; e che invece tramontano il corpo del Cavallo tutto intero, e la coda dell’Uccello, e la testa di Andromeda e del Mostro Marino la parte che va dalla coda fino alla cresta, e di Cefeo “testa, mano e spalle” e la maggior parte del Fiume.[173] Anche in questi versi Arato ha ripreso in qualche modo quanto scritto da Eudosso.

**47.** Dunque le altre cose sono state dette da loro in modo appropriato. Ma riguardo al Centauro si sono sbagliati; infatti non sono la sua coda e in generale le parti posteriori a sorgere per prime, ma la spalla sinistra [ι][174]; essa è infatti molto più settentrionale, e non comincia a sorgere insieme con la Vergine, come sostengono, ma con le Chele;[175] infatti la sua spalla sinistra sorge insieme col 10° grado delle Chele. **48.** Di Andromeda non solo è tramontata la testa quando le Chele stanno per sorgere, ma anche entrambe le braccia.[176] **49.** Il Mostro Marino poi non comincia a tramontare in opposizione alla Vergine, come loro dicono,[177] ma al Leone; infatti la più meridionale delle due stelle brillanti nella coda [Diphda] tramonta in opposi-

171. Qui finisce la citazione di Attalo.
172. Veramente non sembra che Arato dica questo.
173. Quest’ultima è un’illazione di Ipparco: a proposito del Fiume Arato cita lo Scorpione.
174. La ι sorgeva in effetti ai tempi di Ipparco sei minuti prima della δ, la stella di minore longitudine del Centauro catalogata da Tolomeo nell’*Almagesto*. Tuttavia Tolomeo pone la δ sulla coscia destra posteriore e quindi, almeno per Tolomeo la coda, trovandosi più ad ovest, sorgeva prima della ι. Ma probabilmente per Ipparco il Centauro non si estendeva così tanto ad ovest.
175. Ma è quello che dice anche Arato, ai vv. 625-626.
176. Per la verità tutta fino alla cintura e oltre.
177. È un’illazione di Ipparco, ricavata dal fatto che, in opposizione alle Chele, Arato dice che il Mostro tramonta fino alla cresta e, quindi, poteva cominciare a tramontare in opposizione alla Vergine. In effetti Diphda tramontava quando avevano fatto la loro comparsa alcune stelle della Vergine.

zione al 26° grado e mezzo del Leone; e certo il resto del corpo del Mostro Marino non tramonta in opposizione alla Vergine per intero, ma solo fino alla cresta, come dice Arato.[178] **50.** Di Cefeo tramonta solo la testa; le sue spalle giacciono nella parte sempre visibile, come abbiamo già detto. Non solo sono in errore su questo,[179] ma anche quando dicono che la sua testa tramonta in opposizione alla Vergine;[180] infatti non comincia a tramontare in opposizione alla Vergine, ma alle Chele: la più meridionale delle stelle nella testa [ε] tramonta quando sorge il 7° grado e mezzo delle Chele.

**51.** Arato asserisce che quando comincia a sorgere lo Scorpione tramontano le rimanenti parti del Fiume, e le restanti parti di Andromeda e del Mostro Marino, e di Cefeo la parte che va dalle spalle fino alla cintola, e Cassiopea tutta intera, tranne la parte che va dai piedi fino alle ginocchia.[181] **52.** Dice che sorgono la parte restante della Corona, e la coda dell'Idra, e il resto del corpo del Centauro tranne le zampe anteriori. Dice che sorgono anche la Fiera, che il Centauro tiene, e la testa e le mani del Serpentario, e del Serpente, che il Serpentario tiene, la testa fino alla prima spira, e dell'Inginocchiato le parti che restano, tranne la testa e il braccio sinistro. **53.** Eudosso ha esposto le cose restanti allo stesso modo; ma scrive che sorge non la testa del Serpentario, bensì solo la mano sinistra; e che le parti restanti e la Fiera appaiono sopra la Terra non quando sorge lo Scorpione, ma il Sagittario.

**54.** Per tutto il resto scrivono in accordo con i fenomeni. Ma la Fiera non sorge solo con le Chele, come Arato suppone, ma anche con lo Scorpione: comincia a sorgere infatti quando è portato su il 21° grado delle Chele, il resto si leva senza dubbio insieme con il centro dello Scorpione,[182] come afferma anche Eudosso. **55.** É chiaro che, secondo Eudosso, una parte di essa deve sorgere anche con le Chele. Arato poi è in errore in quel che dice sul Serpentario: infatti solo la sua mano sinistra [Yed Prior e Posterior] sorge insieme con le Chele, la testa [Rasalhague] e la mano destra [ν e τ] invece con lo Scorpione.

**56.** Arato poi dichiara che quando sta per sorgere il Sagittario, la testa e il braccio sinistro dell'Inginocchiato sono già sorti,[183] così pure il corpo del Serpentario, e la coda del Serpente,[184] che il Serpentario tiene, e la Lira, e la parte di Cefeo che va dalla testa fino al petto. **57.** Dice che tramontano il Cane tutto intero, e Orione, e la Lepre e l'Auriga dai piedi alla cintola; dice che la Capra e i Capretti, di cui l'una sulla spalla sinistra, gli altri nella mano sinistra, inoltre la testa e la mano destra, tramontano in opposizione al Sagittario, mentre i piedi tramontano in opposizione allo Scorpione.[185] Dice che tramonta anche Perseo, esclusi il ginocchio destro e il piede, e la poppa di Argo.

**58.** Su quanto detto sopra [Arato] è in accordo con i fenomeni. Ma il braccio sinistro

178. In realtà l'ultima stella del Mostro Marino tramontava in coincidenza con la levata delle primissime stelle delle Chele.

179. Vedi nota 79.

180. Arato però dice che Cefeo tramonta con le Chele e lo Scorpione (vedi anche nota 67).

181. Quest'ultima è una supposizione di Ipparco, (male) interpretando i versi 654-656.

182. In realtà con la prima parte dello Scorpione, sia segno che costellazione.

183. Per la verità Arato (vv. 672-673) dice che sorgono assieme al Sagittario.

184. Qui Ipparco equivoca: Arato al v. 665 parla della spira del Serpente, intendendo probabilmente quella centrale, non della coda.

185. Arato, però (versi 673 e 679-685), dice una cosa diversa. Ovvero, quando sorge il Sagittario, la Capra e i Capretti non sono ancora del tutto tramontati, mentre i piedi scendono col Sagittario e il resto della costellazione col Capricorno.

[Maasym] dell'Inginocchiato sorge insieme con le Chele, e non con lo Scorpione,[186] e infatti la sua stessa spalla sinistra [δ] è piuttosto ad ovest del braccio, come se lui fosse con le braccia tese, come anche Arato stesso dice, e sorge col 3° grado dello Scorpione.[187] Non solo la sua testa dunque è portata su con lo Scorpione. Perciò anche nelle affermazioni precedenti sulla levata dello Scorpione è in errore nel dire:

672 la testa con l'altro braccio
sono tratte su dall'arco.

**59.** Anche su questo Arato ha seguito Eudosso. Ma essi sono in disaccordo rispetto al fenomeno anche per quel che riguarda Cefeo; solo la sua testa infatti tramonta e sorge,[188] le spalle e il petto si muovono nel circolo sempre visibile; a parte questo, la occidentale delle stelle nella sua testa [Garnet Star] sorge con il 28° grado dello Scorpione, l'ultima che è portata su, ed è la più meridionale delle stelle nella testa [ε], sorge insieme con il 5° grado e mezzo del Sagittario, sicché la testa di Cefeo non solo sorge insieme con lo Scorpione, ma anche col Sagittario.[189] **60.** Tutto Perseo tramonta in opposizione allo Scorpione e non, come costoro dicono, il piede destro e il ginocchio destro[190] sono portati sotto in opposizione con il Sagittario. **61.** Hanno sbagliato anche le osservazioni riguardo Argo: infatti comincia a tramontare in opposizione non allo Scorpione, ma al centro delle Chele.[191] Si doveva dire allora che la poppa di Argo è tramontata quando lo Scorpione comincia a sorgere, e non quando si alza il Sagittario.

## Cap. III

**1.** Inoltre dicono che, quando il Capricorno è in levata, Argo e Procione tramontano in opposizione al Sagittario, e che sorgono insieme l'Uccello e l'Aquila, e la Freccia e l'Incensiere.
**2.** In effetti le prime affermazioni sono state fatte da loro in accordo con i fenomeni. Ma quanto detto da entrambi sull'Uccello è sbagliato, e ancor di più sbaglia Arato. Infatti Eudosso afferma che l'ala destra dell'Uccello sorge insieme allo Scorpione, il resto del corpo al Sagittario. Da questo è chiaro che, secondo Eudosso, l'ala destra dell'Uccello deve levarsi con le ultime parti dello Scorpione. **3.** Arato invece sostiene

186. Arato non dice che sorge con lo Scorpione (v. nota 183), ma con il Sagittario, e quindi la critica di Ipparco qui è anche troppo tenera.
187. Qui probabilmente c'è stata un'interpolazione, perché Ipparco fa sorgere una stella che si trova più ad ovest assieme ad un segno che si trova più ad est. Infatti la δ sorgeva col 26° grado delle Chele (Maasym col 29° grado).
188. V. nota 79.
189. Ipparco avrebbe dovuto scrivere il contrario: "non solo col Sagittario, ma anche con lo Scorpione", perché è questo che Arato afferma. Comunque, nel 375 a.C. l'ultima stella della testa di Cefeo a sorgere era la δ, e spuntava con già mezzo arco del Sagittario fuori dell'orizzonte.
190. Di nuovo Ipparco sembra confondersi, perché poco prima ha riportato correttamente quanto scritto da Arato, e cioè che il piede e il ginocchio destro *non* s'immergono col Sagittario (quindi, semmai, col Capricorno). Comunque l'osservazione finale di Ipparco è corretta, sia come segno che come costellazione: in opposizione a Perseo sorge lo Scorpione.
191. Prendendo per l'inizio del tramonto di Argo quello di Canopo, sull'orizzonte orientale doveva ancora finire di sorgere la costellazione della Vergine, mentre c'era il 2° grado delle Chele come segno (tuttavia per la latitudine di 36° la costellazione della Vergine era tutta sorta, e sorgeva il 10° grado delle Chele come segno, v. nota 220). In mancanza di precisi riferimenti è impossibile invece identificare l'ultima stella della poppa, anche consultando Tolomeo. E non è neanche chiaro se Arato al verso 689 parli del tramonto della poppa o della nave intera.

che sorge insieme soltanto al Sagittario.[192] In verità le stelle della punta dell'ala destra dell'Uccello sorgono insieme con le ultime parti delle Chele; la stella nella punta dell'ala sinistra [ζ], che è anche l'ultima a sorgere, si leva insieme al 22° grado del Sagittario.
**4.** Dice che quando comincia a sorgere l'Aquario[193] sono già sorti con il Capricorno la testa e le zampe del Cavallo, e tramontano le parti posteriori del Centauro, e la parte che va dalla testa fino alla prima spira dell'Idra. Eudosso dice che sorgono anche Cassiopea e il Delfino.
**5.** Senza dubbio ciò che hanno detto in comune [Arato ed Eudosso] si accorda quasi con i fenomeni, tranne che la testa dell'Idra comincia a tramontare in opposizione con le ultime parti del Sagittario, e non col Capricorno.[194] Non solo si sono levati la testa e le zampe del Cavallo, ma anche le spalle [Scheat e Markab] e il petto [Sadalbari e λ].[195]
**6.** Alcuni si domandano come mai, mentre sugli altri segni dello zodiaco Arato fissa la loro levata con l'inizio della costellazione sull'orizzonte e stabilisce poi così il sorgere e il tramontare delle altre costellazioni, ponga invece per l'Aquario la levata al centro [della costellazione], dicendo così:

693 Il Cavallo, mentre si leva al centro l'Aquario
si proietta sopra l'orizzonte con le zampe e la testa.

**7.** Essendo questa espressione dubbia, Attalo dice che è un errore e che bisogna scrivere così: "il Cavallo, mentre appena si leva l'Aquario." Ma sfugge ad Attalo ed agli altri la volontà del poeta, forse anche il fenomeno. L'Aquario infatti, che è disposto in senso nord sud, ha il petto e la testa che cadono molto fuori del circolo zodiacale verso nord, e i piedi più a sud del circolo zodiacale; nello zodiaco giace la sua parte centrale. **8.** Poiché dunque presuppone la levata dei dodicesimi sugli inizi dei segni, l'affermazione di Arato che mentre l'Aquario leva il corpo al centro "il Cavallo si proietta sopra l'orizzonte con le zampe e la testa" non si riferisce al centro nel senso della lunghezza del dodicesimo, come la maggior parte e Attalo accettano. **9.** Non mi sembra dunque necessario modificare il verso, come Attalo suggerisce, visto oltretutto che in tutte le copie è scritto così.
**10.** Eudosso poi è in errore su Cassiopea e il Delfino: infatti essi non sorgono con il Capricorno; ma Cassiopea con due segni, il Sagittario e il Capricorno: comincia infatti a sorgere con il 21° grado del Sagittario, e l'ultima parte sorge con il 12° grado del Capricorno. Il Delfino intero sorge col Sagittario, dal 19° grado e mezzo fino al 23° grado e mezzo di esso.
**11.** Arato sostiene che quando iniziano a sorgere i Pesci sono tramontate le ultime

192. Al Capricorno, addirittura (vv. 689-691).
193. Arato però parla di "parte centrale dell'Aquario" (v. 693), come del resto puntualizza Ipparco al par. 6. Come in altre parti della sua opera, Ipparco attribuisce affermazioni inesistenti a coloro che critica, e ancora una volta mostra una certa disattenzione nella critica stessa. I fenomeni relativi al Capricorno e al Cavallo sono, infatti, congruenti con il sorgere della parte centrale dell'Aquario, non con l'inizio del suo sorgere, come Ipparco attribuisce ad Arato.
194. Però Arato non dice che l'Idra tramonta col sorgere del Capricorno (vv. 693-698) ma con quello dell'Aquario, e poi parla di testa e spira del collo, non solo della testa. L'ultima stella della spira, ι Hya, tramontava quando sorgeva il 12° grado del segno del Capricorno, era sorta mezza costellazione del Capricorno e già le prime stelle dell'Aquario (Albali, μ, ν), certo, non la parte centrale.
195. È giusto, quando sorgeva la parte centrale dell'Aquario, anche le stelle nelle spalle e nel petto del Cavallo erano sopra l'orizzonte: è quanto sottintende anche Arato, quando dice che in questa fase la testa e i piedi del Cavallo corrono in alto: ergo, devono essere già sorti prima, col Capricorno, come del resto Ipparco riassume efficacemente al par. 4. Non si capisce pertanto il motivo di questo ripensamento.

parti dell'Idra e il Centauro, è sorto il Pesce Australe, non tutto, ma quasi, e di Andromeda la parte destra nel senso della lunghezza. **12.** Eudosso, in quel trattato che anche Arato segue,[196] dice che sorge anche la mano destra di Perseo, invece nello *Specchio* afferma che è sorto completamente, tranne una piccola parte; e dell'Idra che tramonta la parte fino al Corvo.

**13.** A me sembra che Eudosso su Perseo sia in disaccordo col fenomeno solo quando dice che soltanto la sua mano destra sorge con l'Aquario, perché invero con l'Aquario ne sorge quasi la metà. In ogni caso la stella abbastanza brillante [Mirfak] al centro del corpo si leva insieme con il 27° grado dell'Aquario. **14.** Arato invece è in errore supponendo che tutta l'Idra tramonti in opposizione all'Aquario.[197] Infatti la stella posta nell'estremità della coda [π], sopra la testa del Centauro, tramonta al sorgere dell'11° grado dei Pesci. Meglio dunque dice Eudosso, affermando che la coda dell'Idra rimane ancora visibile. **15.** Sul Centauro entrambi sono in errore. Non è infatti tramontato interamente, in opposizione, quando i Pesci cominciano a levarsi su, ma le sue parti anteriori[198] sono ancora sopra l'orizzonte: la testa infatti [1, 2, 3 e 4 Cen], e la spalla destra [Menkent], tramontano in opposizione ai Pesci.[199] **16.** Sono poi in disaccordo entrambi col fenomeno anche sul Pesce Australe, supponendo che esso sorga per intero quasi insieme all'Aquario. Al contrario la maggior parte di esso si leva su con i Pesci:[200] infatti la più meridionale delle stelle nella coda [γ Gru] si leva insieme con il 3° grado dei Pesci, quella luminosa sul muso [Fomalhaut] col 20° grado e mezzo dei Pesci. **17.** Riguardo ad Andromeda si sono completamente sbagliati entrambi: non sorge infatti con l'Aquario e con i Pesci, come essi dicono, ma con il Capricorno e con l'Aquario;[201] per prima infatti sorge la sua mano destra [ι, κ e λ], indubbiamente con la parte centrale del Capricorno, per ultima si leva la sinistra, insieme col 23° grado e mezzo dell'Aquario.

**18.** Riguardo al sorgere dell'Ariete, Arato dice che la parte destra di Andromeda i Pesci:

708 essi stessi trascinano, quella sinistra dal basso tira<br>
l'Ariete che sorge. E quando esso si innalza<br>
a occidente potresti vedere l'Incensiere, ma nell'altro orizzonte<br>
di Perseo che si alza quanto è fra la testa e le spalle;<br>
tuttavia anche la stessa cintura sarebbe dubbio<br>
se appare con l'Ariete che termina [di sorgere] o col Toro.

---

196. Cioè i *Fenomeni*.

197. Veramente Arato dice, ai vv. 693-98, che quando sorge la parte centrale dell'Aquario, dell'Idra tramontano solo la spira del collo e la testa. Comunque la critica di Ipparco è fondata sia riguardo all'affermazione che Arato ha fatto, sia riguardo a quella che Ipparco gli attribuisce.

198. Dovrebbero essere superiori, però: probabilmente un errore di copiatura.

199. Non sappiamo cosa abbia scritto Eudosso, ma Arato, nei versi 700-01, si limita ad affermare che il Centauro tramonta quando sorgono i Pesci, un'osservazione generale, da poeta, che però è corretta e anche perfettamente congruente con quanto scrive Ipparco nella prima parte della sua affermazione. Ipparco, però, si contraddice nella seconda parte. Infatti, se è vero che la spalla destra e la testa del Centauro non sono ancora tramontate quando iniziano a sorgere i Pesci, non si può dire che queste parti tramontino in opposizione ai Pesci: quando due costellazioni sono in opposizione, cioè quando si trovano a 180° di longitudine celeste o a 12 h di ascensione retta o, più in generale, quando si trovano in posizioni opposte l'una all'altra, su due lati opposti dell'orizzonte, ci si deve riferire alle stesse parti delle costellazioni, per esempio l'inizio, il centro o la fine di ciascuna, non a parti diverse.

200. Non sappiamo quanto scrive Eudosso, ma Arato dice che il Pesce Australe (v. 701) sorge dopo i Pesci, il che è corretto dal punto di vista astronomico.

201. Vero sia in termini di costellazioni che di segni.

**19.** É scritto così l'ultimo verso. É probabile tuttavia che il participio tradotto con "che termina" sia errato. Al principio dell'opera infatti [Arato] suppone le levate zodiacali con l'inizio delle costellazioni zodiacali sull'orizzonte orientale, non con il centro o la fine; nel caso delle altre costellazioni[202] alcune parti tramontano o sorgono quando sono a metà [della levata] le costellazioni dello zodiaco, alcune quando iniziano o terminano [di sorgere].[203] **20.** Anche Attalo su questo ha ben riconosciuto l'errore e pensa che debba essere scritto allora, come suggerisce: "se appare con l'Ariete che sorge o col Toro" o, meglio ancora, "se appare terminare [di sorgere] con l'Ariete" sicché "il terminare" è riferito alla cintura. **21.** Tuttavia gli è sfuggita la volontà del poeta anche in questo verso; dice infatti così: "noi in realtà, anche su questo, e in accordo col poeta e in modo coerente con i fenomeni, crediamo sia necessario che i versi siano scritti in questo modo:

712 tuttavia anche la stessa cintura è dubbio
se appare con l'Ariete che sorge o col Toro,
con esso si volge in su per intero.

**22.** Poiché infatti, quando sta per sorgere l'Ariete, Perseo, per consenso comune, è visibile fino alle spalle poi, quando comincia a levarsi l'Ariete, diventa visibile la cintura di Perseo, per il fatto che la sua comparsa certamente si scosta di poco dall'inizio della levata dell'Ariete, così il poeta è in dubbio se ammettere che essa sia già visibile negli istanti in cui l'inizio di Ariete è sul punto di sorgere o se accettare l'idea comunemente riconosciuta che mentre sta per levarsi il Toro, la cintura di Perseo sia visibile con il resto del corpo. **23.** E modificando i versi in questo modo, sia i fenomeni saranno salvati, sia apparirebbe evidente che il poeta descrive la zona intorno alla cintura non solo con competenza, ma anche con precisione."

**24.** Per prima cosa dunque in queste righe Attalo è in errore, pensando che Arato nei *Fenomeni* abbia usato una tale cura meticolosa che sulle stesse stelle che sono nella cintura di Perseo è in dubbio se, essendo l'inizio dell'Ariete in levata, anch'essa sia già visibile con le spalle e con la testa o poco dopo. **25.** Indipendentemente dal fatto che non solo Arato, ma anche Eudosso nei *Fenomeni*, procedono in modo approssimativo, come abbiamo mostrato, Arato dovrebbe essere in dubbio non solo sulla cintura di Perseo, ma anche su tutte le costellazioni che sorgono insieme con due o più segni zodiacali. **26.** Infine sembrerebbe che Attalo ignori anche i fenomeni, supponendo in verità che le cose siano così incerte riguardo alla cintura di Perseo, come ha detto prima. Non solo infatti la cintura di Perseo[204] appare sopra la Terra quando comincia a sorgere l'Ariete, ma anche quasi tutto Perseo tranne il piede sinistro [Atik e ζ] ed il ginocchio [ε]. **27.** Il suo piede destro [58 Per] infatti sorge insieme con l'8°

202. Le costellazioni non zodiacali.

203. Passo probabilmente corrotto e di difficile interpretazione: infatti sembra che Ipparco voglia dire che la cintura di Perseo sorge con l'inizio della levata dell'Ariete, con non la fine della sua levata, ma allora non si capisce perché, e che senso abbia, che dica che alcune parti delle costellazioni non zodiacali, quale è Perseo, sorgono, com'è ovvio, con l'inizio della levata, altre con la fine, altre col centro della costellazione zodiacale sull'orizzonte. Il passo in questione è peraltro trascurabile e abbastanza inutile, visto che nel seguito viene detto che la cintura di Perseo sorge con i Pesci.

204. Probabilmente la cintura di Perseo può essere individuata nella stella Mirfak (che Ipparco in precedenza aveva collocato al centro del corpo e che Tolomeo porrà nel fianco destro). L'identificazione sembra confermata anche dalla figura di Perseo nell'edizione più antica pervenutaci del *Libro delle stelle fisse* di al-Sufi (MS. Marsh 144, 1009-1010, Bodleian Library, Oxford).

grado dei Pesci,[205] la stella nel ginocchio destro [b Per] col 7°; la stella brillante nella testa della Gorgone [Algol] e nella mano sinistra, che precede di poco la coscia sinistra [ν], si leva col 13° grado dei Pesci. Solo la sua gamba sinistra [Menkib] sorge insieme con l'Ariete.[206] **28.** É chiaro dunque che è in errore, dicendo che i fenomeni saranno rispettati e che apparirà chiaro che Arato descrive con competenza e scrupolosa cura le osservazioni sulla cintura di Perseo.

**29.** Mi pare che Arato sia in difficoltà a causa dello stesso motivo, per il quale lo era anche Eudosso, che Arato ha seguito. Infatti [Eudosso] nel trattato sui fenomeni scrive che la parte destra di Perseo sorge insieme ai Pesci sicché, essendo l'inizio dell'Ariete in levata, solo il lato destro di Perseo appare secondo lui sopra la Terra; ma nell'altro trattato che è intitolato *Specchio*, sostiene che esso[207] tutto intero, tranne una piccola parte, sorge insieme ai Pesci. **30.** Dunque essendo i due trattati in accordo fra loro in quasi tutte le osservazioni circa il sorgere delle costellazioni, ma essendo discordante la descrizione su Perseo, giustamente Arato, non sapendo quale indicazione seguire, dubita se la cintura di Perseo appaia nel cielo con le spalle e con la testa quando sorge l'Ariete o quando ormai si porta su il Toro, come afferma uno dei due trattati di Eudosso.[208] **31.** Non dunque perché sia difficile da percepire per la piccolezza della differenza, come Attalo suppone, [Arato] afferma che è dubbio se la cintura di Perseo appaia già sopra la Terra quando c'è l'inizio dell'Ariete sull'orizzonte, o se, quando ormai comincia a sorgere il Toro, allora anch'essa si leva, ma per non essere in grado di dire per quale motivo sia stata tramandata in due modi diversi.

**32.** Essendo scritto in due maniere, in alcune copie "anche sarebbe dubbio"[209] in altri "anche è dubbio",[210] bisogna scrivere "sarebbe",[211] e non, come fa Attalo, "è"[212]. Infatti la congiunzione ἄν[213] si lega correttamente con "sarebbe". Né vi può essere incertezza a causa del fatto che il termine "dubbie" è espresso al plurale, perché questa forma di costruzione infatti è comune.

**33.** Sia Arato che Eudosso poi sostengono che, quando inizia a sorgere il Toro, Perseo appare tutto sopra la Terra, e pure la mano sinistra dell'Auriga, nella quale si trovano i Capretti, e il piede sinistro, e la parte del Mostro Marino che va dalla coda fino alla cresta. Dice[214] poi che è tramontata una parte del Bovaro.

**34.** Sono in errore tuttavia pensando che solo il lato sinistro dell'Auriga sorga insieme all'Ariete;[215] infatti la sua spalla destra [Menkalinan] si leva insieme col 22° grado dell'Ariete; la più settentrionale delle stelle sulla testa [ξ] sorge ancora prima

205. In realtà col 7° grado dell'Ariete (probabilmente un errore di copiatura).
206. Precisamente col 7° grado e mezzo dell'Ariete.
207. Perseo.
208. Però il dubbio dovrebbe essere fra Pesci ed Ariete, non fra Ariete e Toro.
209. Καὶ κ' ἀμφήριστα πέλοιτο.
210. Καὶ κ' ἀμφήριστα πέλονται.
211. Πέλοιτο.
212. Πέλονται.
213. Ipparco scrive qui ἄν come equivalente del κε di Arato (v. 712). Il κε, soggetto a elisione nel testo (κ'), usato da poeti epici e lirici, modifica il verbo cui è riferito, nel senso di precisare il modo in cui l'azione è presentata, ed ha valori diversi in relazione ai tempi verbali e ai modi con cui si accompagna. Nella lingua greca risulta inappropriato l'uso con l'indicativo presente (πέλονται, da cui il rilievo mosso da Ipparco) mentre con l'ottativo presente (πέλοιτο, corrispondente, in questo caso, al condizionale presente italiano) ha valore potenziale, cioè esprime una possibilità nel presente.
214. Strano passaggio al singolare, tanto più che subito dopo riprende il plurale: è presumibile comunque che si riferisca al solo Arato, il principale bersaglio delle critiche di Ipparco.
215. Veramente hanno parlato del Toro, come ha anche detto Ipparco poco prima. Addirittura Arato dice che Auriga termina di sorgere con i Gemelli (vv. 716-717).

con i Pesci, sicché anche su questo essi sbagliano, in quanto pensano che l'Auriga inizi a sorgere con l'Ariete.

**35.** Ugualmente si sono sbagliati anche sulle descrizioni riguardo al Mostro Marino, in due modi distinti. Infatti non comincia a sorgere, come dicono, con l'Ariete,[216] ma con i Pesci, né si leva con l'Ariete fino alla cresta soltanto, ma quasi interamente, con l'eccezione, delle stelle che sono nella testa, di quella brillante nell'estremità della mascella [Menkar].[217] **36.** Abbiamo detto sopra che in realtà neppure il Bovaro comincia a tramontare in opposizione all'Ariete, come sostengono, ma al Toro.[218]

**37.** Arato poi afferma che, quando cominciano a sorgere i Gemelli, i piedi del Serpentario tramontano "fino alle stesse ginocchia", e che sorgono il Mostro Marino[219] e gli inizi del Fiume.

**38.** Questo è stato detto da loro in accordo con i fenomeni.

216. Anche qui Arato però parla del Toro (vv. 719-720) e dice che completa il sorgere con i Gemelli (vv. 726-727).
217. Che sorge con il Toro.
218. Vedi note 147 e seguenti.
219. Per la verità è tutto sopra l'orizzonte.

## Seconda parte

### Cap. IV

**1.** Queste sono le cose che, fra quelle scritte da Arato ed Eudosso nei *Fenomeni*, mi sembrava utile presentare ed esaminare. **2.** Di seguito proporrò per sommi capi, per ciascuna delle costellazioni fisse, insieme a quale dei 12 segni zodiacali sorga e tramonti, e con quale grado dello zodiaco cominci e finisca di sorgere e tramontare nei luoghi della Grecia, e in generale dove il giorno più lungo è di 14½ ore equinoziali.[220] **3.** Abbiamo raccolto altrove le loro esposizioni specifiche[221] in modo che in quasi ogni località dell'ecumene[222] sia possibile seguire le differenze delle levate e dei tramonti simultanei.
**4.** Per prima cosa dunque esporremo le levate e i tramonti simultanei delle costellazioni a nord del circolo zodiacale, poi quelli delle costellazioni a sud, infine delle 12 zodiacali; intendo le costellazioni zodiacali, dal momento che alcune occupano più spazio del corrispondente dodicesimo, altre meno.[223] Inoltre poiché alcune sono molto più settentrionali, altre più meridionali del circolo zodiacale, possono anticipare e ritardare di molto nelle levate e nei tramonti sui dodicesimi in sé.
**5.** Esporremo inoltre esattamente, per ogni singola [costellazione], il segno zodiacale che è in meridiano, con l'indicazione dei gradi, le stelle fisse che sono in meridiano all'inizio e alla fine delle levate e dei tramonti di ogni costellazione, e in quante ore equinoziali[224] ciascuna costellazione sorge o tramonta. **6.** Riusciremo a specificare esattamente ciascuno di questi punti fino a raggiungere una differenza trascurabile dal vero. Penso sia per te molto facile da capire che tale trattazione è molto più utile di quelle composte dagli antichi[225] e sta nella relazione più stretta con molti principi di astronomia.

### Cap. V

**1.** Il Bovaro dunque sorge insieme allo zodiaco dall'inizio della Vergine fino al 27° grado della Vergine; è in meridiano, quando esso sorge, l'arco di zodiaco che va dal

220. Ovvero a 36° lat. N. A questo punto, in teoria, c'è un importante cambio di prospettiva. Fino a qui Ipparco ha dichiarato di aver esposto le proprie critiche ad Arato ed Eudosso assumendo una latitudine di 37° e una durata del giorno di 14 ore e 36 minuti (v. I, 3 12). Sembra poca cosa la differenza di sei minuti nella durata del giorno, e stranamente anche Ipparco tende a minimizzarla adombrando che si tratti della stessa cosa, ma non è così, come abbiamo già detto, soprattutto quando si ha a che fare con le levate e i tramonti, in particolar modo di astri di particolare declinazione. Anzi, è proprio con le verifiche astronomiche effettuate su questa parte finale dell'opera che ci si rende pienamente conto che Ipparco ha utilizzato un globo celeste tarato per la latitudine di 36° che, ricordiamo, è quasi mezzo grado inferiore a quella della capitale dell'isola di Rodi, da dove quasi certamente l'astronomo osservava. Sembra anche improbabile che Ipparco disponesse di due globi differenti, uno per la latitudine di 37° e uno per quella di 36°: infatti, anche i dati e le osservazioni della prima parte, ad un esame più approfondito, appaiono realizzati a 36° di latitudine. Naturalmente, essendo le considerazioni e i confronti della prima parte più di tipo qualitativo che quantitativo, essi non ne vengono sostanzialmente influenzati. Tuttavia, almeno in un caso, in II, 2, 61, come abbiamo visto, è possibile trovare una discrepanza sensibile (v. nota 191).
221. Si riferisce ovviamente ai fenomeni delle levate simultanee.
222. L'ecumene è la parte abitata della Terra. Per Ipparco, e gran parte degli scrittori antichi, comprendeva l'Europa, il Vicino Oriente e il Nord Africa.
223. Perché, come detto, i dodicesimi sono segmenti uguali di eclittica, larghi ciascuno 30° in longitudine.
224. Anche se lo specifica solo in alcuni punti, è sottinteso che anche altrove Ipparco intende ore di questo tipo.
225. Perché è molto più dettagliata.

26° grado e mezzo del Toro fino al 27° grado dei Gemelli. E come prima stella del Bovaro sorge quella nella testa [Nekkar],[226] ultima quella nel piede destro [ζ].
Quando comincia a sorgere il Bovaro, delle altre stelle sono in meridiano la spalla sinistra [Bellatrix] e il piede sinistro [Rigel] di Orione, che sono ad ovest del circolo meridiano di circa mezzo cubito. Quando esso termina [di sorgere] è in meridiano la stella brillante sulle anche del Cane [Wezen].
Il Bovaro sorge per intero all’incirca in 2 ore equinoziali.
**2.** Quando sorge la Corona sorge insieme ad essa la sezione dello zodiaco che va dal 27° grado della Vergine fino al 4° grado e mezzo delle Chele; è in meridiano dal 25° grado e mezzo dei Gemelli fino al 4° grado e mezzo del Cancro. E prima stella della Corona sorge quella ad ovest della stella più brillante [Nusakan], ultima la più settentrionale [ε] di quelle poste a nord est della più brillante.
É in meridiano, quando comincia a sorgere [la Corona], la stella sulle anche del Cane [Wezen], per ultime[227] la stella ben visibile [β Cnc] ad ovest della testa dell’Idra, nelle zampe meridionali del Cancro, e la più settentrionale delle due stelle [Talitha][228] nelle zampe anteriori dell’Orsa;[229] prossime [al passaggio in meridiano] sono anche le stelle occidentali di quelle intorno alla nebulosa del Cancro [η e θ].
La Corona sorge in ⅔ di ora.
**3.** Quando sorge l’Inginocchiato, sorge insieme lo zodiaco dall’11° grado della Vergine fino al 7° grado e mezzo dello Scorpione; è in meridiano dal 7° grado e mezzo dei Gemelli fino al 14° grado del Leone. E prime stelle dell’Inginocchiato sorgono quella nel piede destro [ν Boo] e quella nel ginocchio destro [τ], ultima quella nella mano sinistra.[230] Delle altre stelle, quando sorge l’Inginocchiato, sono in meridiano per prima[231] la seconda da ovest [ν] delle quattro stelle nei piedi dei Gemelli, per ultima la più meridionale delle due stelle poste intorno a quella brillante sulle reni del Leone [60 Leo].
L’Inginocchiato sorge in 4 ore e 3/5 all’incirca.
**4.** Quando sorge il Serpentario sorge insieme lo zodiaco dal 29° grado delle Chele fino al 23° grado dello Scorpione; è in meridiano dal 3° grado del Leone al 3° grado della Vergine. E prime stelle del Serpentario sorgono quelle nella mano sinistra [Yed Prior e Posterior], in realtà poste anche nel Serpente, ultima la seconda da occidente [θ] delle quattro poste nel suo piede destro.
Per prima è in meridiano la seconda stella da nord [Algieba] delle brillanti nel collo e nel petto del Leone, per ultima quella nella testa del Corvo [Alchiba].
Il Serpentario sorge in 2 ore.
**5.** Quando sorge il Serpente, che il Serpentario tiene, sorge insieme a questo lo zodiaco dall’8° grado delle Chele fino a mezzo grado del Sagittario; è in meridiano dal

226. Prima di questa sorge la stella nella spalla sinistra [Seginus]. Ma probabilmente qui, come anche altre volte in seguito, Ipparco non intende citare proprio le stelle estreme di ogni costellazione ma, per comodità, alcune delle più brillanti, o più importanti, vicine a questi estremi.
227. Qui, e in tutti i luoghi simili successivi, col significato di: “passano in meridiano quando la costellazione finisce di sorgere”.
228. Qui Ipparco si è convertito alla rappresentazione “estesa”, moderna, dell’Orsa Maggiore, dopo averci intrattenuto a lungo nella prima parte del quinto capitolo del primo libro su quella minimalista.
229. La Grande Orsa.
230. Tolomeo presso la mano (nel polso) sinistra dell’Inginocchiato non pone una stella sola, ma tre, ο, ν e ξ. I fenomeni citati si adattano bene alla levata della più orientale, la ο.
231. Qui e in tutti i luoghi simili successivi, col significato di: “passano in meridiano quando la costellazione inizia a sorgere”.

7° grado e mezzo del Cancro fino al 14° grado della Vergine. E per prima stella sorge la più settentrionale delle occidentali nella sua testa [ι], per ultima quella nella punta della coda [Alya].
É in meridiano come prima stella quella brillante nell'acrostolio di Argo [Tureis], che è circa mezzo cubito ad est del circolo meridiano, come ultime l'Annunciatore della vendemmia e la spalla più settentrionale della Vergine [δ], entrambe circa un cubito ad est del circolo meridiano.
Il Serpente sorge in 4 ore e ½.
**6.** Quando sorge la Lira, sorge insieme a questa lo zodiaco dall'8° grado e mezzo dello Scorpione fino al 18° grado dello Scorpione; è in meridiano dal 12° grado e mezzo del Leone fino al 26° grado del Leone. E come prima stella della Lira sorge quella posta a nord accanto alla più brillante [ε], per ultima quella orientale delle due brillanti nella barra [Sulafat].
Sono in meridiano, come prime stelle, quella più meridionale delle due poste da ciascun lato accanto a quella brillante sulle reni del Leone [60], come ultime la stella brillante nella punta della coda del Leone [Denebola], e quella nell'estremità dell'ala sinistra della Vergine [Zavijava], che è circa mezzo cubito ad est del circolo meridiano.
La Lira sorge in 4/5 di ora.
**7.** Quando sorge l'Uccello sorge insieme lo zodiaco dal 26° grado e mezzo delle Chele fino al 22° grado del Sagittario; è in meridiano dall'inizio del Leone fino al 9° grado e mezzo delle Chele. E per prima stella sorge la più settentrionale di quelle nell'ala destra [κ]; per ultima la più meridionale delle stelle nell'ala sinistra [ζ].
Prima stella in meridiano è quella brillante nel cuore del Leone [Regolo], ultime la più meridionale delle stelle sotto la spalla destra del Centauro [η],[232] e Arturo, che è circa mezzo cubito ad est del meridiano.
L'Uccello sorge in 4 ore e 2/5.
**8.** Di Cefeo sorgono solo le parti della testa. Lo zodiaco sorge insieme a lui dal 26° grado e mezzo dello Scorpione fino al 5° grado e mezzo del Sagittario; è in meridiano dall'8° grado e mezzo della Vergine al 21° grado della Vergine. E come prima stella sorge la occidentale delle tre nella testa [Garnet Star],[233] come ultima la più meridionale di quelle nella testa [ε].
É in meridiano come prima stella quella luminosa nella coda del Corvo [Algorab],[234] che è circa mezzo cubito ad est del meridiano, come ultima il gomito sinistro della Vergine [θ], che è circa mezzo cubito ad ovest del meridiano.
Cefeo sorge in circa ¾ di ora.
**9.** Quando sorge Cassiopea, sorge lo zodiaco dal 22° grado del Sagittario al 12° grado del Capricorno; è in meridiano dall'11° grado delle Chele al 2° grado e mezzo dello Scorpione. E prima stella sorge quella brillante nel seggio [κ],[235] ultima quella nella testa [ζ].
Sono in meridiano, come prime stelle, la più meridionale delle stelle sotto la spalla destra del Centauro [η], circa mezzo cubito ad ovest del meridiano, e Arturo, come

232. In I, 8, 23 e in I, 11, 20 Ipparco la identifica come la mano destra del Centauro.
233. Per Tolomeo questa stella non fa parte di Cefeo, ma appartiene alle "non figurate".
234. Per Tolomeo, che non pone stelle sulla coda del Corvo, questa stella è nell'ala orientale.
235. "Quella brillante nel seggio" dovrebbe essere Caph, che per Tolomeo è al centro del seggio. In altri tre luoghi Ipparco cita la stella debole, o piccola, nel seggio, che è sempre identificabile, grazie ai fenomeni, con la κ. Anche qui i fenomeni si adattano benissimo alla κ, e per nulla a Caph. Forse si tratta di un errore di copiatura dei codici.

ultime la stella centrale nella fronte dello Scorpione [Dschubba], e la stella della Corona ad ovest di quella luminosa [Nusakan].
Cassiopea sorge in 1 ora e ⅓.
**10.** Quando sorge Andromeda sorge insieme a lei lo zodiaco dal 15° grado[236] del Capricorno fino al 23° grado e mezzo dell'Aquario; è in meridiano dal 5° grado e mezzo dello Scorpione al 7° grado e mezzo del Sagittario. E come sua prima stella sorge la più settentrionale di quelle nella mano destra [λ], come ultima la stella nella mano sinistra [ζ][237].
É in meridiano come prima stella quella della Corona contigua alla brillante [Nusakan], come ultima il gomito sinistro dell'Inginocchiato [μ].
Andromeda sorge in 2 ore e ⅛.
**11.** Quando sorge il Cavallo sorge insieme ad esso lo zodiaco dall'inizio del Capricorno fino al 21° grado dell'Aquario; è in meridiano dal 20° grado e mezzo delle Chele al 5° grado e mezzo del Sagittario. E prima sua stella sorge la più meridionale delle zampe anteriori [κ], ultima la brillante sulle reni [Algenib].
Sono in meridiano, come prime stelle, quella al centro della chela meridionale [ι] delle Chele, come ultime le tre stelle del Serpentario [66, 67 e 68] poste in linea retta presso la spalla destra, che sono esterne ad esso, e la stella brillante [π] nella coscia sinistra dell'Inginocchiato, poco ad ovest del meridiano.
Il Cavallo sorge in 3 ore.
**12.** Quando sorge la Freccia sorge insieme ad essa lo zodiaco dal 5° grado del Sagittario fino al 9° grado e mezzo del Sagittario; è in meridiano dal 19° grado della Vergine fino al 25° grado della Vergine. E per prime sorgono le stelle nella cocca [α e β], per ultima quella nella punta [γ].
É in meridiano come prima stella quella nel gomito sinistro della Vergine [θ], poco ad est del meridiano, come ultime Spiga, poco ad ovest del meridiano, e la stella sotto la spalla sinistra del Centauro [HIP 65 936].
La Freccia sorge in ⅓ d'ora.
**13.** Quando sorge l'Aquila, sorge con essa  lo zodiaco dal 9° grado del Sagittario fino al 13° grado e mezzo del Sagittario; è in meridiano dal 24° grado della Vergine fino al 29° grado e mezzo della Vergine. E prima stella dell'Aquila sorge la più settentrionale delle due piccole nelle ali [μ], ultima la più meridionale delle tre brillanti nel corpo [Alshain].
É in meridiano come prima stella Spiga, come ultima la più settentrionale delle stelle nella testa del Centauro [4 Cen].
L'Aquila sorge in 2/5 di ora.[238]
**14.** Quando sorge il Delfino sorge con esso lo zodiaco dal 19° grado e mezzo del Sagittario fino al 23° grado e mezzo del Sagittario; è in meridiano dal 7° grado e mezzo delle Chele fino al 13° grado delle Chele. E prime stelle sorgono le occidentali [Rotanev e Sualocin] delle quattro nel rombo, ultima la più meridionale delle stelle nella coda [κ].
É in meridiano come prima stella il piede sinistro [λ] della Vergine, come ultime la stella più settentrionale [HIP 72010] delle stelle nel tirso del Centauro, e la stella [φ]

236. 10° grado del Capricorno.
237. Per Tolomeo è nel braccio.
238. L'Aquila di Ipparco era una costellazione molto più piccola di quella attuale, come sarà del resto per Tolomeo, che classificherà le stelle della costellazione "estesa" come slegate intorno all'Aquila.

posta a nord del ginocchio e del piede destro della Vergine, che è ad ovest del meridiano di quasi mezzo cubito.
Il Delfino sorge in ¼ d'ora.
**15.** Quando sorge Perseo sorge insieme a lui lo zodiaco dal 25° grado del Capricorno al 13° grado e mezzo dell'Ariete; è in meridiano dal 15° grado e mezzo dello Scorpione fino al 7° grado e mezzo del Capricorno. E per prima sorge la nebulosa nel falcetto [NGC 869-884], per ultime le stelle poste nel piede sinistro sopra la Pleiade [Atik e ζ].[239]
Sono in meridiano come prime stelle quella brillante al centro dell'Incensiere [α], e quella nel braccio dell'Inginocchiato [γ], ad ovest della spalla destra, come ultime la più settentrionale di quelle nel ginocchio del Capricorno [ψ], e la più settentrionale [κ] delle stelle nell'ala destra dell'Uccello, che è ad ovest del meridiano di circa mezzo cubito.
Perseo sorge in 3 ore e 5/6 di ora.
**16.** Quando sorge l'Auriga sorge insieme lo zodiaco dal 10° grado e mezzo dei Pesci fino al 15° grado e mezzo del Toro; è in meridiano dal 20° grado del Sagittario fino al 29° grado del Capricorno. E per prime sorgono le stelle nella testa [δ e ξ], per ultima quella nel piede destro [Elnath].
Sono in meridiano, per prime, la occidentale delle stelle nel mantello [43 Sgr] e la centrale di quelle nel dorso del Sagittario [ψ], e la seconda, contata dall'estremità, della coda del Serpente, che il Serpentario tiene [Alya];[240] per ultime sono in meridiano la stella brillante nel muso del Cavallo [Enif], la stella orientale di quelle nella zampa sinistra dell'Uccello [ξ], e la più meridionale [η][241] delle due occidentali ben visibili nella spalla destra di Cefeo.
L'Auriga sorge in 3 ore.

## Cap. VI

**1a.** Riguardo dunque alle levate delle costellazioni a nord del circolo zodiacale questo è ciò che accade alla latitudine considerata; simile è il discorso in relazione ai tramonti.
**1b.** Quando tramonta il Bovaro, tramonta insieme a lui lo zodiaco dal 6° grado dello Scorpione al 18° grado e mezzo del Capricorno; è in meridiano dal 22° grado del Capricorno al 4° grado dell'Ariete. E prima stella tramonta la più meridionale di quelle nel piede sinistro [υ], ultima la più settentrionale di quelle nella mazza [ν].
Sono in meridiano, delle altre fisse, per prime quella brillante nel centro della coda dell'Uccello [Deneb], e la occidentale delle stelle che giacciono a sud vicino a quelle nella coda del Capricorno [Nashira], per ultime la nebulosa [NGC 869-884] nel falcetto di Perseo, la orientale delle tre nella testa dell'Ariete [Hamal], e il Nodo del cordone dei Pesci [Alrescha].

239. Qui al singolare, mentre in tutto il primo libro Ipparco usa il plurale.
240. Alya, come detto da Ipparco in altri cinque punti del trattato, è l'ultima stella della coda del Serpente, non la penultima, ma d'altra parte il fenomeno si adatta molto meglio ad essa che non alla η, la penultima stella: Alya è solo 1¾° ad est del meridiano, mentre η ben 8° ad ovest. Manitius avanza l'ipotesi che in questo luogo Ipparco consideri come ultima stella della coda del Serpente δ Aquilae (che Manitius cita erroneamente come δ Alcinoi, invece che come δ Antinoi, con riferimento ad Antinoo, costellazione adiacente all'Aquila, già citata da Tolomeo e ancora usata in qualche atlante stellare ai suoi tempi), ma non ne spiega il motivo.
241. Per Tolomeo è sopra il gomito.

Il Bovaro tramonta in 4 ore equinoziali e ⅔ di ora.
**2.** Quando tramonta la Corona, tramonta insieme lo zodiaco dal 23° grado del Sagittario fino al 3° grado e mezzo del Capricorno; è in meridiano dall'inizio dei Pesci fino al 13° grado e mezzo dei Pesci. E prima stella tramonta la più luminosa di quelle nella Corona [Alphecca], ultima la più debole [ι], che è anche l'ultima della parte orientale del semicerchio.
Delle altre stelle sono in meridiano, come prime, quella luminosa nella punta della coda del più meridionale dei Pesci [ω], e le stelle occidentali [27 e 30] del parallelogramma posto a sud di questo, come ultime la stella al centro del corpo di Cassiopea [Navi], e quella nella mano sinistra di Andromeda [ζ].
La Corona tramonta in circa 1 ora.
**3.** Quando tramonta l'Inginocchiato tramonta insieme lo zodiaco dal 14° grado del Sagittario fino al 16° grado dell'Aquario; è in meridiano dal 23° grado dell'Aquario fino al 7° grado e mezzo del Toro. E come prima stella tramonta quella nella mano destra [γ Ser],[242] come ultima quella nel piede sinistro.[243]
Sono in meridiano, delle altre stelle, per prima la orientale delle stelle nella quarta curva [101 Aqr] dell'acqua dell'Aquario, la più settentrionale [τ] di quelle vicine nel corpo del Cavallo, e la spalla sinistra di Cefeo [ι],[244] che si trova circa mezzo cubito ad est del meridiano; per ultime sono in meridiano la più meridionale delle stelle brillanti orientali [τ] del quadrilatero nel Mostro Marino, e la stella posta a sud, senza nome e luminosa [υ Cet].[245]
L'Inginocchiato tramonta in 4 ore e 3/5 all'incirca.
**4.** Quando tramonta il Serpentario, tramonta insieme lo zodiaco dall'11° grado dello Scorpione fino al 2° grado del Capricorno; è in meridiano dal 25° grado del Capricorno al 10° grado e mezzo dei Pesci. E prima stella tramonta quella nel piede sinistro [ψ],[246] ultima quella nella testa [Rasalhague].
Sono in meridiano, delle altre stelle, per prime la stella orientale di quelle nella coda del Capricorno [Deneb Algedi], e la più meridionale delle stelle nell'ala sinistra dell'Uccello [ζ], per ultime la stella nella testa [ζ] e quella piccola nel seggio[κ] di Cassiopea, la più settentrionale di quelle nel petto di Andromeda [δ],[247] e la più meridionale delle stelle nella coda del Mostro Marino [Diphda], che resta un po' ad est del meridiano.
Il Serpentario tramonta in circa 3 ore.
**5.** Quando tramonta il Serpente che il Serpentario tiene, tramonta insieme ad esso lo zodiaco dal 25° grado dello Scorpione fino al 9° grado del Capricorno; è in meridiano dall'8° grado dell'Aquario fino al 19° grado e mezzo dei Pesci. E come prime

242. Seguendo il suggerimento di Manitius, indichiamo questa stella, pur se Tolomeo la assegna alla costellazione adiacente, sia perché i fenomeni vi si adattano bene, sia perché va a completare la stessa descrizione di Tolomeo, nel senso che la γ Ser sta proprio sul prolungamento della linea che comprende Kornephoros, γ Her e Marsic, che costituiscono rispettivamente per Tolomeo la spalla, il braccio e il gomito della figura.
243. Nel piede sinistro Tolomeo pone tre stelle, la 74, la 77 e la 82, ma per l'uso del singolare e perché esse sono molto deboli è difficile che Ipparco intendesse indicare queste; è più probabile che intendesse invece la ι, molto più luminosa, e che Tolomeo pone nella gamba.
244. Per Tolomeo è nel braccio.
245. Tuttavia la υ e la τ si trovavano rispettivamente 28° e 34° ad ovest del meridiano; probabilmente si tratta di errori nei manoscritti.
246. Per Tolomeo, che ne ha altre due sul piede sinistro, la ω e la ρ, è la più meridionale della gamba, ma è al suo tramonto che corrispondono bene i fenomeni citati.
247. Per Tolomeo è nella parte superiore del dorso.

stelle tramontano quelle nel corpo, che sono in comune anche con la mano sinistra del Serpentario [Yed Prior e Posterior], come ultima quella che è nell'estremità della coda [Alya].
Delle altre stelle sono in meridiano, per prime quella nel piede destro di Cefeo [κ], e quella al centro della brocca dell'Aquario [ζ],[248] per ultime la più meridionale delle stelle nella coda del Pesce settentrionale [χ], e la stella nel ginocchio di Cassiopea [Ruchbah].
Il Serpente tramonta in circa 3 ore.
**6.** Quando tramonta la Lira, tramonta insieme lo zodiaco dal 4° grado dell'Aquario al 12° grado dell'Aquario; è in meridiano dal 22° grado e mezzo dell'Ariete al 3° grado del Toro. E prima stella tramonta la occidentale delle due brillanti nella barra [Sheliak], ultima quella che giace a nord accanto alla brillante [ε].
Sono in meridiano, delle altre stelle, per prima quella centrale nella coda dell'Ariete [ζ], per ultime la più settentrionale e più visibile delle stelle fra la più vivida della Pleiade e le Iadi [37 Tau], che sono sulla parte sinistra della fronte del Toro, circa ⅔ di cubito ad est del meridiano, la seconda e luminosa delle stelle da nord [δ] nella grande curva del Fiume che parte da Orione, che è mezzo cubito ad est del meridiano, e la stella fra il taglio del Toro e quella nella scapola [30 Tau], che è ad ovest del meridiano di circa ⅔ di cubito.
La Lira tramonta in ⅔ di ora.
**7.** Quando tramonta l'Uccello, tramonta con esso lo zodiaco dal 4° grado e mezzo dell'Aquario fino al 14° grado dei Pesci; è in meridiano dal 23° grado e mezzo dell'Ariete fino al 12° grado dei Gemelli. E prima stella tramonta quella nel becco [Albireo], ultima la più settentrionale di quelle nell'estremità dell'ala destra [κ].
Sono in meridiano, delle altre stelle, per prime la stella brillante nella coscia destra di Perseo [δ], e la stella orientale di quelle nella coda dell'Ariete [63 Ari], per ultime la terza stella da ovest di quelle che si trovano nei piedi [Alhena], e la occidentale delle tre nelle ginocchia [Mebsuta] dei Gemelli, che sono circa mezzo cubito ad ovest del meridiano.
L'Uccello tramonta in 3 ore e ⅙.
**8.** Di Cefeo tramonta soltanto la regione della testa. E tramonta con lui lo zodiaco dal 7° grado e mezzo dell'Ariete al 14° grado dell'Ariete; è in meridiano dal 9° grado del Cancro fino al 14° grado del Cancro. E prima stella tramonta la più meridionale delle tre nella testa [ε], ultima la più settentrionale di esse [λ].
Delle altre stelle sono in meridiano, come prime, la più meridionale delle stelle orientali intorno alla nebulosa del Cancro [Asellus Australis], un po' ad est del meridiano, e la stella brillante nell'acrostolio di Argo [Tureis], come ultime la stella brillante nelle ginocchia anteriori dell'Orsa[249] [α Lyn],[250] un po' ad est del meridiano, la centrale delle tre della chela meridionale del Cancro [Acubens], la stella sull'inizio del collo dell'Idra [ζ],[251] e la stella brillante al centro della fiancata di Argo [Regor], un po' ad est del meridiano.
Cefeo tramonta in ⅓ d'ora.
**9.** Quando tramonta Cassiopea, tramonta con lei lo zodiaco dal 21° grado dell'Ariete

248. Per Tolomeo è nella mano destra, il che tuttavia non osta all'identificazione di Ipparco.
249. La Grande.
250. I fenomeni si adattano, anche se meno bene, pure alla θ UMa, che Tolomeo pone sul ginocchio sinistro, tuttavia nel libro III, 5, 1b, le coordinate, assai più precise, riferite a questa come "stella oraria" tolgono ogni ragionevole dubbio. Ricordiamo che la costellazione della Lince ancora non esisteva, all'epoca.
251. Ma per Tolomeo è "quasi sulla mascella".

fino al 23° grado e mezzo del Toro; è in meridiano dal 23° grado del Cancro fino al 5° grado della Vergine. E prima stella tramonta quella nella testa [ζ], ultima quella nei piedi [ε].
Delle altre stelle sono in meridiano come prime la zampa anteriore [Subra] e la più settentrionale delle stelle orientali nella testa del Leone [Rasalas], e la stella più luminosa dell'Idra [Alphard], come ultime la penultima di quelle luminose nella coda del Drago [κ], e quelle nel becco [Alchiba] e nella testa del Corvo [ε].
Cassiopea tramonta in 2 ore e ⅔ di ora.
**10.** Quando tramonta Andromeda, tramonta insieme a lei lo zodiaco dal 21° grado e mezzo dei Pesci fino al 27° grado e mezzo dell'Ariete; è in meridiano dal 20° grado e mezzo dei Gemelli fino al 2° grado del Leone. E prima stella tramonta quella nella testa [Alpheratz], ultima la più settentrionale delle stelle nel piede destro [φ Per].
Delle altre stelle sono in meridiano all'inizio[252] quella nel muso della Grande Orsa [Muscida], quella nella testa del Gemello occidentale [Castore], e la più settentrionale di quelle nella testa del Cane [θ]; alla fine sono in meridiano la stella nella punta della coda del Drago [Giausar], la centrale delle tre in linea retta[253] nel collo del Leone [Algieba], e dell'Idra la terza delle quattro verso est dopo quella brillante [$\upsilon^1$].
Andromeda tramonta all'incirca in 3 ore.
**11.** Quando tramonta il Cavallo, con esso tramonta lo zodiaco dal 12° grado e mezzo dell'Aquario al 13° grado dei Pesci; è in meridiano dal 3° grado e mezzo del Toro al 10° grado e mezzo dei Gemelli. E prima stella tramonta quella brillante nella sua bocca [Enif], ultima quella brillante sulle reni [Algenib].
Sono in meridiano, delle altre stelle, come prime la più settentrionale delle stelle fra la vivida della Pleiade e le Iadi [37 Tau], che sono sulla parte sinistra della fronte del Toro, e la stella nella scapola del Toro [λ],[254] un po' ad est del meridiano; come ultime sono in meridiano il terzo piede da ovest [Alhena] e la occidentale delle tre stelle nelle ginocchia dei Gemelli [Mebsuta].
Il Cavallo tramonta in 2 ore e ½.
**12.** Quando tramonta la Freccia, tramonta con questa lo zodiaco dal 26° grado e mezzo del Capricorno fino al 1° grado e mezzo dell'Aquario; è in meridiano dal 14° grado dell'Ariete fino al 19° grado e mezzo dell'Ariete. E come prime stelle tramontano quelle nella cocca [α e β], come ultima quella nella punta [γ].
Delle altre stelle sono in meridiano, come prime, la stella nella spalla destra di Perseo [γ], la più settentrionale delle tre senza nome in linea retta sopra la coda dell'Ariete [39 Ari],[255] e quella nella sua schiena [ν], circa mezzo cubito ad ovest del meridiano, e la occidentale delle tre brillanti nella mascella meridionale del Mostro Marino [δ]; come ultime sono in meridiano la stella brillante al centro del corpo di Perseo [Mirfak], poco ad ovest del meridiano, e la orientale delle stelle nella mascella meridionale del Mostro Marino [Menkar], che resta un po' ad est del meridiano.
La Freccia tramonta in ⅓ di ora.
**13.** Quando tramonta l'Aquila, tramonta con questa lo zodiaco dal 16° grado e mezzo del Capricorno fino al 23° grado del Capricorno; è in meridiano dal 2° grado

252. Qui e in tutti i luoghi simili successivi, con lo stesso significato di "per prime", "come prime", dei luoghi precedenti, ovvero passano in meridiano all'inizio del tramonto della costellazione.
253. Non si capisce dove si trovi questa linea retta nel popolare "falcetto" del Leone.
254. Per Tolomeo è nel petto.
255. In realtà questa stella, non catalogata da Tolomeo, è ben a nord della coda, ed è il vertice settentrionale di un triangolo scaleno.

dell'Ariete fino all'8° grado dell'Ariete. E come prime stelle tramontano le due piccole nelle ali [μ e σ], come ultima la più settentrionale delle tre brillanti nel corpo [Tarazed].
Delle altre stelle sono in meridiano, per prime, la stella nel piede sinistro di Andromeda [Almach], la centrale delle stelle nella testa dell'Ariete [Sheratan], e la più meridionale delle orientali nel quadrilatero del Mostro Marino [τ], per ultime la orientale di quelle nella base del Triangolo [γ], circa mezzo cubito ad ovest del meridiano, e la occidentale delle stelle nella mascella settentrionale [$\xi^1$] del Mostro Marino.[256]
L'Aquila tramonta in ⅓ di ora.
**14.** Quando tramonta il Delfino, tramonta insieme lo zodiaco dal 2° grado dell'Aquario fino al 7° grado e mezzo dell'Aquario; è in meridiano dal 19° grado e mezzo dell'Ariete fino al 29° grado dell'Ariete. E come prima stella tramonta la occidentale di quelle nella coda [ε], come ultima la più settentrionale delle stelle nel lato orientale del rombo [γ].
Delle restanti stelle sono in meridiano per prime la stella brillante al centro del corpo di Perseo [Mirfak], poco ad ovest del meridiano, e la orientale delle stelle nella mascella meridionale [Menkar] del Mostro Marino, poco ad est del meridiano, per ultime la stella nel ginocchio sinistro di Perseo [ε], e l'estremità settentrionale del lato occidentale della Pleiade [Taigete].
Il Delfino tramonta in ½ ora.
**15.** Quando tramonta Perseo, tramonta con lui lo zodiaco dal 2° grado del Toro fino al 29° grado del Toro; è in meridiano dal 5° grado e mezzo del Leone fino al 9° grado e mezzo della Vergine. E prima stella tramonta la più settentrionale delle occidentali nella testa della Gorgone [π], ultima quella nel ginocchio destro [b Per].
Delle altre stelle sono in meridiano, all'inizio, la stella nel ventre del Leone [ρ],[257] circa mezzo cubito ad est del meridiano, la quarta delle quattro verso est dopo quella brillante dell'Idra [λ], che si trova circa mezzo cubito ad ovest del meridiano, e l'ultima delle stelle nella chiglia di Argo [N Vel]; alla fine sono in meridiano la stella nella coda del Corvo [Algorab][258] e quella nelle sue zampe [β].
Perseo tramonta in 2 ore e ⅓.
**16.** Quando tramonta l'Auriga, tramonta con lui lo zodiaco dal 23° grado del Toro fino al 1° grado e mezzo del Cancro; è in meridiano dal 2° grado della Vergine fino al 22° grado delle Chele. E prima stella tramonta quella nel piede sinistro [ι], ultime quelle nella testa [ξ e δ].
Delle altre stelle sono in meridiano, all'inizio, la stella nella punta della coda del Drago [Giausar],[259] e la più settentrionale della Corona,[260] alla fine la stella al centro della chela meridionale delle Chele [ι].
L'Auriga tramonta in poco più di 3 ore.
**17.** Dopo aver trattato tutte le costellazioni a nord dello zodiaco, sulle restanti daremo conto nel successivo [libro] ai fini di sviluppare la trattazione in modo simmetrico.

256. Per Tolomeo è "quasi sulla cresta".
257. Per Tolomeo è sull'ascella sinistra.
258. Vedi nota 234.
259. Si trovava tuttavia oltre 5° a ovest del meridiano, mentre il fenomeno si adatterebbe alla κ Dra.
260. Qui ci dev'essere un errore dei copisti, perché la Corona si trovava da tutt'altra parte, circa 45° ad oriente del meridiano. Forse si può ipotizzare che Ipparco intendesse la Chioma di Berenice, la "Treccia" di Tolomeo, la cui stella settentrionale delle tre più luminose, la γ, si trovava a 4½° dalla culminazione.

# Libro III

## Cap. I

**1a.** Dopo aver parlato, o Escrione, nel libro precedente, delle costellazioni a nord del circolo zodiacale, insieme a quale dei 12 segni dello zodiaco ciascuna di esse sorga e tramonti, e quale segno zodiacale sia in meridiano quando ciascuna sorge e tramonta, e quali delle stelle fisse siano in meridiano quando ciascuna comincia a sorgere o a tramontare e anche quando termina [di sorgere e tramontare], e inoltre in quante ore equinoziali ciascuna sorga o tramonti, daremo ora le stesse indicazioni su ciascuna delle costellazioni a sud del circolo zodiacale e sulle stesse 12 costellazioni zodiacali.
**1b.** Delle costellazioni a sud del circolo zodiacale, quando sorge l'Idra sorge insieme lo zodiaco dal 18° grado e mezzo del Cancro fino al 15° grado e mezzo delle Chele; è in meridiano dal 2° grado e mezzo dell'Ariete al 18° grado del Cancro.[261] E prima stella sorge la più settentrionale di quelle nelle fauci [δ],[262] ultima quella nell'estremità della coda [π].
Sono in meridiano, come prime stelle, la nebulosa nel falcetto di Perseo [NGC 869-884], il Nodo dei Pesci [Alrescha], che è circa mezzo cubito ad est del meridiano, e la orientale delle tre stelle luminose nella testa dell'Ariete [Hamal], che è circa mezzo cubito ad est del meridiano; come ultime sono in meridiano la più meridionale delle stelle occidentali nella testa del Leone [Alterf],[263] e la più settentrionale delle stelle al centro dell'albero di Argo [α Pyx].
L'Idra sorge in 7 ore e 1/15.
**2.** Quando sorge la Coppa, sorge insieme ad essa lo zodiaco dal 26° grado e mezzo del Leone al 10° grado e mezzo della Vergine; è in meridiano dal 20° grado e mezzo del Toro fino al 7° grado e mezzo dei Gemelli. E prima stella sorge la più settentrionale delle quattro nella base [ν Hya],[264] ultima la più meridionale delle sei nella concavità [η].
Sono in meridiano come prime stelle quelle al centro della testa dell'Auriga [δ e ξ], come ultime la seconda stella da ovest delle quattro nei piedi dei Gemelli [ν], e la stella brillante sotto la Lepre, che non ha un nome [Phact].
La Coppa sorge in 1 ora e ¼.
**3.** Quando sorge il Corvo, sorge con esso lo zodiaco dal 16° grado della Vergine al 23° grado della Vergine; è in meridiano dal 14° grado dei Gemelli fino al 22° grado dei Gemelli. E prima stella sorge quella sulle reni [Gienah],[265] ultima quella nelle zampe [β].
Sono in meridiano come prime la stella nel braccio occidentale del Gemello occidentale [τ], e il piede destro del Gemello orientale [ξ], come ultime la terza stella da ovest delle tre nelle spalle dei Gemelli [υ], e la stella nel punto d'inizio delle parti anteriori del Cane,[266] un po' ad est del meridiano.
Il Corvo sorge in 3/5 d'ora.
**4.** Quando sorge il Centauro, sorge con esso lo zodiaco dal 10° grado delle Chele

261. In realtà il 13° grado.
262. Per Tolomeo è al di sopra dell'occhio.
263. Tuttavia era più ad est del meridiano di oltre 5½°.
264. Per Tolomeo questa stella è nell'Idra. D'altra parte Tolomeo ha solo una stella sulla base, Alkes.
265. Per Tolomeo è nell'ala destra.
266. La stella che più si avvicina, come posizione, è la π, che Tolomeo pone nel petto.

al 4° grado del Sagittario;[267] è in meridiano dal 12° grado del Cancro al 16° grado e mezzo[268] della Vergine. E prima stella sorge quella nella spalla sinistra [ι], ultima la orientale delle zampe anteriori [Rigil Kentaurus].
Sono in meridiano per prime la stella ad ovest delle brillanti nel ponte di Argo [Naos], e la orientale della mascella meridionale dell'Idra [η], per ultima l'Annunciatore della vendemmia, che è ad ovest del meridiano di circa mezzo cubito.[269]
Il Centauro sorge in 4 ore e ⅓ di ora.
**5.** Quando sorge la Fiera che il Centauro tiene, sorge con essa lo zodiaco dal 23° grado dello Scorpione[270] al 21° grado del Sagittario;[271] è in meridiano dal 3° grado della Vergine[272] fino al 10° grado[273] della Bilancia.[274] E prima stella sorge quella nella zampa posteriore occidentale [ψ Cen]: è questa la più settentrionale delle stelle sotto la spalla destra del Centauro;[275] ultima la più meridionale di tutte, e posta nell'estremità del corpo, sulle reni [ζ].
Sono in meridiano, come prime stelle, la più settentrionale delle orientali nella testa del Leone [Rasalas], e quella posta nelle sue zampe anteriori [Subra], come ultime la stella brillante sulle reni della Grande Orsa,[276] e quella che giace sull'angolo retto del triangolo rettangolo sotto la Coppa [o Hya].
La Fiera sorge in 2 ore e ¼.
**6.** Quando sorge l'Incensiere, sorge insieme lo zodiaco dal 15° grado del Sagittario[277] fino al 23° grado del Sagittario; è in meridiano dal 2° grado delle Chele[278] fino all'11° grado delle Chele. E prima stella sorge quella sul focolare dell'Incensiere [$\varepsilon^1$], ultima la più meridionale di quelle nella base [θ].
Sono in meridiano, come prime stelle, la più meridionale di quelle nel piede sinistro del Bovaro [υ], e quella brillante nelle zampe anteriori del Centauro [Hadar],[279] come ultima quella a nord del ginocchio destro e del piede destro della Vergine [φ], che è ad est del meridiano di circa mezzo cubito.
L'Incensiere sorge in ½ ora.

267. 28° grado dello Scorpione.
268. 10° grado.
269. In realtà oltre 7° ad est del meridiano. Gli errori di questa parte sono senz'altro dovuti alla difficoltà di osservare una stella che rimaneva sempre molto bassa sull'orizzonte di Rodi, poco più di 3°, come Rigil Kentaurus.
270. 22° grado delle Chele. Vista l'entità degli errori (fino ad uno ed anche due segni interi) è da ritenere che questa parte del testo sia stata pesantemente rimaneggiata.
271. 15° grado dello Scorpione.
272. 25° grado del Cancro.
273. 22° grado e mezzo del Leone.
274. È questo l'unico luogo dell'opera in cui Ipparco chiama in questo modo le Chele, ed è il primo passo in ordine cronologico in cui appare questo appellativo nella letteratura occidentale. Forse Ipparco è stato influenzato dalla sua conoscenza dell'astronomia babilonese, nella quale la costellazione *zibanitu*, "bilancia", comprendeva le Chele (la Bilancia moderna) e parte della Vergine. Oppure, più probabilmente, si tratta del frutto di un rimaneggiamento medievale.
275. Questa stella è catalogata da Tolomeo nel Centauro.
276. È impossibile dire con certezza di quale stella si tratti, data la genericità della descrizione. Basandosi su Tolomeo (che la pone sul dorso) quella che più si attaglia alla descrizione è Dubhe, una delle tre stelle più brillanti della Grande Orsa (più o meno pari ad Alioth e Alkaid), ma che si trovava 7° ad ovest del meridiano quando sorgeva ζ Lupi. Secondo Manitius si tratta di Megrez (che però per Tolomeo è all'inizio della coda), che si trovava 2⅓° ad est del meridiano. L'unica stella davvero in culminazione era Phecda (posta però da Tolomeo sulla coscia posteriore sinistra).
277. In realtà 10° grado.
278. 25° grado della Vergine. Errori dovuti probabilmente alla declinazione molto meridionale delle stelle dell'Incensiere.
279. La descrizione corrisponde a Rigil Kentaurus, ma la posizione si adatta a questa stella, che Tolomeo pone sul ginocchio sinistro.

**7.** Quando sorge il Pesce Australe, sorge con questo lo zodiaco dal 16° grado e mezzo dell'Aquario al 20° grado e mezzo dei Pesci;[280] è in meridiano dal 4° grado del Sagittario fino al 24° grado e mezzo del Sagittario. E prima stella sorge la più settentrionale di quelle nella coda [ι], ultima la brillante nella punta del muso [Fomalhaut]. Sono in meridiano, come prime stelle, quella al centro dell'arco del Sagittario [Kaus Media], e la più meridionale delle stelle nella spalla destra del Serpentario [γ], che è circa mezzo cubito ad est del meridiano, come ultime, la stella luminosa nelle zampe posteriori del Sagittario [ι], quella nella punta della coda del Serpente, che il Serpentario tiene [Alya], e la occidentale delle due brillanti nella barra della Lira [Sheliak], che è ad ovest del meridiano di circa mezzo cubito.
Il Pesce sorge in 1 ora e 2/5 di ora.
**8.** Quando sorge il Mostro Marino, sorge insieme lo zodiaco dal 20° grado dei Pesci fino al 7° grado del Toro; è in meridiano dal 23° grado e mezzo del Sagittario fino al 21° grado e mezzo del Capricorno. E prima stella sorge la più settentrionale di quelle nella coda [ι], ultima la più meridionale delle stelle orientali brillanti nel quadrilatero [τ].
É in meridiano come prima stella la occidentale delle stelle nella barra della Lira [Sheliak], come ultima la luminosa nella coda dell'Uccello [Deneb].
Il Mostro Marino sorge in 2 ore.
**9.** Quando sorge Orione, sorge con lui lo zodiaco dal 27° grado e mezzo del Toro fino al 3° grado del Cancro; è in meridiano dal 9° grado dell'Aquario fino al 13° grado dei Pesci. E prima stella sorge quella nella mano sinistra [$o^2$], ultima quella nel piede destro [Saiph]. Sono in meridiano delle altre stelle, come prime, il piede destro di Cefeo [κ], e la centrale delle stelle nella brocca dell'Aquario [ζ], come ultime la stella piccola nel seggio di Cassiopea [κ], e la più settentrionale delle stelle nel petto di Andromeda [π].[281]
Orione sorge in 2 ore e 1/6.
**10.** Quando sorge il Fiume [che inizia] da Orione, sorge con questo lo zodiaco dal 13° grado del Toro fino al 10° grado del Cancro; è in meridiano dal 27° del Capricorno fino al 22° grado dei Pesci. E prima stella sorge la stella di nord ovest del parallelogramma posto nella grande ansa e vicino al Mostro Marino [ρ Cet],[282] ultima la più luminosa e occidentale e più meridionale di tutte [Acamar].
Prime fra le altre stelle sono in meridiano la orientale delle stelle poste a nord di quelle nella coda del Capricorno [μ], e la più meridionale di quelle nella zampa sinistra dell'Uccello [68 Cyg], ultime la stella nelle ginocchia di Cassiopea [Ruchbah], e quella sulla punta della coda del Pesce settentrionale [ρ].[283]
Il Fiume sorge in 3 ore e 3/5.
**11.** Quando sorge la Lepre, sorge con essa lo zodiaco dal 27° grado dei Gemelli fino all'11° grado e mezzo del Cancro; è in meridiano dal 4° grado dei Pesci fino al 25° grado dei Pesci. E prima stella sorge la occidentale delle quattro più settentrionali nelle orecchie [ι], ultima la più meridionale delle stelle nelle zampe posteriori [γ].

---

280. 13° grado.
281. Per Tolomeo è nella spalla destra, ma Bayer la raffigura proprio al centro del petto.
282. Per Tolomeo è nel petto del Mostro Marino.
283. Dalla descrizione sembrerebbe trattarsi della stessa stella di II, 6, 5 poiché è evidente (anche dall'identificazione della seconda stella culminante, Ruchbah) che Ipparco riteneva che il tramonto di Alya e il sorgere di Acamar fossero contemporanei. In effetti però i due fenomeni erano separati da quasi 20 minuti di tempo. Mentre per Ruchbah che, essendo una stella molto settentrionale, non si muove moltissimo in questo torno di tempo sul meridiano, l'identificazione appare obbligata, non appare invece decidibile quale delle due stelle dei Pesci volesse indicare Ipparco. Anche qui, l'errore è dovuto alla posizione molto meridionale di Acamar.

Sono in meridiano delle altre, come prime, la stella brillante nelle reni del Cavallo [Algenib], e il piede sinistro di Cefeo [Errai], come ultima la più settentrionale delle stelle occidentali poste nel quadrilatero del Mostro Marino [θ].
La Lepre sorge in 1 ora e 1/5.
**12.** Quando sorge il Cane, sorge insieme ad esso lo zodiaco dal 15° grado del Cancro fino al 4° grado e mezzo del Leone; è in meridiano dal 28° grado dei Pesci fino al 23° grado e mezzo dell'Ariete. E prima stella sorge quella più settentrionale nella punta della zampa anteriore [Mirzam], ultima quella nella punta della coda [Aludra].
Sono in meridiano delle altre, per prime, la stella nei piedi di Cassiopea [ε], e la più settentrionale di quelle nel piede destro di Andromeda [φ Per], che è un po' ad ovest del meridiano, per ultime, quella brillante nella coscia destra di Perseo [δ], e la orientale delle stelle nella coda dell'Ariete [63 Ari].
Il Cane sorge in 1 ora e ⅔ di ora.
**13.** Quando sorge Procione sorge insieme a questo lo zodiaco dal 3° grado e mezzo del Cancro fino al 9° grado del Cancro; è in meridiano dal 15° grado dei Pesci fino al 19° grado e mezzo dei Pesci. E prima stella sorge la occidentale e doppia [Gomeisa],[284] ultima la orientale e brillante [Procione].
Sono in meridiano, delle altre, per prime la stella al centro del corpo di Cassiopea [Navi], e le orientali delle quattro stelle nella coda del Mostro Marino [$\varphi^3$ e $\varphi^4$]; per ultime sono in meridiano la stella orientale di quelle nella cintura di Andromeda [Mirach], e la più meridionale delle occidentali nel quadrilatero del Mostro Marino [η].
Procione sorge in ⅓ di ora.
**14.** Quando sorge Argo, sorge insieme a questa lo zodiaco dal 6° grado del Leone fino al 3° grado e mezzo delle Chele[285]; è in meridiano dal 25° grado e mezzo dell'Ariete fino al 3° grado e mezzo del Cancro.[286] E prima sorge quella tripla[287] sotto la stella nella punta della coda del Cane, ultima la più meridionale, brillante, delle stelle nel taglio della nave [Aspidiske].
Sono in meridiano, delle altre, per prime la stella nella coscia sinistra di Perseo [ν], e la più meridionale di quelle nel taglio del Toro [o], per ultima quella occidentale, abbastanza ben visibile, della testa dell'Idra, che è fra le zampe meridionali del Cancro [δ].
Argo sorge in 2 ore e ⅔ di ora.

## Cap. II

**1.** Di seguito diremo del tramonto delle [costellazioni] a sud dello zodiaco.
Quando dunque tramonta l'Idra, tramonta con questa lo zodiaco dal 29° grado dei Gemelli fino all'11° grado della Vergine;[288] è in meridiano dal 18° grado e mezzo delle Chele fino al 18° grado e mezzo del Sagittario. E prima stella tramonta la più

284. Con γ CMi Gomeisa costituisce una doppia larga (separazione 40') visibile a occhio nudo.
285. 8° grado.
286. 9° grado. Questo e l'errore precedente sono indubbiamente dovuti alla posizione estremamente meridionale di Aspidiske, che non si sollevava più di 3° dall'orizzonte di Rodi.
287. Si tratta di HIP 35 957, 35 795, 35 855. Stelle non catalogate da Tolomeo, tutte molto deboli (magnitudini 5,34, 5,37 e 5,40), separate da 18', 23' e 28'. Riuscire a vedere singolarmente tre stelle così fioche, per di più mai alte più di 24° neppure da Rodi, non è un'impresa facile, e dimostra che Ipparco doveva essere dotato di vista molto acuta.
288. 16° grado della Vergine.

meridionale di quelle nelle fauci [σ],[289] ultima quella nell'estremità della coda [π].
Sono in meridiano, delle altre stelle, all'inizio la terza brillante del Drago a partire dalla coda [Thuban], che è ad ovest del meridiano di circa mezzo cubito, la stella brillante nella cintura del Bovaro [Izar], e quella ad occidente della stella brillante nell'estremità della chela settentrionale delle Chele [δ]; alla fine sono in meridiano la tempia più meridionale del Drago [Eltanin], la stella nella punta della coda del Serpente, che il Serpentario tiene in mano [Alya], e la occidentale delle stelle nel mantello del Sagittario [43].
L'Idra tramonta in 4 ore.
**2.** Quando tramonta la Coppa, tramonta con questa lo zodiaco dal 21° grado del Cancro fino al 12° grado e mezzo del Leone; è in meridiano dall'11° grado e mezzo dello Scorpione fino all'inizio del Sagittario. E prima stella tramonta la più settentrionale di quelle nella base [ν Hya], ultime la più meridionale e la più settentrionale di quelle nella concavità [η e θ].[290]
Sono in meridiano delle altre stelle, all'inizio, il terzo e il quarto e il quinto segmento dello Scorpione dopo il petto [ε, μ e ζ], e la terza stella posta nel braccio dell'Inginocchiato [Marsic], a partire dalla spalla destra, circa mezzo cubito ad ovest del meridiano, alla fine, la stella all'inizio della coscia sinistra dell'Inginocchiato [HIP 83 838], e la punta [della freccia] del Sagittario [Alnasl].
La Coppa tramonta in 1 ora e ⅓.
**3.** Quando tramonta il Corvo, tramonta con questo lo zodiaco dal Leone…[291]

**5.** Quando tramonta la Fiera, che il Centauro tiene in mano, tramonta con essa lo zodiaco dal 28° grado e mezzo del Leone fino all'11° grado delle Chele; è in meridiano dal 10° grado del Sagittario fino al 5° grado e mezzo del Capricorno. E prima stella tramonta quella nelle cosce posteriori,[292] ultima la orientale delle stelle nella testa [ξ].
Sono in meridiano delle altre stelle, all'inizio, la stella nelle fauci del Drago [Alrakis], che è ad ovest del meridiano di quasi mezzo cubito, e il gomito sinistro dell'Inginocchiato [μ], che è ad ovest del meridiano di circa mezzo cubito, alla fine, la occidentale [ψ] delle tre stelle brillanti nella prima spira del Drago, la stella nella punta dell'ala destra dell'Uccello [κ], che resta un po' ad est del meridiano, quella sulla punta della Freccia [γ], e la più meridionale delle stelle nel muso del Capricorno [ο].
La Fiera tramonta in 2 ore.
**6.** Quando tramonta l'Incensiere, tramonta insieme lo zodiaco dal 18° grado e mezzo del Leone[293] fino al 9° grado delle Chele; è in meridiano dal 5° grado del Sagittario[294] al 4° grado e mezzo del Capricorno. E prima stella tramonta la più meridionale di quelle nel bordo e doppia [β e γ], ultima la più settentrionale delle stelle nella base [σ].
Sono in meridiano, delle altre stelle, all'inizio la stella brillante nella coscia sinistra

289. Per Tolomeo è nelle narici.
290. Le due stelle che tramontano assieme sono in verità la più settentrionale e la seconda più meridionale, la più meridionale essendo la ζ.
291. Qui c'è una lacuna nel testo.
292. Siamo indecisi fra la α e la π. Alla prima, che per Tolomeo è nel poplite, si adattano meglio i fenomeni in meridiano, ma c'è un errore superiore ai 5° all'orizzonte. Per la seconda, che in effetti Tolomeo pone nella coscia, vi sono errori di 2°, 2½° e 4° in meridiano, e di 1½° all'orizzonte. D'altra parte la π non è certo la prima stella a tramontare della Fiera.
293. 29° grado.
294. 11° grado e mezzo.

dell'Inginocchiato [π],[295] e la centrale di quelle nell'arco del Sagittario [Kaus Media],[296] alla fine la più meridionale delle stelle brillanti nelle corna del Capricorno [Dabih].
L'Incensiere tramonta in 2 ore e 1/6.
**7.** Quando tramonta il Pesce Australe, tramonta con questo lo zodiaco dal 24° grado del Sagittario fino al 17° grado e mezzo del Capricorno; è in meridiano dal 3° grado dei Pesci fino al 2° grado dell'Ariete. E prima stella tramonta la più meridionale di quelle brillanti nella coda [γ Gru], ultima la stella molto brillante nel muso [Fomalhaut].
Sono in meridiano, delle altre stelle, all'inizio la stella nella punta della coda del più meridionale dei due Pesci [ω], alla fine il piede sinistro di Andromeda [Almach], la centrale delle stelle nella testa dell'Ariete [Sheratan], e la più meridionale delle stelle orientali nel quadrilatero nel Mostro Marino [τ].
Il Pesce tramonta in 1 ora e 4/5.
**8.** Quando tramonta il Mostro Marino, tramonta con esso lo zodiaco dal 26° grado e mezzo dell'Aquario fino al 14° grado dell'Ariete; è in meridiano dal 22° grado del Toro fino al 14° grado e mezzo del Cancro. E prima stella comincia a tramontare la più meridionale di quelle nella coda [Diphda], ultima la orientale delle stelle nella parte settentrionale del cranio [λ].
Sono in meridiano, delle altre stelle, all'inizio la più meridionale di quelle nella testa dell'Auriga [δ] e quella nel piede destro [Elnath], alla fine la stella brillante nelle zampe anteriori dell'Orsa [θ], e la stella brillante al centro della fiancata di Argo [Regor], che è ad ovest del meridiano di circa mezzo cubito.
Il Mostro Marino tramonta in 4 ore meno ⅛.
**9.** Quando tramonta Orione tramonta con lui lo zodiaco dal 7° grado del Toro al 30° grado del Toro; è in meridiano dal 12° grado del Leone al 13° grado della Vergine. E prima stella tramonta quella nel piede sinistro [Rigel], ultime le più settentrionali delle stelle nella mazza [$\chi^1$ e $\chi^2$].
Sono in meridiano, delle altre stelle, all'inizio la stella nelle ginocchia posteriori della Grande Orsa [ψ], e la stella posta a sud accanto a quella brillante sulle reni del Leone [60], alla fine la spalla sinistra della Vergine [Porrima].
Orione tramonta all'incirca in 2 ore.
**10.** Quando tramonta il Fiume [che inizia] da Orione, tramonta insieme a questo lo zodiaco dal 7° grado dei Pesci[297] al 4° grado e mezzo del Toro; è in meridiano dal 4° grado dei Gemelli[298] al 9° grado e mezzo del Leone. E per prima stella tramonta la occidentale e più brillante di tutte [Acamar], per ultima la prima e posta a sud accanto al piede di Orione [λ].
Sono in meridiano delle altre stelle, all'inizio del tramonto [del Fiume], Propus dei Gemelli, che è un po' ad est del meridiano,[299] e le due stelle al centro del corpo della Lepre [Arneb e Nihal];[300] alla fine sono in meridiano la stella orientale di quelle nel dorso del Leone [51 Leo],[301] e quella nelle ginocchia posteriori dell'Orsa [ψ], che è

295. Che si trovava però oltre 7° ad ovest del meridiano.
296. Che si trovava a 7½° ad ovest del meridiano. Gli errori di questa sezione sono dovuti alla forte declinazione meridionale di β e γ Arae, sui -50°.
297. Inizio dei Pesci.
298. 26° grado del Toro.
299. Si trovava in realtà 8° ad est del meridiano.
300. Si trovavano rispettivamente a 6° e a 5½° ad est del meridiano. In questa sezione gli errori sono imputabili alla forte declinazione meridionale di Acamar all'epoca, -50°.
301. La descrizione non si attaglia a questa stella, ma a Zosma, che rimaneva però 6° ad est del meridiano; tenendo conto della buona precisione dei posizionamenti relativi al tramonto della λ Eri, possiamo pensare che ci si riferisca circa mezzo cubito ad est del meridiano.
Il Fiume tramonta in 4 ore e 7/10 di ora.
**11.** Quando tramonta la Lepre, tramonta lo zodiaco dal 26° grado e mezzo dell'Ariete al 14° grado del Toro; è in meridiano dal 30° grado del Cancro al 21° grado del Leone. E prima stella tramonta quella nelle zampe anteriori [ε], ultima quella nella punta della coda [η].
Sono in meridiano delle altre stelle, all'inizio, la stella sulle spalle dell'Orsa [Dubhe], la stella nella punta della coda del Drago [Giausar], quella nel cuore del Leone [Regolo], e la più settentrionale delle stelle nella troncatura di Argo [Markeb], poco ad ovest del meridiano, alla fine sono in meridiano la zampa posteriore del Leone [σ], la stella nelle sue cosce posteriori [ι], e la più meridionale delle stelle nella base della Coppa [β].
La Lepre tramonta in 1 ora e ⅓.
**12.** Quando tramonta il Cane tramonta con esso lo zodiaco dall'11° grado del Toro fino al 29° grado del Toro; è in meridiano dal 17° grado del Leone all'11° grado della Vergine. E prima stella tramonta quella brillante nelle zampe posteriori [Furud], ultima la più meridionale delle stelle ben visibili nella testa [Muliphein].
Sono in meridiano, delle altre stelle, all'inizio la occidentale delle stelle nelle zampe posteriori dell'Orsa [Alula Borealis], e la stella sulle reni del Leone [Zosma]; alla fine sono in meridiano la stella nelle zampe del Corvo [β], e la spalla più a sud della Vergine [Porrima], che è ad est del meridiano di circa ⅔ di cubito.
Il Cane tramonta in 1 ora e ½.
**13.** Quando tramonta Procione, tramonta con esso lo zodiaco dal 15° grado dei Gemelli al 18° grado dei Gemelli; è in meridiano dall'inizio delle Chele fino al 4° grado delle Chele. E prima stella tramonta quella occidentale e doppia [Gomeisa], ultima quella orientale e brillante [Procione].
Sono in meridiano, delle altre stelle, all'inizio la centrale delle stelle nel piede sinistro del Bovaro [τ], che è un po' ad est del meridiano, e la più settentrionale delle stelle nella testa del Centauro [4 Cen], che è circa mezzo cubito ad ovest del meridiano; alla fine sono in meridiano la stella nella punta della coda dell'Orsa [Alkaid], che resta un po' ad est del meridiano, la più settentrionale delle stelle nel piede sinistro del Bovaro [Muphrid], la stella nella punta della coda dell'Idra [π], e la spalla destra del Centauro [Menkent].
Procione tramonta in 1/5 di ora.
**14.** Quando tramonta Argo, tramonta con essa lo zodiaco dal 16° grado dell'Ariete[302] fino al 18° grado dei Gemelli; è in meridiano dal 17° grado del Cancro[303] fino al 5° grado delle Chele. E prima stella tramonta la più brillante e più meridionale di quelle nel timone, che alcuni chiamano Canopo, ultima la più settentrionale delle stelle al centro dell'albero [γ Pyx].[304]
Sono in meridiano, delle altre stelle, all'inizio la più meridionale delle stelle occidentali nella testa del Leone [Alterf], che è circa mezzo cubito ad est del meridiano,[305] e la stella piccola alla radice del collo dell'Idra [ω];[306] alla fine sono

---

alla 51 Leo, pur se molto debole (magnitudine 5,50).
302. 10° grado.
303. 11° grado.
304. Per Tolomeo è la α Pyx che corrisponde alla descrizione di Ipparco, ma i fenomeni trovano buona corrispondenza con questa stella.
305. In realtà 7½°.
306. 5¾° ad est del meridiano. In questa sezione l'imprecisione dei fenomeni riportati è sicuramente dovuta alle

in meridiano la stella nella punta della coda della Grande Orsa [Alkaid], la più settentrionale delle stelle nel piede sinistro del Bovaro [Muphrid], che è circa mezzo cubito ad ovest del meridiano, la stella nella punta della coda dell'Idra [π], e la spalla destra del Centauro [Menkent], entrambe poco ad ovest del meridiano.
Argo tramonta all'incirca in 5 ore.

## Cap. III

**1a.** Dopo aver esposto i dati riguardanti la levata ed il tramonto di ciascuna delle costellazioni poste esternamente al circolo zodiacale, ora daremo conto delle levate simultanee e dei tramonti delle 12 costellazioni dello zodiaco.
**1b.** Quando, dunque, sorge il Cancro, sorge con esso l'arco zodiacale dal 23° grado dei Gemelli fino al 18° grado del Cancro; è in meridiano dal 5° grado dei Pesci fino al ½° grado dell'Ariete. E prima stella sorge quella nell'estremità della chela settentrionale [$\sigma^3$],[307] ultima quella nell'estremità della chela meridionale [Acubens].
Sono in meridiano, delle altre stelle, per prima la stella brillante nella testa di Andromeda [Alpheratz], per ultime la stella occidentale delle tre brillanti nella testa dell'Ariete [Mesarthim], la stella senza nome e brillante posta a sud del Mostro Marino sotto il centro del corpo [υ], la stella più meridionale delle orientali nel quadrilatero del Mostro Marino [τ], e il piede sinistro di Andromeda [Almach], che resta un po' ad est del meridiano.
Il Cancro sorge in 1 ora e ⅔ di ora.
**2.** Quando sorge il Leone, sorge insieme con esso lo zodiaco dal 7° grado e mezzo del Cancro fino al 18° grado e mezzo del Leone; è in meridiano dal 20° grado dei Pesci fino all'11° grado e mezzo del Toro. E prima stella sorge, delle stelle nella testa, la più settentrionale delle occidentali [κ], ultima la stella nelle zampe posteriori [σ].
Sono in meridiano, delle altre stelle, come prime la stella orientale della cintura di Andromeda [Mirach], e la più meridionale delle occidentali del quadrilatero nel Mostro Marino [η], come ultime la più luminosa delle Iadi [Aldebaran], e la stella nel gomito sinistro dell'Auriga [ε], che resta circa mezzo cubito ad est del meridiano.
Il Leone sorge in 3 ore e ¼.
**3.** Quando sorge la Vergine, sorge con lei lo zodiaco dal 22° grado del Leone fino all'8° grado delle Chele; è in meridiano dal 14° grado e mezzo del Toro fino al 9° grado del Cancro. E prima stella sorge, delle quattro nella testa, la più settentrionale delle occidentali [ξ], ultima quella nel piede destro [μ].
Sono in meridiano, delle altre stelle, per prima la stella brillante sulla spalla sinistra dell'Auriga [Capella], per ultime la stella brillante nell'acrostolio di Argo [Tureis], e la più settentrionale delle stelle nelle fauci dell'Idra [δ], che è ad est del meridiano di circa mezzo cubito.
La Vergine sorge in 3 ore e 4/5.

---

difficoltà di osservazione di Canopo (v. nota 120).
307. Questa stella non è catalogata da Tolomeo, che sulla chela settentrionale riporta solo la ι, e che nel complesso elenca solo nove stelle nel Cancro. Tuttavia Eratostene in *Catasterismi* ne cita 18, e dice che sulla chela destra ce ne sono tre. Nei dintorni ci sono diverse stelline debolmente visibili a occhio nudo, ma effettivamente solo due sono decisamente più brillanti, la $\rho^2$ e la $\sigma^3$, di magnitudine 5,21. Solo la seconda, per la sua posizione, più lontana dal centro del corpo del Cancro, può essere però presa come estremità della chela. Così la rappresenterà il Bayer nella sua *Uranometria*. E, per inciso, i fenomeni citati da Ipparco si adattano perfettamente alla sua levata.

**4.** Quando sorgono le Chele, sorge con esse lo zodiaco dal 16° grado delle Chele fino al 6° grado dello Scorpione; è in meridiano dal 18° grado del Cancro fino all'11° grado del Leone. E prima stella sorge la più meridionale di quelle brillanti nelle estremità delle Chele [Zubenelgenubi], ultima la più meridionale di quelle nella fronte dello Scorpione [π Sco].[308]
Sono in meridiano, delle altre stelle, come prima la più meridionale delle occidentali delle stelle nella testa del Leone [Alterf], e la più settentrionale di quelle al centro dell'albero di Argo [α Pyx], come ultima la orientale delle due stelle nel dorso del Leone [54 Leo].
Le Chele sorgono in 1 ora e 3/5.
**5.** Quando sorge lo Scorpione, sorge insieme lo zodiaco dal 3° grado dello Scorpione fino al 9° grado del Sagittario; è in meridiano dal 7° grado e mezzo del Leone fino al 22° grado e mezzo della Vergine. E prima stella sorge la più settentrionale di quelle nella fronte [Acrab], ultima [quella che rappresenta] il terzo segmento a partire dal pungiglione, che è il sesto dopo quelli nel petto [$\iota^1$].
Sono in meridiano, delle altre stelle, per prime la stella brillante nel petto della Grande Orsa [Merak],[309] e la terza da occidente delle quattro stelle dopo quella più luminosa dell'Idra [μ], per ultime Spiga e la spalla sinistra del Centauro [ι], che è ad ovest del meridiano di circa mezzo cubito.
Lo Scorpione sorge in 3 ore meno 1/10.
**6.** Quando sorge il Sagittario, sorge insieme ad esso lo zodiaco dal 5° grado e mezzo del Sagittario fino al 18° grado del Capricorno; è in meridiano dal 19° grado e mezzo della Vergine all'8° grado e mezzo dello Scorpione. E prima stella sorge quella nella punta [della freccia, Alnasl], ultima quella luminosa nelle zampe posteriori [ι].
Sono in meridiano, delle altre stelle, per prima la stella nel gomito sinistro della Vergine [θ], che è poco ad ovest del meridiano, per ultime la stella centrale e più brillante di quelle nel petto dello Scorpione [Antares], la stella occidentale nella mano sinistra del Serpentario [Yed Prior], e la terza verso est a partire da quella brillante della Corona, che è circa mezzo cubito ad ovest del meridiano [δ].
Il Sagittario sorge in 3 ore.
**7.** Quando sorge il Capricorno, sorge con esso lo zodiaco dal 28° grado e mezzo del Sagittario fino al 27° grado del Capricorno; è in meridiano dal 18° grado e mezzo delle Chele fino al 18° grado dello Scorpione. E prima stella sorge la più settentrionale di quelle brillanti nelle corna [Algedi], ultima la orientale delle stelle brillanti nella coda [Deneb Algedi].
Sono in meridiano, delle altre, per prime la più brillante delle stelle nella cintura del Bovaro [Izar], la occidentale dell'estremità della chela settentrionale delle Chele [δ], e la terza contando dall'estremità della coda del Serpente che è fra le Orse [Thuban], brillante, circa mezzo cubito ad ovest del meridiano; per ultime sono in meridiano la spalla destra [Kornephoros] dell'Inginocchiato e la stella posta a nord accanto a quelle nella sua gamba destra [υ].
Il Capricorno sorge in 1 ora e 5/6.
**8.** Quando sorge l'Aquario, sorge con lui lo zodiaco dal 6° grado del Capricorno al 20° grado e mezzo dell'Aquario; è in meridiano dal 27° grado delle Chele fino al 7°

308. Non è del tutto chiaro perché Ipparco qui si riferisca ad una stella della costellazione adiacente. È vero che le Chele facevano iconograficamente parte, in cielo, della figura dello Scorpione, ma è anche vero che esse costituivano un segno zodiacale e una costellazione autonomi. Forse si tratta del frutto di un rimaneggiamento.
309. Per Tolomeo è nel fianco.

grado del Sagittario. E prima stella sorge la occidentale delle stelle nella mano sinistra [Albali], ultima la stella brillante nel piede destro [Skat].[310]

Sono in meridiano, delle altre, per prima la stella nella testa del Bovaro [Nekkar], per ultime la più settentrionale delle stelle nell'arco del Sagittario [Kaus Borealis],[311] le tre stelle senza nome in linea retta presso la spalla destra del Serpentario [66, 67 e 68], e la stella brillante nella coscia sinistra dell'Inginocchiato [π], che è circa mezzo cubito ad ovest del meridiano.

L'Aquario sorge in 2 ore e ⅔ di ora.

**9.** Quando sorgono i Pesci, sorge con essi lo zodiaco dal 7° grado dell'Aquario fino al 16° grado dell'Ariete; è in meridiano dal 25° grado e mezzo dello Scorpione al 9° grado del Capricorno. E prima stella sorge quella nell'estremità del muso del Pesce più meridionale [β], ultima quella nel Nodo dei fili [Alrescha].

Sono in meridiano, delle altre stelle, per prime la stella posta nel punto dove inizia la coscia destra dell'Inginocchiato [η], e la stella posta a nord di quelle nel pungiglione dello Scorpione [M 7][312]; la stella nella testa dell'Inginocchiato [Rasalgethi] resta ad est del meridiano di circa ⅔ di cubito; per ultime sono in meridiano la più meridionale delle stelle nelle ginocchia del Capricorno [ω], e quelle nella gola [η] e nella curvatura dell'ala destra dell'Uccello [δ].

I Pesci sorgono in 3 ore e 1/10.

**10.** Lo zodiaco sorge con l'Ariete dal 18° grado e mezzo dei Pesci fino al 21° grado dell'Ariete; è in meridiano dal 23° grado e mezzo del Sagittario fino al 14° grado del Capricorno. E prima stella sorge quella sulla zampa anteriore [η Psc],[313] ultima la orientale di quelle nella coda [63 Ari].

Sono in meridiano, delle altre stelle, per prime la occidentale delle stelle nella barra della Lira [Sheliak], e la orientale di quelle nel dorso del Sagittario [52 Sgr],[314] che è ad ovest del meridiano di circa ⅔ di cubito; per ultime sono in meridiano la orientale delle stelle nel braccio sinistro dell'Aquario [μ],[315] la stella nel petto del Capricorno [η], e la occidentale delle stelle nella coda del Delfino [ε].

L'Ariete sorge in 1 ora e 2/5.

**11.** Quando sorge il Toro, sorge insieme lo zodiaco dal 7° grado del Toro al 29° grado del Toro; è in meridiano dal 21° grado e mezzo del Capricorno al 9° grado dell'Aquario. E prima stella sorge la più meridionale delle quattro nel taglio [o],[316] ultima quella nella punta del corno destro [ζ].

Sono in meridiano, delle altre, per prima la stella brillante al centro della coda dell'Uccello [Deneb], per ultime la occidentale delle tre stelle nella testa di Cefeo [μ] e il suo piede destro [κ], che è circa mezzo cubito ad ovest del meridiano, e la centrale delle stelle nella brocca dell'Aquario [ζ], che è circa mezzo cubito ad ovest del meridiano.

Il Toro sorge in 1 ora e ⅛.

**12.** Quando sorgono i Gemelli sorge con questi lo zodiaco dal 1° grado e mezzo dei Gemelli fino al 30° grado dei Gemelli; è in meridiano dal 10° grado e mezzo dell'A-quario fino all'8° grado e mezzo dei Pesci. E prima stella sorge quella nella mano sinistra del Gemello occidentale [θ], ultima quella nella mano destra[317] del Gemello orientale.
Sono in meridiano, delle altre stelle, per prime quella brillante nel corpo di Cefeo [Alfirk], quella brillante nel piede destro dell'Aquario [Skat], e quella brillante nel muso del Pesce Australe [Fomalhaut]; per ultima è in meridiano la stella centrale delle tre nella spalla destra di Andromeda [ρ].[318]
I Gemelli sorgono in 1 ora e 5/6.

---

310. Ipparco ha escluso dal novero quelle appartenenti al fiotto d'acqua.
311. Per Tolomeo ce n'è un'altra più settentrionale, la μ.
312. L'ammasso stellare aperto che Tolomeo classificherà come stella nebulosa.
313. Vedi nota 61.
314. Per Tolomeo è nel gomito destro.
315. Per Tolomeo è la centrale del braccio.
316. A dire la verità a sorgere per prima è la più settentrionale nel taglio, la 5; comunque i fenomeni sono corretti per la o.

## Cap. IV

**1.** Quando tramonta il Cancro, tramonta con questo lo zodiaco dal 26° grado e mezzo dei Gemelli al 19° grado e mezzo del Cancro; è in meridiano dal 17° grado delle Chele al 12° grado dello Scorpione. E prima stella tramonta quella ben visibile nelle zampe meridionali del Cancro [β], posta in linea retta con le stelle ad ovest di quelle intorno alla nebulosa del Cancro, ultima la stella nell'estremità della chela settentrionale del Cancro [ι].[319]
Sono in meridiano, delle altre stelle, all'inizio il piede destro del Bovaro [ζ], e la stella brillante nell'estremità della chela meridionale delle Chele [Zubenelgenubi]; alla fine sono in meridiano la stella nel braccio destro dell'Inginocchiato [Marsic], che è la terza a partire dalla spalla destra, e il terzo, quarto, quinto segmento dello Scorpione di quelli dopo il petto [ε, μ e ζ].
Il Cancro tramonta in 1 ora e 3/5.
**2.** Quando il Leone tramonta, tramonta con esso lo zodiaco dal 20° grado e mezzo del Cancro fino al 14° grado della Vergine; è in meridiano dal 13° grado dello Scorpione fino al 20° grado e mezzo del Sagittario. E per prima stella tramonta quella nelle zampe anteriori del Leone [Subra], per ultima quella nella punta della coda [Denebola].
Sono in meridiano, delle altre stelle, all'inizio del tramonto, il piede destro dell'Inginocchiato [χ],[320] il ginocchio sinistro del Serpentario [ζ], che è ad est del meridiano di circa mezzo cubito, e il primo segmento dello Scorpione [ε]; alla fine sono in meridiano la tempia settentrionale del Drago [Grumium], e la stella al centro della schiena del Sagittario [$\chi^1$],[321] che è ad ovest del meridiano di circa mezzo cubito.
Il Leone tramonta in 2 ore e ⅔ di ora.
**3.** Quando tramonta la Vergine, tramonta con lei lo zodiaco dal 27° grado del Leone fino al 26° grado delle Chele; è in meridiano dall'11° grado del Sagittario al 16°

317. Tolomeo non pone alcuna stella sulla mano destra del Gemello orientale, così pure al-Sufi e Bayer, autori dei due atlanti più fedeli a Tolomeo. D'altra parte, a suffragio, abbiamo la testimonianza di Eratostene che dice che ce n'è una. Adottiamo quindi il suggerimento di Manitius, che propone la μ Cnc, anche se la stella è catalogata da Tolomeo nella costellazione zodiacale adiacente, perché si tratta dell'unica della regione la cui levata e tramonto si accordano in modo soddisfacente con i fenomeni.
318. Vedi note 97 e 98.
319. Per qualche strano motivo, Ipparco sembra qui indicare un'altra stella all'estremo della chela a nord rispetto a III, 3, 1b. E, del resto, i fenomeni qui citati si confanno proprio alla ι.
320. Anche qui sembra che Ipparco intenda un'altra stella rispetto a II, 5, 3, che per Tolomeo è però nella gamba. D'altra parte la ν Boo rimane piuttosto distante dal meridiano, 3½°, mentre la χ Her detiene un errore molto piccolo, inferiore a ½°, simile a quello delle altre due stelle citate.
321. Per Tolomeo è sulla spalla destra.

grado del Capricorno. E prima stella tramonta quella nella punta dell'ala sinistra [Zavijava], ultima quella nel piede settentrionale [μ].
Sono in meridiano, delle altre stelle, all'inizio la seconda stella dall'estremità della coda del Serpente [η], che il Serpentario tiene in mano, e la occidentale delle stelle più deboli situate agli angoli opposti del quadrilatero del Sagittario [φ]; alla fine sono in meridiano la stella brillante al centro del corpo dell'Uccello [Sadr], e la più meridionale delle stelle orientali nel rombo del Delfino [δ].
La Vergine tramonta in 2 ore e $\frac{8}{15}$.
**4.** Quando tramontano le Chele, tramonta con queste lo zodiaco dal 16° grado e mezzo delle Chele fino al 15° grado dello Scorpione; è in meridiano dal 10° grado del Capricorno fino al 27° grado e mezzo del Capricorno. E prima stella tramonta quella brillante nell'estremità della chela meridionale [Zubenelgenubi], ultima la stella al centro della chela settentrionale [ξ Sco].[322]
Sono in meridiano delle altre stelle, all'inizio, la più meridionale delle stelle nell'ala destra dell'Uccello [θ], alla fine la stella brillante nella bocca del Cavallo [Enif].
Le Chele tramontano in 1 ora e ¼.
**5.** Quando tramonta lo Scorpione, tramonta con questo lo zodiaco dal 12° grado e mezzo delle Chele fino al 6° grado dello Scorpione; è in meridiano dal 6° grado e mezzo del Capricorno fino al 22° grado del Capricorno. E prima stella tramonta quella posta sul terzo segmento [ζ], ultima la più settentrionale di quelle sulla fronte [Acrab].
Sono in meridiano, delle altre stelle, all'inizio la stella centrale di quelle nell'ala destra dell'Uccello [ι], che resta ad est del meridiano di circa mezzo cubito, e la più settentrionale di quelle sulle ginocchia del Capricorno [ψ]; alla fine sono in meridiano la stella luminosa al centro della coda dell'Uccello [Deneb], e la occidentale delle stelle luminose nella coda del Capricorno [Nashira], che è ad est del meridiano di circa ⅔ di cubito.
Lo Scorpione tramonta in 1 ora.
**6.** Quando tramonta il Sagittario, tramonta con esso lo zodiaco dal 3° grado[323] dello Scorpione fino al 26° grado e mezzo del Sagittario; è in meridiano dal 20° grado del Capricorno fino al 4° grado e mezzo dei Pesci. E prima stella tramonta la brillante nella zampa anteriore [Arkab], ultima la più settentrionale delle stelle nel mantello [υ].
Sono in meridiano, delle altre stelle, all'inizio la stella nella mano destra di Cefeo [θ],[324] la stella nella curvatura dell'ala sinistra dell'Uccello [ζ], e la orientale delle stelle nel dorso del Capricorno [ι], che è ad ovest del meridiano di circa mezzo cubito;[325] alla fine sono in meridiano il piede sinistro di Cefeo [Errai], e la stella nell'ombelico del Cavallo [Alpheratz].
Il Sagittario tramonta in 3 ore.
**7.** Quando tramonta il Capricorno, tramonta con esso lo zodiaco dal 2° grado del Capricorno fino al 23° grado e mezzo del Capricorno; è in meridiano dal 10° grado e mezzo dei Pesci fino al 9° grado dell'Ariete. E prima stella tramonta la più meridionale di quelle nel ginocchio [ω], ultima la orientale delle stelle nella coda [Deneb Algedi].

322. Per Tolomeo la stella al centro della chela settentrionale è la γ Lib, ma con essa i fenomeni risultano completamente errati. Accettiamo perciò la proposta di Manitius, che trova fondamento fra l'altro nel disegno di Bayer, con la quale le cose tornano a posto.
323. In realtà 8° grado e mezzo.
324. Per Tolomeo è sotto il gomito destro.
325. Per la verità oltre 5°. Questo errore e il precedente sono indubbiamente il risultato della forte declinazione meridionale (-45°) della stella tramontante Arkab.

Sono in meridiano, delle altre stelle, all'inizio la più debole di quelle nel seggio di Cassiopea [κ], e la più settentrionale di quelle nel petto di Andromeda [δ]; alla fine sono in meridiano la spalla sinistra di Perseo [θ], che resta ad est del meridiano di circa ⅔ di cubito, e la stella sulla cresta del Mostro Marino [$\xi^1$].
Il Capricorno tramonta in 1 ora e ⅔ di ora.
**8.** Quando tramonta l'Aquario, tramonta insieme a lui lo zodiaco dal 17° grado e mezzo del Capricorno fino al 15° grado dell'Aquario; è in meridiano dal 3° grado dell'Ariete fino al 6° grado e mezzo del Toro. E prima stella tramonta la occidentale delle stelle nella mano sinistra [Albali], ultima la orientale di quelle nel vaso [η].
Sono in meridiano, delle altre stelle, all'inizio la stella nel falcetto di Perseo che è nebulosa [NGC 869-884], il piede sinistro di Andromeda [Almach], che è poco ad ovest del meridiano, ugualmente anche la centrale delle stelle nella testa dell'Ariete [Sheratan], alla fine la stella nella scapola del Toro [λ],[326] che è circa mezzo cubito ad ovest del meridiano, e quella nel muso delle Iadi [Hyadum Prima], che è ad est del meridiano di circa mezzo cubito.
L'Aquario tramonta in 2 ore e ⅛.
**9.** Quando tramontano i Pesci, tramonta con essi lo zodiaco dal 23° grado dell'Aquario fino al 5° grado dell' Ariete; è in meridiano dal 17° grado del Toro fino al 6° grado del Cancro. E prima stella tramonta quella nell'estremità del muso del Pesce meridionale [β], ultime quelle nella punta del muso del Pesce settentrionale [σ e 82].
Sono in meridiano, delle altre stelle, all'inizio la stella nel punto dove comincia il corno destro del Toro [97 Tau], e la spalla sinistra dell'Auriga [Capella], che è ad ovest del meridiano di circa mezzo cubito; alla fine sono in meridiano la più settentrionale delle stelle nelle zampe anteriori dell'Orsa [Talitha], le due stelle occidentali [η e θ] delle quattro poste intorno alla nebulosa che si trova nel Cancro, e la stella nelle sue zampe meridionali [β].
I Pesci tramontano in 3 ore e ½.
**10.** Quando tramonta l'Ariete, tramonta con esso lo zodiaco dal 29° grado dei Pesci fino al 26° grado dell'Ariete; è in meridiano dal 29° grado dei Gemelli al 29° grado del Cancro. E prima stella tramonta quella nelle zampe anteriori [η Psc], ultima la orientale di quelle nella coda [63 Ari].
Sono in meridiano, delle altre stelle, all'inizio la stella nella punta della coda del Cane [Aludra], che è ad est del meridiano di circa mezzo cubito, e la stella al centro del fianco meridionale del timone [τ Pup]; alla fine la occidentale di quelle nelle zampe posteriori dell'Orsa [Tania Borealis], che è circa mezzo cubito ad ovest del meridiano, e la terza stella da nord di quelle brillanti nel collo e nel petto del Leone [η], che è ad est del meridiano di circa ⅔ di cubito.
L'Ariete tramonta in 2 ore.
**11.** Quando tramonta il Toro, tramonta con esso lo zodiaco dal 20° grado dell'Ariete fino al 26° grado del Toro; è in meridiano dal 21° grado e mezzo del Cancro al 7° grado della Vergine. E prima stella tramonta la più meridionale delle quattro nel taglio [o], ultima quella nella punta del corno sinistro [Elnath].
Sono in meridiano, delle altre stelle, all'inizio la più meridionale delle tre stelle nel collo dell'Idra [$\tau^1$], poste a nord di quella brillante, che è circa mezzo cubito ad est del meridiano; alla fine sono in meridiano la stella al centro dell'ala sinistra della Vergine [Zaniah],

326. Per Tolomeo è nel petto.

che è poco ad ovest del meridiano, e la piccola stella al centro del corpo del Corvo [ζ].
Il Toro tramonta in 3 ore.
**12.** Quando tramontano i Gemelli, tramonta con loro lo zodiaco dal 4° grado dei Gemelli fino al 1° grado e mezzo del Cancro; è in meridiano dal 17° grado della Vergine fino al 21° grado e mezzo delle Chele. E prima stella tramonta quella chiamata Propus, ultima quella nella mano destra del Gemello orientale [μ Cnc].[327]
Sono in meridiano, delle altre stelle, all'inizio l'Annunciatore della vendemmia e la spalla destra della Vergine [δ], che è circa mezzo cubito ad ovest del meridiano; alla fine è in meridiano la stella brillante nell'estremità della chela settentrionale [Zubeneschamali], che è circa ⅔ di cubito ad est del meridiano.
I Gemelli tramontano in 2 ore e 1/6.

## Cap. V

**1a.** Oltre che prestare attenzione all'osservazione delle levate e dei tramonti simultanei, penso sia utile anche che noi consideriamo quali stelle fisse distino le une dalle altre ad intervalli di un'ora equinoziale. Questo, infatti, ci sembra utile per calcolare esattamente l'ora della notte, per registrare i tempi delle eclissi di Luna, e per altri programmi astronomici.
**1b.** Dunque sul coluro solstiziale giace una stella, quella nell'estremità della coda del Cane [Aludra], posta nel semicerchio dove si trova il punto tropicale estivo. Da questa stella dista un intervallo di un'ora la stella situata nel punto dove inizia il collo dell'Idra [ζ], ed è vicinissima[328] anche la stella brillante nelle ginocchia anteriori dell'Orsa [α Lyn].
**2.** Determina il secondo intervallo orario, intorno all'inizio del Leone, la piccola stella occidentale del Leone [ν], a poco meno di un cubito dalla stella luminosa nel cuore. E questa stellina è meno di un dito ad ovest del cerchio passante per i poli che determina il secondo intervallo orario.
**3.** Determina il terzo intervallo orario, nei pressi del centro del Leone, la stella più meridionale delle due, sempre del Leone, poste da entrambi i lati accanto a quella luminosa sulle reni [60 Leo].
**4.** Determina il quarto intervallo orario, intorno all'inizio della Vergine, la stella posta sull'angolo retto del triangolo rettangolo sotto la Coppa [o Hya]; la stella brillante [Megrez] sulle reni della Grande Orsa è ad est del circolo che delimita la quarta ora meno di 1/20 di ora.
**5.** Nessuna stella delimita esattamente la quinta ora intorno al centro della Vergine, perché sia l'Annunciatore della vendemmia, che la stella luminosa [δ] posta sulla spalla destra della Vergine, sono ad est [del circolo] più di 1/10 di ora.
**6.** Determina il più esattamente possibile la sesta ora sul coluro equinoziale la stella del Centauro [φ] contigua a quelle brillanti nella parte più meridionale del tirso, distanti circa mezzo cubito e che stanno pressoché nel centro del petto del Centauro; la centrale delle stelle [τ] nel piede sinistro del Bovaro resta circa 1/20 di ora ad est del circolo che delimita le sei ore.

327. Vedi nota 317.
328. In termini di coordinate.

**7.** Dei sei intervalli orari fra il punto equinoziale d'autunno e il tropico invernale, determinano abbastanza esattamente il primo intervallo orario, sul circolo intorno al centro delle Chele, la spalla sinistra [Seginus] del Bovaro, e delle Chele la stella luminosa [Zubenelgenubi] nell'estremità della chela meridionale, e più esattamente la spalla è più ad est di un nonnulla, la chela è circa $1/30$ di ora ad ovest del suddetto circolo tra i poli.
**8.** Determina il secondo intervallo orario, intorno all'inizio dello Scorpione, la stella più settentrionale [Unukalhai] delle due brillanti che giacciono nel corpo del Serpente; la stella della Corona ad ovest di quella brillante [Nusakan] resta circa $1/30$ di ora ad est del circolo [che passa] attraverso i poli, similmente anche la centrale [Dschubba] delle tre stelle luminose nella fronte dello Scorpione.
**9.** Determinano la terza ora, intorno al centro dello Scorpione, la stella dell'Inginocchiato [Kornephoros] posta sulla spalla destra e quella [φ] al centro della sua gamba destra, che è spostata circa $1/30$ di ora ad est del circolo.
**10.** Determinano la quarta ora, intorno all'inizio del Sagittario, la stella brillante [π] nella coscia sinistra dell'Inginocchiato e, della Piccola Orsa, le stelle più brillanti ed occidentali (Kochab e Pherkad] delle quattro nel quadrilatero, e la più settentrionale [Cebalrai] delle stelle nella spalla destra del Serpentario.
**11.** Determina la quinta ora, intorno al centro del Sagittario, la stella orientale [π] delle tre nella testa del Sagittario, spostata di circa $1/30$ di ora ad est del circolo.
**12.** Determina il più esattamente possibile la sesta ora, sul circolo [che passa] attraverso i punti tropicali, la stella più settentrionale [Tarazed] delle tre luminose nel corpo dell'Aquila, che è circa $1/20$ di ora ad est del circolo.
**13.** Dei sei intervalli orari fra il punto del tropico d'inverno e l'equinozio di primavera, determinano il primo intervallo orario, sul circolo intorno al centro del Capricorno, la più settentrionale [Sualocin] delle stelle occidentali nel rombo del Delfino e la occidentale delle stelle [θ] nel dorso del Capricorno.
**14.** Determinano il secondo intervallo orario, intorno all'inizio dell'Aquario, la stella brillante [Enif] nella bocca del Cavallo, e la più meridionale delle stelle [η] ad ovest della spalla destra di Cefeo.
**15.** Determina il terzo intervallo orario, intorno al centro dell'Aquario, la centrale [ζ] delle tre stelle, poste in linea retta, nella testa di Cefeo, e la più settentrionale delle due stelle [ξ] nel collo del Cavallo, che è $1/30$ di ora ad ovest del cerchio [che passa] attraverso i poli.
**16.** Determina il quarto intervallo orario, intorno all'inizio dei Pesci, la più meridionale delle stelle [ι] del braccio destro di Andromeda, che è ad est del circolo [che passa] attraverso i poli di circa $1/20$ di ora.
**17.** Determina il quinto intervallo orario, intorno al centro dei Pesci, la stella al centro del corpo di Cassiopea [η] e terza contando dalla testa.
**18.** Determinano il sesto intervallo orario, proprio sul circolo tracciato fra i punti equinoziali e i poli, la stella [Mothallah] posta sul vertice del Triangolo, che giace sopra l'Ariete, e la stella occidentale delle tre brillanti [Mesarthim] nella testa dell'Ariete, che è circa $1/20$ di ora ad est del circolo.
**19.** Dei sei intervalli orari fra il punto equinoziale di primavera e il tropico d'estate, determina il primo intervallo orario, sul circolo presso il centro dell'Ariete, la più brillante delle stelle [Algol] nella testa della Gorgone, che Perseo tiene nella mano sinistra.

**20.** Determina il secondo intervallo orario, intorno all'inizio del Toro, la stella orientale [53 Per] delle tre in linea retta, delle cinque intorno al ginocchio destro di Perseo.
**21.** Determinano il terzo intervallo orario, intorno al centro del Toro, la quarta [$\pi^1$] e la settima [$\pi^4$] delle stelle nella pelle che Orione tiene nella mano sinistra, e la stella ben visibile [ι] al centro delle corna del Toro, che forma quasi un triangolo equilatero con le stelle brillanti nelle estremità delle corna.
**22.** Determina il quarto intervallo orario, intorno all'inizio dei Gemelli, la orientale [$\chi^2$ Ori] delle due stelle, piccole ma ben visibili, poste in linea retta con la punta del corno destro del Toro verso est; queste stanno anche nella mazza di Orione, sul bordo settentrionale. Determina in modo quasi preciso la quarta ora anche la più meridionale [Nihal] delle due stelle brillanti al centro del corpo della Lepre.
**23.** Determinano il quinto intervallo orario, intorno al centro dei Gemelli, la centrale [Mekbuda] delle tre stelle brillanti nelle ginocchia dei Gemelli, che è ad ovest del circolo [che passa] attraverso i poli di circa $1/30$ di ora, e la stella brillante [Furud] nelle zampe posteriori del Cane.